\documentclass[twocolumn,english,superscriptaddress]{revtex4-2}
\usepackage[T1]{fontenc}
\usepackage{babel}
\usepackage{dsfont}
\usepackage{amsmath}
\usepackage{amssymb}
\usepackage{graphicx}
\usepackage{setspace}
\usepackage{esint}
\usepackage[unicode=true]
 {hyperref}

\makeatletter
\renewcommand{\fnum@figure}{FIG.~\thefigure}
\usepackage[caption=false]{subfig}
\usepackage{hyperref}

\makeatother

\begin{document}
\title{A non-Hermitian optical atomic mirror}
\author{Yi-Cheng Wang}
\email{r09222006@ntu.edu.tw}

\affiliation{Department of Physics, National Taiwan University, Taipei 10617, Taiwan}
\affiliation{Institute of Atomic and Molecular Sciences, Academia Sinica, Taipei
10617, Taiwan}
\author{Jhih-Shih You}
\email{jhihshihyou@ntnu.edu.tw}

\affiliation{Department of Physics, National Taiwan Normal University, Taipei 11677,
Taiwan}
\author{H. H. Jen}
\email{sappyjen@gmail.com}

\affiliation{Institute of Atomic and Molecular Sciences, Academia Sinica, Taipei
10617, Taiwan}
\date{\today}
\begin{abstract}
Explorations of symmetry and topology have led to important breakthroughs
in quantum optics, but much richer behaviors arise from the non-Hermitian
nature of light-matter interactions. A high-reflectivity, non-Hermitian
optical mirror can be realized by a two-dimensional subwavelength
array of neutral atoms near the cooperative resonance associated with
the collective dipole modes. Here we show that exceptional points
develop from a nondefective degeneracy by lowering the crystal symmetry
of a square atomic lattice, and dispersive bulk Fermi arcs that originate
from exceptional points are truncated by the light cone. We also find,
although the dipole-dipole interaction is reciprocal, the geometry-dependent
non-Hermitian skin effect emerges. Furthermore, skin modes localized
at a boundary show a scale-free behavior that stems from the long-range
interaction and whose mechanism goes beyond the framework of non-Bloch
band theory. Our work opens the door to the study of the interplay
among non-Hermiticity, topology, and long-range interaction.
\end{abstract}
\maketitle
The exquisite control of light-matter interactions is centrally important
to construct new quantum optical setups and attain new functionalities.
Recent research has shown that under the control of the cooperative
response of dipole modes, an atomic array with subwavelength spacings can
be characterized as a high-reflectivity optical atomic mirror~\cite{Shahmoon2017,Rui2020}.
A number of interesting predictions for this optical mirror include enhanced
photon storage~\cite{Asenjo-Garcia2017}, topological quantum optics~\cite{Perczel2017,Bettles2017},
and quantum information processing~\cite{Bekenstein2020}. These
phenomena can be identified from the band structures of collective
atomic excitations, where the quasimomenta modes inside and outside
the light cone exhibit distinct behaviors, respectively~\cite{Shahmoon2017,Asenjo-Garcia2017}.
Specifically, the former delineate non-Hermitian physics. Due to the
intrinsically loss processes associated with free-space emission,
the atomic array with photon-mediated dipole-dipole interactions opens
the door to the observation of a wide range of outstanding non-Hermitian
phenomena that would be challenging in condensed matter.

Recent progress in non-Hermitian physics~\cite{Ashida2020,El-Ganainy2018}
reveals two phenomena that have no Hermitian counterparts. One is
the exceptional points~\cite{Moiseyev2011}~(EPs), at which both
the complex eigenvalues and eigenstates of a non-Hermitian matrix
coalesce. This results in Riemann surface topologies of bulk bands
that have been demonstrated in diverse physical systems, including
photonic crystals~\cite{Zhen2015,Zhou2018,Cerjan2019}, topolectrical
circuits~\cite{Hofmann2020}, and exciton-polariton systems~\cite{Gao2018,Su2021}.
This relates to the nontrivial winding in the complex energy plane
that induces the other intriguing non-Hermitian phenomenon---non-Hermitian
skin effect~\cite{Borgnia2020,Okuma2020,Zhang2020}.

The non-Hermitian skin effect~\cite{Yao2018,MartinezAlvarez2018,Kunst2018}~(NHSE)
means that an extensive number of exponentially localized eigenstates
can pile at the boundaries under open boundary conditions~(OBCs).
This indicates the breakdown of the conventional bulk-boundary correspondence~\cite{Kunst2018},
which has triggered an avalanche of research aimed at reestablishing
the correspondence in non-Hermitian systems~\cite{Bergholtz2021}.
Various complementary approaches have been proposed in this vein,
including the celebrated non-Bloch band theory~\cite{Yao2018,Yokomizo2019,Kawabata2020}.
This band theory successfully interprets NHSE and the exponential
localization of skin modes in one-dimensional~(1D) systems with finite-range
couplings. Yet, little is known about the interplay between the long-range
interaction and these non-Hermitian phenomena.

\begin{figure*}[!tp]
\centering{}\includegraphics{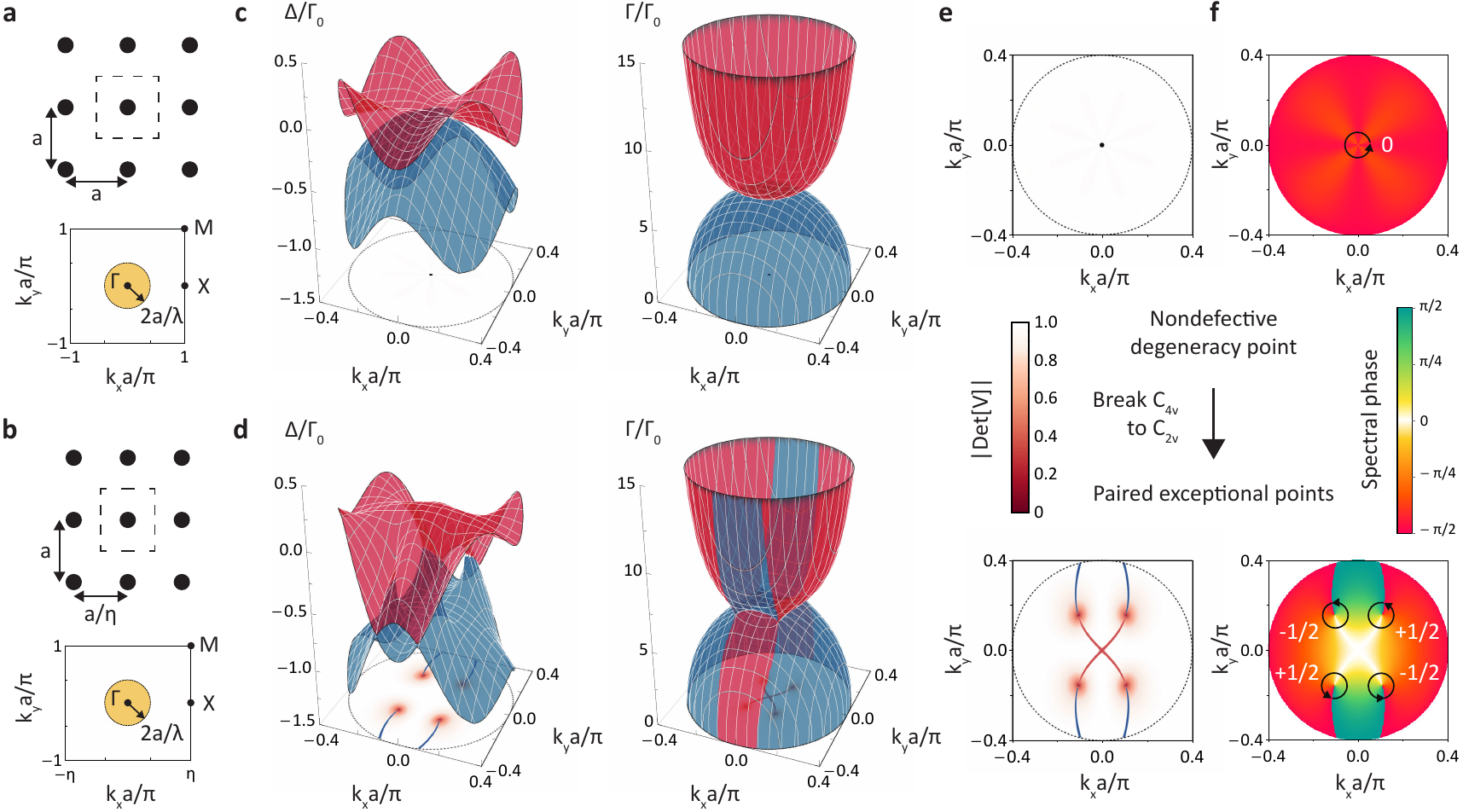}\caption{\label{fig:1}\textbf{Paired exceptional points split from a nondefective
degeneracy point.} \textbf{a},\textbf{b},~Schematics of square~(\textbf{a})
and rectangular~(\textbf{b}) atomic lattices and their irreducible
Brillouin zones in the subwavelength regime. The yellow circular region
represents the light cone, wherein the system is non-Hermitian. \textbf{c},\textbf{d},~Collective
frequency shift $\Delta_{\mathbf{k}}$~(left) and overall decay rate
$\Gamma_{\mathbf{k}}$~(right) of infinite square~(\textbf{c}) and
rectangular~(\textbf{d}) lattices within the light cone~(black dashed
circle). Two energy bands $E_{1,2}(\mathbf{k})=\hbar(\omega_{0}+\Delta_{\mathbf{k}})-\frac{i}{2}\hbar\Gamma_{\mathbf{k}}$
are colored in red and blue, respectively. \textbf{e},~A non-Hermitian
degeneracy point can be identified as NDP or EP by calculating $\text{det}[V(\mathbf{k})]$.
The upper panel shows that the non-Hermitian degeneracy point at the
high symmetry point $\Gamma$ in \textbf{c} is an NDP, and the lower
panel shows that four non-Hermitian degeneracy points in \textbf{d}
corresponding to the coalescence of two eigenstates~(i.e., $\text{det}[V(\mathbf{k})]$
approaches zero) are EPs. These EPs are joined by the degeneracy of
the real and imaginary parts of $E_{1,2}(\mathbf{k})$ in \textbf{d},
known as real~(blue) and imaginary~(red) Fermi arcs in the $k_{x}$-$k_{y}$
plane. \textbf{f},~The spectral phase that reflects the winding of
bulk energy bands. The vorticity of NDP in the upper panel is zero,
while that of each EP in the lower panel is a half-integer. The plots
are obtained with a subwavelength lattice constant $a=0.2\lambda$
and $\eta=1.1$ for rectangular lattice.}
\end{figure*}

In this work, we consider two-dimensional~(2D) atomic lattices with
resonant dipole-dipole interactions~\cite{Lehmberg1970,Novotny2006}~(RDDIs),
which serve as a high-reflectivity optical mirror at cooperative resonance~\cite{Shahmoon2017,Rui2020}.
To efficiently calculate the bulk band structures of infinite 2D atomic
lattices, we develop a model-independent generalization of Euler-Maclaurin
formula~\cite{Abramowitz1972}~(see Methods). Our numerical method
is applicable to systems with different types of long-range interactions.
Due to the inherent non-Hermiticity of dipole-dipole interaction,
there is no Hermitian limit of our system, such that a scenario in
the previous studies~\cite{Zhen2015,Zhou2018,Cerjan2019,Hofmann2020,Gao2018,Su2021},
i.e., EPs can be split from the Hermitian degeneracies such as Dirac
or Weyl points, does not work in this case. Here we demonstrate that
paired EPs can be split from a symmetry-protected nondefective degeneracy
point~(NDP) by a symmetry-breaking perturbation. We find that a ribbon
geometry exhibits extensive geometry-dependent skin modes. In particular,
these modes show a scale-free behavior that stems from the long-range
interaction and the mechanism responsible for this behavior goes beyond
the framework of non-Bloch band theory. Furthermore, we show that
the skin modes can emerge in 2D finite atomic arrays by manipulating
the orientations of open boundaries and the lattice configurations.
Possible experimental observations are also discussed.\\
\\
\textbf{Non-Hermitian degeneracy points.} We consider a 2D rectangular
atomic lattice spanned by two direct lattice vectors $\mathbf{a}_{1}=a/\eta\mathbf{e}_{x}$
and $\mathbf{a}_{2}=a\mathbf{e}_{y}$ with a lattice constant ratio
$|\mathbf{a}_{2}|/|\mathbf{a}_{1}|=\eta$ in free space~(FIG.~\ref{fig:1}a,b).
Each atom has a V-type energy level composed of one ground state $\left|g\right\rangle $
and two circularly-polarized excited states $\left|\pm\right\rangle =\mp(\left|x\right\rangle \pm i\left|y\right\rangle )/\sqrt{2}$
such that the system supports two in-plane polarizations with an atomic
transition wavelength $\lambda$ and decay rate $\Gamma_{0}$. In
the circularly-polarized basis, the non-Hermitian dynamics of a single
excitation is described by the following two-band effective Hamiltonian
kernel~(see Supplementary Information)

\begin{singlespace}
\begin{align}
\mathcal{H}{}_{\text{eff}}(\mathbf{k}) & =\hbar(\omega_{0}+\Omega_{\mathbf{k}})\begin{pmatrix}1 & 0\\
0 & 1
\end{pmatrix}+\hbar\Gamma_{0}\begin{pmatrix}0 & \kappa_{+-}(\mathbf{k})\\
\kappa_{-+}(\mathbf{k}) & 0
\end{pmatrix},\nonumber \\
\label{eq:1}
\end{align}
where $\mathbf{k}$ is the Bloch momentum in the irreducible Brillouin
zone and $\omega_{0}=2\pi c/\lambda$ is the atomic transition frequency.
Here Eq.~(\ref{eq:1}) has the identical momentum-dependent interacting
energy $\hbar\Omega_{\mathbf{k}}$ for $\left|\pm\right\rangle $,
and $\kappa_{+-(-+)}(\mathbf{k})$ describes the couplings between
two circularly-polarized states. As a result, the 2D Brillouin zone
exhibits two kinds of distinct collective excitations separated by
the light cone $|\mathbf{k}|=2\pi/\lambda$~\cite{Shahmoon2017}~(FIG.~\ref{fig:1}a,b),
wherein the dissipative modes couple to far-field radiation, while
the modes with $|\mathbf{k}|>2\pi/\lambda$ related to evanescent
wave confined to the atomic lattice plane are dissipationless. Here,
we focus on the non-Hermitian physics within the light cone.
\end{singlespace}

The bulk eigenenergy spectrum can be obtained as $E_{1,2}(\mathbf{k})=\hbar(\omega_{0}+\Omega_{\mathbf{k}})\pm\hbar\Gamma_{0}\sqrt{\kappa_{+-}(\mathbf{k})\kappa_{-+}(\mathbf{k})}$,
and the existence of degeneracy corresponds to $\kappa_{+-}(\mathbf{k})\kappa_{-+}(\mathbf{k})=0$.
If $\kappa_{+-}(\mathbf{k})$ and $\kappa_{-+}(\mathbf{k})$ are simultaneously
zero, the corresponding eigenstates are linearly independent such
that $\mathcal{H}{}_{\text{eff}}(\mathbf{k})$ is diagonalizable,
and these degeneracies are called nondefective degeneracy points.
If only one of $\kappa_{+-(-+)}(\mathbf{k})$ is zero, $\mathcal{H}{}_{\text{eff}}(\mathbf{k})$
is nondiagonalizable, and the defective degeneracies whose eigenstates
coalesce are known as exceptional points. Therefore, for a square
lattice, $\kappa_{+-(-+)}(\mathbf{k})$ vanish at high symmetry point
$\Gamma$~($\mathbf{k}=0$) due to the $C_{4}$ rotational symmetry~(see
Supplementary Information), which ensures a symmetry-protected NDP.
Accordingly, EPs can emerge by breaking this symmetry, and we note
that Ref.~\cite{Yang2021} has proved that NDP in a two-band system
is unstable, i.e., it can be deformed into EPs by a generic perturbation.

We perform a 2D generalization of the Euler-Maclaurin formula to determine
the photonic band structures of infinite square~(FIG.~\ref{fig:1}c)
and rectangular~(FIG.~\ref{fig:1}d) lattices with in-plane polarizations
in free space. Here we diagonalize the effective Hamiltonian as $\mathcal{H}{}_{\text{eff}}(\mathbf{k})=V(\mathbf{k})E(\mathbf{k})V^{-1}(\mathbf{k})$,
where $E(\mathbf{k})$ is the diagonal matrix composed of its eigenvalues
$E_{1,2}(\mathbf{k})$ and $V(\mathbf{k})$ is formed by two normalized
right eigenstates. The coalescence of eigenstates happens when $\text{det}[V(\mathbf{k})]$
approaches zero in the vicinity of EPs. The degeneracy point at $\Gamma$
in FIG.~\ref{fig:1}c is identified as a NDP due to the corresponding
nonzero $\text{det}[V(\mathbf{k})]$ in FIG.~\ref{fig:1}e. In FIG.~\ref{fig:1}d,
we break the $C_{4}$ rotational symmetry by tuning the lattice constant
ratio~($\eta\neq1$), and we find that four EPs are split from the
NDP.

In the lower panels of FIG.~\ref{fig:1}e, we observe that these
four EPs are joined by dispersive bulk Fermi arcs~\cite{Zhou2018},
along which the real parts of two eigenenergies are degenerate. We
note that our bulk Fermi arcs do not lie on the isofrequency surface
due to the momentum dependent $\hbar\Omega_{\mathbf{k}}$. These open-end
bulk Fermi arcs usually terminate at EPs; however, they are truncated
by the light cone in our case. Importantly, these dispersive bulk
Fermi arcs and EPs are topologically stable and associated with a
non-Hermitian topological invariant called the vorticity~\cite{Leykam2017,Gong2018,Shen2018}
\begin{equation}
v=-\oint_{C}\frac{d\mathbf{k}}{2\pi}\cdot\nabla_{\mathbf{k}}\text{arg}[E_{1}(\mathbf{k})-E_{2}(\mathbf{k})],\label{eq:2}
\end{equation}
where $C$ is a counterclockwise closed loop that encloses a degeneracy
point. The vorticity can be determined by the spectral phase $\text{arg}[E_{1}(\mathbf{k})-E_{2}(\mathbf{k})]$
that acquires a $\pm\pi$ change around each EP but $0$ around NDP.
In FIG.~\ref{fig:1}f, we find that the vorticity of NDP~(EPs) in
a square~(rectangular) lattice is zero~(half-integer). We also check
the stabilities of these non-Hermitian degeneracy points via the Zeeman
splitting arising from a magnetic field. We find that EPs persist,
in stark contrast to the symmetry-protected NDP and Dirac point~(see
Supplementary Information).\\
\\
\textbf{Non-Hermitian skin effect.} Recently, it was revealed that
a 1D system under OBC exhibits NHSE as long as the energy spectrum
of the corresponding bulk Hamiltonian $\mathcal{H}_{\text{1D}}(k)$
encloses a nonzero spectral area in the complex energy plane. If a
system has the reciprocity and mirror symmetry, $\mathcal{H}_{\text{1D}}(k)=\mathcal{H}_{\text{1D}}^{T}(-k)$,
the bulk spectrum forms a doubly degenerate spectral arc. As a result,
there is no NHSE under OBC for a 1D system with reciprocal couplings,
but NHSE can emerge by invoking nonreciprocal couplings to break such
a constraint on $\mathcal{H}_{\text{1D}}(k)$. Since RDDI is reciprocal,
a 1D atomic chain with RDDI does not exhibit NHSE. Therefore we turn to a 2D atomic lattice with reciprocal and anisotropic
RDDIs to investigate NHSE.
\begin{figure*}[!tp]
\centering{}\includegraphics{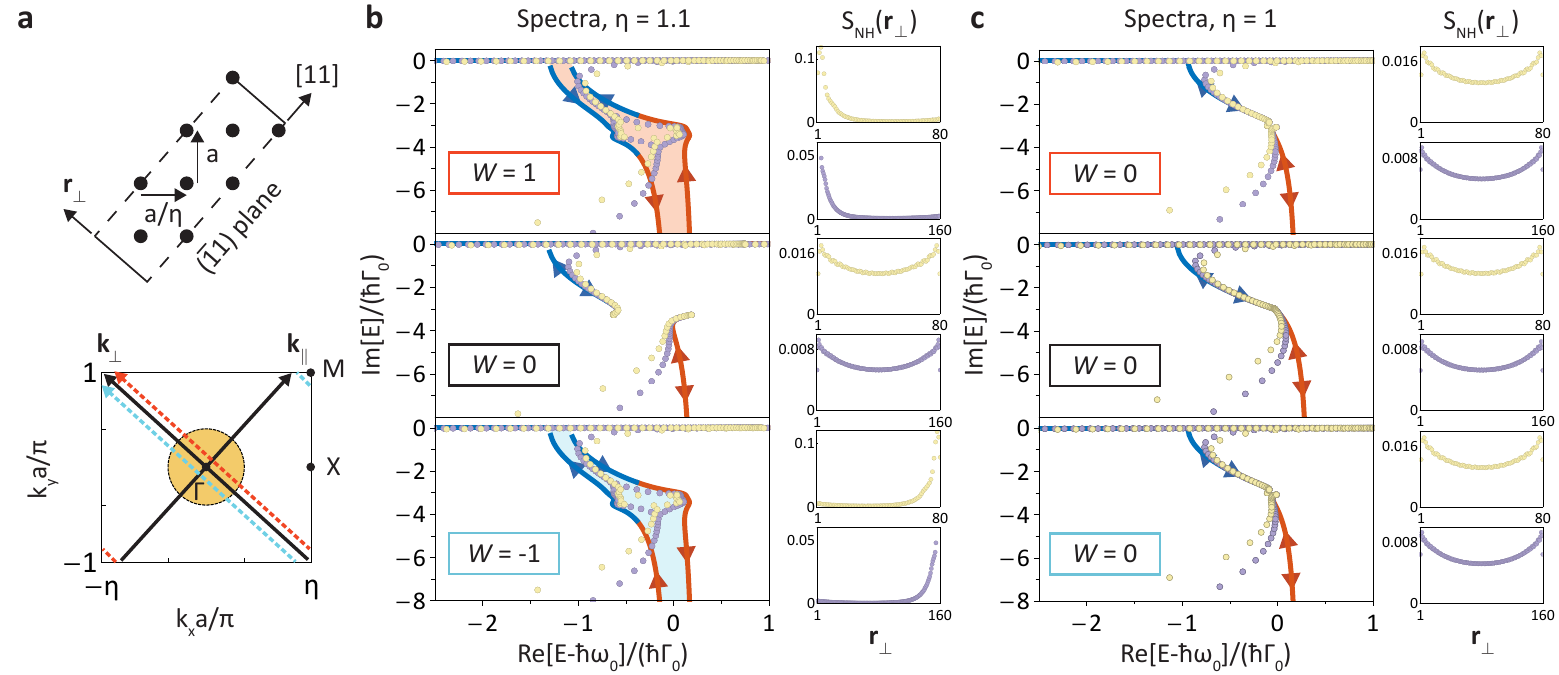}\caption{\label{fig:2}\textbf{Geometry-dependent non-Hermitian skin effect
in a ribbon geometry.} \textbf{a},~Illustration of the ribbon geometry
for a rectangular atomic lattice. The boundaries are open on the~$(\bar{1}1)$
plane~(dashed line) and extend infinitely in the~$[11]$ direction~(solid
line). \textbf{b},\textbf{c},~Open boundary eigenenergy spectra $\sigma[\mathcal{H}_{L}(\mathbf{k}_{\parallel},\mathbf{r}_{\perp})]$
of rectangular~(\textbf{b}, $\eta=1.1$) and square~(\textbf{c},
$\eta=1$) lattices in ribbon geometries with a width of $80$ and
$160$ unit cells~(light yellow dots and purple dots, respectively)
and the corresponding bulk spectra $\sigma[\mathcal{H}_{\text{eff}}(\mathbf{k}_{\parallel},\mathbf{k}_{\perp})]=\{E_{1,2}(\mathbf{k}_{\parallel},\mathbf{k}_{\perp})\}$~(curves
in red and blue) at fixed $\mathbf{k}_{\parallel}=0.1\pi/a$~(top
panels, orange dashed line in \textbf{a}) and $\mathbf{k}_{\parallel}=0$~(middle
panels, $\mathbf{k}_{\perp}$ axis in \textbf{a}) and $\mathbf{k}_{\parallel}=-0.1\pi/a$~(bottom
panels, cyan dashed line in \textbf{a}). The non-Hermitian parts of
$\sigma[\mathcal{H}_{\text{eff}}(\mathbf{k}_{\parallel},\mathbf{k}_{\perp})]$
result in the nontrivial winding~(top and bottom panels in \textbf{b}),
and the corresponding spatial distributions~(right columns) demonstrate
that there are extensive skin modes localized at the edge normal to
$\mathbf{r}_{\perp}$ axis in a ribbon geometry. Otherwise, inversion
and mirror symmetries lead to doubly degenerate spectral arcs with
zero winding numbers, which suppress the NHSE. The plots are obtained
with the same parameters in FIG.~\ref{fig:1}.}
\end{figure*}

NHSE in 1D systems has a clear picture, while a general
description of NHSE in two and higher-dimensional systems remains
unclear. To study NHSE in 2D atomic lattices, we start by considering a ribbon
geometry, which extends infinitely along the parallel direction~(the~$[11]$
direction in FIG.~\ref{fig:2}a) and has OBC with $L$ unit cells
in the perpendicular direction. Due to the translational symmetry
in the parallel direction, the Bloch momentum $\mathbf{k}_{\parallel}$
is a good quantum number, such that the Hamiltonian of the ribbon
geometry at a fixed $\mathbf{k}_{\parallel}$ reduces to the Hamiltonian
kernel $\mathcal{H}_{L}(\mathbf{k}_{\parallel},\mathbf{r}_{\perp})$
of an effective 1D finite lattice under OBC~(over the~$(\bar{1}1)$
plane in FIG.~\ref{fig:2}a). Consequently,  the
NHSE can be understood in a simple effective 1D picture.

In FIG.~\ref{fig:2}b,c, we numerically calculate the respective
OBC spectra $\sigma[\mathcal{H}_{L}(\mathbf{k}_{\parallel},\mathbf{r}_{\perp})]$
of rectangular and square lattices in ribbon geometries with a width
of $L$ unit cells at $\mathbf{k}_{\parallel}=\pm0.1\pi/a$ and $\mathbf{k}_{\parallel}=0$.
For these given $\mathbf{k}_{\parallel}$, the corresponding bulk spectra
$\sigma[\mathcal{H}_{\text{eff}}(\mathbf{k}_{\parallel},\mathbf{k}_{\perp})]=\{E_{1,2}(\mathbf{k}_{\parallel},\mathbf{k}_{\perp})\}$
are shown for comparison and can be viewed as the spectra of infinite 1D lattices with the normal momentum $\mathbf{k}_{\perp}$ being a good quantum number. First we note a deviation between the
OBC spectra and the bulk spectra in the vicinity of the light cone:
the bulk spectra show a divergent behavior, which arises from the RDDI
in the infinite system, while the OBC spectra remain finite. 

The emergence of NHSE in ribbon geometries is manifest in the the bulk spectra of infinite 1D lattices.
 In the square lattice,
the bulk spectra become doubly degenerate spectral arcs due to the mirror symmetry~\cite{Kawabata2019}, indicating that
the OBC eigenstates are delocalized. In the rectangular lattice, we note that OBC eigenstates
can also be delocalized at $\mathbf{k}_{\parallel}=0$. This is due to the inversion symmetry at this special momentum. 
In the absence of these lattice symmetries, the bulk spectra enclose
nonzero spectral areas, as shown in the upper and lower panels of
FIG.~\ref{fig:2}b. This signifies the hallmark of geometry-dependent skin
modes.

To visualize NHSE, we consider the average spatial distribution of OBC
normalized right eigenstates $\bigl|\psi_{n}^{R}(\mathbf{r}_{\perp})\bigr\rangle$
\begin{equation}
S_{\text{NH}}(\mathbf{r}_{\perp})=\frac{1}{N'}\sum_{n=1}^{N'}\sum_{j=\pm}\left|\psi_{nj}^{R}(\mathbf{r}_{\perp})\right|^{2},\label{eq:3}
\end{equation}
where $\pm$ represents the in-plane polarization and $n$ is in the
ascending order of the imaginary parts of eigenenergies. Here we only
consider first $N'$ right eigenstates $\bigl|\psi_{n}^{R}(\mathbf{r}_{\perp})\bigr\rangle$
with largest decay rates. In FIG.~\ref{fig:2}b,c, $N'/2L$ corresponds
to the fraction of bulk eigenmodes within the light cone at a given
$\mathbf{k}_{\parallel}$ and we show the average spatial distribution $S_{\text{NH}}(\mathbf{r}_{\perp})$ for both $L=80$ and $L=160.$ It is evident that  for a rectangular lattice in a ribbon geometry with a positive~(negative) $\mathbf{k}_{\parallel},$ $S_{\text{NH}}(\mathbf{r}_{\perp})$ shows the emergence of extensive skin modes localized at the $\mathbf{r}_{\perp}=1$~($\mathbf{r}_{\perp}=L$)
boundary. For other cases, the OBC eigenstates are
delocalized.

As a topological phenomenon, the emergence of extensive skin modes
is predicted by the point-gap topology of the bulk bands~\cite{Okuma2020}.
For a fixed $\mathbf{k}_{\parallel}$ and a reference point $E_{r}$
that is not covered by the bulk spectra~(point-gap) in the complex
energy plane, the integer-valued spectral winding number can be defined
as~\cite{Okuma2020,Yang2021}

\begin{equation}
W(\mathbf{k}_{\parallel},E_{r})=\oint_{C_{\perp}}\frac{d\mathbf{k}_{\perp}}{2\pi i}\cdot\nabla_{\mathbf{k}_{\perp}}\log\det[\mathcal{H}_{\text{eff}}(\mathbf{k}_{\parallel},\mathbf{k}_{\perp})-E_{r}\mathds{1}],\label{eq:4}
\end{equation}
where $C_{\perp}$ forms a closed loop at a fixed $\mathbf{k}_{\parallel}$
in the 2D Brillouin zone~(e.g., three paths in FIG.~\ref{fig:2}a).
When the bulk spectra enclose a nonzero spectral area in the complex
energy plane, $W(\mathbf{k}_{\parallel},E_{r})$ for $E_{r}$ within
the spectral area is nonzero, where we denote the corresponding spectral
winding numbers by the shaded regions in FIG.~\ref{fig:2}b~($+1$
for orange and $-1$ for cyan shaded regions, respectively). Accordingly, the
skin modes that lie within the interiors of bulk spectra localize
at the left or right boundaries when $W(\mathbf{k}_{\parallel},E_{r})=\pm1$,
which coincides with other systems with finite-range couplings~\cite{Okuma2020}
even if we have long-range interactions here. In addition, by tuning
the lattice constant ratio $\eta$ in the top panels of FIG.~\ref{fig:2}b,c,
a point-gap opening in the ribbon geometry at nonzero $\mathbf{k}_{\parallel}$
represents a non-Hermitian topological phase transition~\cite{Gong2018}.

We note that, although the divergent bulk spectrum shows up near the
light cone in FIG.~\ref{fig:2}b,c, the corresponding spectral winding
numbers are quantized~(see Supplementary Information) and can be
applied to characterize the point-gap topology. In addition, this
NHSE is sensitive to the orientation of the open boundary. For instance, both
the square and rectangular lattices can support NHSE when the OBC is imposed
on the~$(\bar{1}2)$ plane in a ribbon geometry since there is no
mirror symmetry along the~$[21]$ direction. In general, the effective
coupling in a ribbon geometry at a fixed $\mathbf{k}_{\parallel}$
is nonreciprocal unless the original coupling is isotropic or there
is a mirror symmetry over the open boundary~\cite{Scheibner2020}.
Therefore, a so-called geometry-dependent NHSE~\cite{Zhang2021}
can emerge when the orientation of the open boundary in two and higher
dimensional systems is modified. We emphasize that non-Hermiticity and
anisotropy are the essential ingredients for such a geometry-dependent
nonreciprocal coupling.\\
\\
\textbf{Scale-free localization.} In addition to the deviation between
the OBC and bulk spectra, we observe the scale-free behavior of spatial
distributions $LS_{\text{NH}}(\mathbf{r}_{\perp},L)=L'S_{\text{NH}}(\mathbf{r}_{\perp},L')$
for system size $L=80$ and $L'=160$ in FIG.~\ref{fig:2}b,c. This
implies that the number of skin modes in a ribbon geometry and their
characteristic length $\xi$ of the exponentially decreasing probability
$\left|\psi_{nj}^{R}(\mathbf{r}_{\perp})\right|^{2}\sim e^{-|\mathbf{r}_{\perp}|/\xi}$
are proportional to the system size $L$. We note that similar scale-free
behavior arises in critical NHSE~\cite{Li2020,Yokomizo2021} in the
systems with finite-range couplings. However, the scale-free localization
in the atomic array stems from the long-range interaction and the
mechanism underpinning this behavior goes beyond the framework of
the non-Bloch band theory.
\begin{figure}[!t]
\centering{}\includegraphics{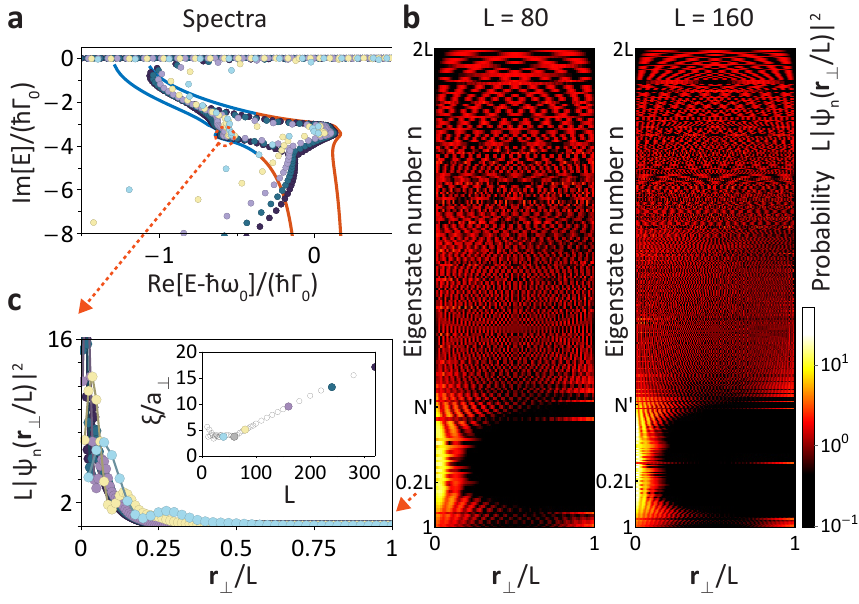}\caption{\label{fig:3}\textbf{Size dependence of non-Hermitian skin effect
in a ribbon geometry.} \textbf{a},~The OBC spectra at $L=40$, $80$,
$160$, $240$, and $320$ unit cells~(light blue, light yellow,
purple, dark green, and black, respectively) gradually approach bulk
spectrum as $L$ increases. \textbf{b},~Rescaled probability distributions
of normalized right eigenstates at $80$ and $160$ unit cells in
the ascending order of the imaginary part of eigenenergy. $N'$ represents
the number of localized modes in spatial distributions in FIG.~\ref{fig:2}b,c.
\textbf{c},~Rescaled probability distributions of the $n=0.2L$ eigenstates.
The inset: a crossover from a constant to scale-free characteristic
length as system size increases, and $a_{\perp}$ is the lattice constant
for this ribbon geometry in the $\mathbf{r}_{\perp}$ direction. The
plots are obtained with the same parameters in the top panel of FIG.~\ref{fig:2}b.}
\end{figure}

Here we briefly introduce the non-Bloch band theory in 1D systems
with finite-range couplings. By considering the analytic continuation
of Bloch momentum $k\rightarrow k+i\kappa(k)$~\cite{Yao2018,Yokomizo2019,Lee2019},
the OBC energy spectrum~$\{E_{\text{OBC}}\}$ in the thermodynamic
limit~$(L\gg1)$ can be obtained from the non-Bloch Hamiltonian $\mathcal{H}_{\text{1D}}(k+i\kappa(k))$.
In addition, each OBC eigenstate of a 1D chain with lattice constant
$a$, $\psi_{L\rightarrow\infty}(ma)$ at $m$th site, can be decomposed
into $M$ possible $\beta_{i}^{m}=e^{i(k_{i}+i\kappa(k_{i}))\cdot ma}$
and is dominated by two non-Bloch modes $\beta_{r,s}^{m}$ with the
same modulus $|\beta|=|\beta_{r}|=|\beta_{s}|$ that corresponds to
the decay length $-a/\text{log}|\beta|$. These $\beta_{i}$ are solutions
to the following characteristic equation~(see Supplementary Information)

\begin{equation}
\det[\mathcal{H}_{\text{1D}}(\beta)-E_{\text{OBC}}\mathds{1}]=0\label{eq:5}
\end{equation}
subject to OBCs, and $M$ is determined by the coupling range. These
results are derived from the asymptotic behavior of the OBC eigenstate
$\psi_{L}(ma)$ in the finite system for large $L$. We note that
the above statement is valid when the coupling range is finite. It
is because in this case the finite systems with different sizes are
governed by the \textquoteleft same\textquoteright{} Hamiltonian $\mathcal{H}_{\text{1D}}(\beta)$.
\begin{figure*}[!tp]
\centering{}\includegraphics{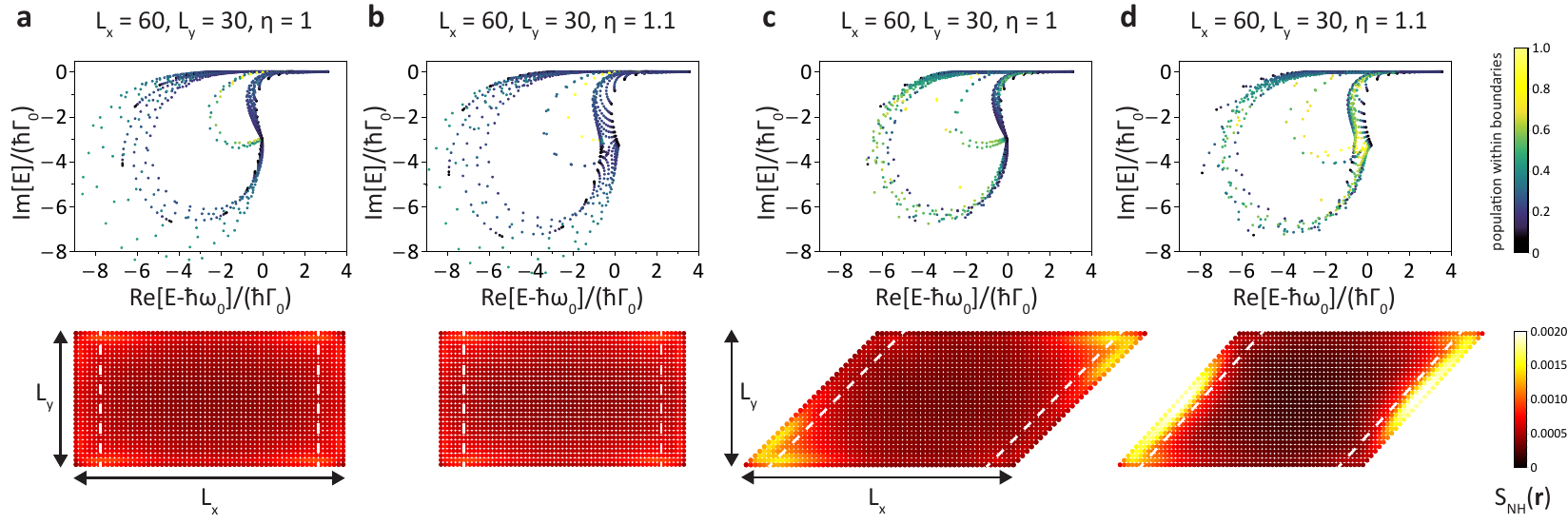}\caption{\label{fig:4}\textbf{Geometry-dependent non-Hermitian skin effect
in rectangle and parallelogram-shaped boundaries.} \textbf{a\protect\nobreakdash-d},~Spectra~(upper)
and spatial distribution of non-Hermitian right eigenstates~(lower)
of 2D square~(\textbf{a},\textbf{c}) and rectangular~(\textbf{b},\textbf{d})
atomic lattices~($a=0.2\lambda$) with rectangle~(\textbf{a},\textbf{b})
and parallelogram~(\textbf{c},\textbf{d}) shaped boundaries. At a
fixed system size $L_{x}\times L_{y}=60\times30,$ only the
geometry in \textbf{d} exhibits extensive skin modes localized at
left and right boundaries since mirror symmetries of the other geometries
suppress NHSE. Each mode in all top panels is colored according to
its population within the given boundaries~(out
of white dashed lines, leftmost and rightmost $6$ sites), and those extensive skin
modes in the top panel of \textbf{d} are located around the exceptional
points in the corresponding bulk spectra.}
\end{figure*}

In contrast, in the presence of long-range RDDI the finite systems
with different sizes are governed by \textquoteleft different\textquoteright{}
$\mathcal{H}_{L}(\beta)$, such that the characteristic equation becomes
size-dependent and has $M=4(L-1)$ solutions of $\beta$. As a result,
both the OBC spectrum~(FIG.~\ref{fig:3}a) and $\beta$ depend on
$L$, and each OBC eigenstate in the thermodynamic limit is dominated
by several $\beta$ with no fixed scale. To show the size dependence
of eigenstate, we numerically fit the characteristic length $\xi$
of the probability of each OBC eigenstate in FIG.~\ref{fig:3}c,
which presents the crossover from a constant to scale-free characteristic
length.\\
\\
\textbf{Non-Hermitian skin effect in 2D finite-size systems.} We further
explore NHSE in a 2D finite atomic array with parallelogram-shaped
boundaries by investigating the spatial distribution of skin modes~(Eq.~(\ref{eq:3})).
In FIG.~\ref{fig:4}, we choose $N'=[\pi(a/\lambda)^{2}/\eta]\times2L_{x}L_{y}$
that corresponds to the fraction of bulk eigenmodes within the light
cone in 2D Brillouin zone. In FIG.~\ref{fig:4}d, there are extensive
skin modes localized at the oblique~($[11]$ direction) but not at the
horizontal~($[10]$ direction) boundaries, and there is no skin mode
in FIG.~\ref{fig:4}a-c. To understand the NHSE here, we can extend
a finite system infinitely along one open boundary to reconstruct
a ribbon geometry. This allows us to determine which boundary extensive skin
modes are localized at by identifying its energy spectrum topology. For
instance, there are two possible ribbon geometries for the finite
system in FIG.~\ref{fig:4}d: one extends infinitely along the oblique
direction and the other extends infinitely along the horizontal direction.
The skin modes emerge in the former case, while the NHSE in the latter
case is suppressed by the mirror symmetry. Therefore, the population
distributions of non-Hermitian eigenstates rely on the orientations
of open boundaries and the lattice configurations.

In the upper panels of FIG.~\ref{fig:4}, the OBC spectra with the
same lattice configuration~(FIG.~\ref{fig:4}a,c for square lattice
and b,d for rectangular lattice) are insensitive to the orientations
of open boundaries. However, when the lattice configuration changes
from square to rectangular lattice, the $E_{\text{OBC}}$ around NDP
in FIG.~\ref{fig:4}a,c are deformed into $E_{\text{OBC}}$ around
EP in FIG.~\ref{fig:4}b,d. We further explore the relationship between
OBC and bulk spectra by the light scattering from this finite system~(see Supplementary Information). We only need two detunings
to extract the bulk Hamiltonian from the scattering matrices in the
finite system. This allows us to identify the bulk band structure
and its non-Hermitian degeneracy points from light scattering.

The frequency shift and the linewidth are comparable in the upper
panels of FIG.~\ref{fig:4}, which smears the optical response from
each skin mode and its anomalous transport behavior~\cite{Yi2020}.
However, when an incident light shines on a 2D atomic array that hosts
skin modes at oblique incidence, the dynamical property of induced
dipoles shows the asymmetric spatial distribution, which reflects
the nonreciprocal effective coupling induced by the nonzero in-plane
momentum of the incident light. This provides an experimental signature
of NHSE in 2D finite atomic arrays.

We have shown that a 2D atomic array with the inherent non-Hermiticity
and the anisotropy of RDDI presents nontrivial topologies that have
no Hermitian counterparts. Our results here can be manifested in a
general setting of coupled 2D quantum emitters~\cite{Solntsev2021}.
Besides these distinct non-Hermitian degeneracies and NHSE we uncover
here, the interplay between non-Hermiticity and topology~\cite{Bergholtz2021,Coulais2021,Lu2014,Ozawa2019}
may further generate topologically protected edge states robust to
not only Hermitian but also non-Hermitian defects. In addition, one
can couple the atomic arrays with different electromagnetic environments
to further modify the non-Hermiticity of RDDI and to enhance the optical
response of some selective skin modes. Our work paves the way towards
the exploration of exotic many-body states in two and higher-dimensional
systems and opens up new opportunities in manipulating topological
properties by tailoring long-range interactions in an atomic array.\\
\\
\textbf{Methods}\\
\textbf{2D generalization of Euler-Maclaurin formula.}\\
In the calculation of photonic band structures, we encounter an infinite
summation of RDDI $\sum_{\mathbf{R}\neq\mathbf{0}}e^{-i\mathbf{k}\cdot\mathbf{R}}\mathbf{G}_{0}(\mathbf{R})$
excluding the self-energy $\mathbf{G}_{0}(\mathbf{0})$, where $\mathbf{G}_{0}(\mathbf{R})$
is the free-space dyadic Green's function and the summation runs over
all direct lattice vectors except for $\mathbf{R}=\mathbf{0}$. Since
the oscillating terms result in the slow convergence, we can rewrite
this infinite summation in the reciprocal space as~$(\mathbf{a}_{1}\times\mathbf{a}_{2})^{-1}\sum_{\mathbf{G}}\mathbf{g}_{0}(\mathbf{G}+\mathbf{k})-\mathbf{G}_{0}(\mathbf{0})$,
where the summation runs over all reciprocal lattice vectors $\mathbf{G}$
and $\mathbf{g}_{0}(\mathbf{G}+\mathbf{k})$ is Green's function in
reciprocal space. Here we note that the self interaction $\mathbf{G}_{0}(\mathbf{0})$
is just the integral of $\mathbf{g}_{0}(\mathbf{G}+\mathbf{k})$,
i.e., $\mathbf{G}_{0}(\mathbf{0})=\int(2\pi)^{-2}d^{2}G\mathbf{g}_{0}(\mathbf{G}+\mathbf{k})$.
Therefore, an infinite summation of RDDI becomes the difference between
the summation $\sum_{\mathbf{G}}$ and the integration $\int(2\pi)^{-2}(\mathbf{a}_{1}\times\mathbf{a}_{2})d^{2}G$
of the same function~$(\mathbf{a}_{1}\times\mathbf{a}_{2})^{-1}\mathbf{g}_{0}(\mathbf{G}+\mathbf{k})$.
While both the summation and the integration are divergent due to
the self-interaction, their difference is physically meaningful and
convergent. Thus, we can compute this difference via the generalization
of Euler-Maclaurin formula since it is in the form of the difference
between the summation and the integration of the same function.

The original Euler-Maclaurin formula is for a single summation on
a 1D lattice. The difference between the summation and the integration
of the same function over any interval $I$ composed of unit cell
sandwiched between lattice sites $j$ and $j+1$ can be approximated
by the function and its higher-order derivatives on two ends of the
interval $\partial I$. Here we extend this idea to the double summation
on a 2D lattice, where a 1D interval $I$ with two ends becomes a
2D region $R$ whose boundary has a variety of shapes, such as a simply-connected
region whose boundary is a closed loop and a hollow region with the
inner and outer boundaries.

By means of Euler-Maclaurin formula, we can estimate this difference
over any region composed of the periodic unit cell in the reciprocal
space by the correction terms consisting of higher-order derivates
of~$(\mathbf{a}_{1}\times\mathbf{a}_{2})^{-1}\mathbf{g}_{0}(\mathbf{G}+\mathbf{k})$
at the boundaries. In avoid of the singularity arising from the light
cone, the periodically-tiled hollow region $R$ we use to perform
the Euler-Maclaurin formula should enclose the light cone in the reciprocal
space. When the outer boundary of $R$ extends infinitely, the correction
terms at this boundary become negligible due to the presence of ultraviolet
frequency cutoff. Thus, the rest correction terms only lie on the
inner boundary $\partial R$ of $R$, and the infinite summation reduces
to the finite summation and integration of~$(\mathbf{a}_{1}\times\mathbf{a}_{2})^{-1}\mathbf{g}_{0}(\mathbf{G}+\mathbf{k})$
over the finite region enclosed by $\partial R$ and several correction
terms on $\partial R$ in the reciprocal space. We note that this
method is applicable for the infinite summation due to other long-range
interactions. Further details of the explicit formula and application
are presented in the Supplementary Information.\textbf{}\\
\textbf{}\\
\textbf{Acknowledgments}\\
We thank Hui Liu for useful discussions.
Y.-C.W. and H.H.J. acknowledge support from the Ministry of Science
and Technology~(MOST), Taiwan, under the Grant No. MOST-109-2112-M-001-035-MY3. Y.-C.W. and J.-S.Y. are supported by the Ministry of Science and Technology, Taiwan~(Grant No. MOST-110-2112-M-003-008-MY3). J.-S.Y. and H.H.J. is also grateful for support from National Center for Theoretical
Sciences in Taiwan.\textbf{}\\
\textbf{}\\
\textbf{Author contributions}\\
Y.-C.W. conducted the analytical and numerical calculations. J.-S.Y. and H.H.J. conceived the idea and supervised the project. All
authors contributed to the writing of the manuscript.\textbf{}\\
\\

\end{document}


\title{Supplementary Information for ``A non-Hermitian optical atomic mirror''}
\author{Yi-Cheng Wang}
\email{r09222006@ntu.edu.tw}

\affiliation{Department of Physics, National Taiwan University, Taipei 10617, Taiwan}
\affiliation{Institute of Atomic and Molecular Sciences, Academia Sinica, Taipei
10617, Taiwan}
\author{Jhih-Shih You}
\email{jhihshihyou@gmail.com}

\affiliation{Department of Physics, National Taiwan Normal University, Taipei 11677,
Taiwan}
\author{H. H. Jen}
\email{sappyjen@gmail.com}

\affiliation{Institute of Atomic and Molecular Sciences, Academia Sinica, Taipei
10617, Taiwan}
\date{\today}
\begin{abstract}
The supplementary information is organized as follows. Section~1
reviews the general formalism of a two-dimensional atomic array in
free space. Section~2 presents the model-independent generalization
of Euler-Maclaurin formula we develop to calculate the infinite summation
in the bulk spectrum. Section~3 discusses the non-Hermitian degeneracy
points of a two-dimensional infinite atomic array. Section~4 discusses
atomic lattices in ribbon geometries with infinite and finite widths.
Finally, section~5 presents the optical responses of realistic finite
atomic arrays.
\end{abstract}
\maketitle

\section{Optical atomic mirror}

\subsubsection{General formalism}

We first consider a coupled singly-excited of two-dimensional~(2D)
atomic lattice in free space. Each atom is a four-level system formed
by one ground state $\left|g\right\rangle $ and three degenerate
excited states $\left|x\right\rangle $, $\left|y\right\rangle $,
and $\left|z\right\rangle $ with resonant frequency $\omega_{0}$
and decay rate $\Gamma_{0}$. Within the dipole and Markov approximations,
we eliminate the reservoir photons in light-matter interactions, and
the $N$-atom non-Hermitian effective Hamiltonian is given by~\cite{Perczel2017PRAS}
\begin{eqnarray}
\hat{{\cal H}}_{\text{eff}} & = & \sum_{m=1}^{N}\sum_{\nu=x,y,z}\hbar\left(\omega_{0}-i\frac{\Gamma_{0}}{2}\right)\hat{\sigma}_{m\nu}^{\dagger}\hat{\sigma}_{m\nu}\nonumber \\
 &  & -\frac{3\pi\hbar\Gamma_{0}c}{\omega_{0}}\sum_{m\neq n}\sum_{\mu,\nu=x,y,z}G_{\mu\nu}(\mathbf{r}_{m}-\mathbf{r}_{n})\hat{\sigma}_{m\mu}^{\dagger}\hat{\sigma}_{n\nu},\nonumber \\
\label{eq:S1}
\end{eqnarray}
where $\hat{\sigma}_{n\nu}=\left|g_{n}\right\rangle \left\langle \nu_{n}\right|$
is the lowering operator of $n$th atom and $G_{\mu\nu}(\mathbf{r})$
is the dyadic Green's function representing the resonant dipole-dipole
interaction~(RDDI) between two atoms separated by a distance $\mathbf{r}$
with dipoles aligned along $\mu$ and $\nu$ directions. For an infinite
periodic 2D Bravais lattice, single excitation can be well described
in reciprocal space, and the eigenstates of Eq.~(\ref{eq:S1}) are
Bloch states according to the translational symmetry. The Fourier
transformations of atomic operators is defined as
\begin{equation}
\hat{\sigma}_{n\nu}=\sum_{\mathbf{k}}e^{i\mathbf{k}\cdot\mathbf{r}_{n}}\hat{\sigma}_{\mathbf{k}\nu},\label{eq:S2}
\end{equation}
where the sum over wave vector $\mathbf{k}$ is restricted to the
first Brillouin zone~(BZ). Defining the Bloch state with a quasimomentum
$\mathbf{k}_{B}$ in the Cartesian basis, $\left|\psi_{\mathbf{k}_{B}}\right\rangle =\bigl(c_{x}\hat{\sigma}_{\mathbf{k}_{B}x}^{\dagger},c_{y}\hat{\sigma}_{\mathbf{k}_{B}y}^{\dagger},c_{z}\hat{\sigma}_{\mathbf{k}_{B}z}^{\dagger}\bigr)^{T}\left|g\right\rangle ^{\otimes N\rightarrow\infty}$,
the eigenvalue problem of Eq.~(\ref{eq:S1}) reads
\begin{eqnarray}
E_{\mathbf{k}_{B}} & = & \left\langle \psi_{\mathbf{k}_{B}}\right|{\cal \hat{H}_{\text{eff}}}\left|\psi_{\mathbf{k}_{B}}\right\rangle =\sum_{\nu}\hbar\left(\omega_{0}-i\frac{\Gamma_{0}}{2}\right)c_{\nu}^{*}c_{\nu}\nonumber \\
 &  & -\frac{3\pi\hbar\Gamma_{0}c}{\omega_{0}}\sum_{\mu,\nu}\sum_{\mathbf{R}\neq0}e^{-i\mathbf{k}_{B}\cdot\mathbf{R}}G_{\mu\nu}(\mathbf{R})c_{\mu}^{*}c_{\nu},\label{eq:S3}
\end{eqnarray}
where $\mathbf{R}$ is real space lattice vectors. Equation~(\ref{eq:S3})
is the same as diagonalization of a 3 $\times$ 3 matrix and we can
use it to define the Hamiltonian kernel ${\cal H}_{\text{eff}}(\mathbf{k}_{B})$.
The challenge in the determination of dispersion relation described
in Eq.~(\ref{eq:S3}) is the calculation on 2D Fourier transformation
of dyadic Green's function since it's an infinite summation with oscillatory
terms. The standard procedure to deal with this infinite sum is Ewald's
method~\cite{Zhen2008S}, which computes the slowly convergent part
in reciprocal space to avoid the oscillatory terms in real space.
However, an artificial truncation or the ultraviolet frequency cutoff~\cite{Perczel2017PRAS}
is necessary to reach the desired precision. Here, we exploit the
2D generalization of Euler-Maclaurin formula to give a model-independent
approach to calculate the infinite sum in reciprocal space without
the ultraviolet frequency cutoff, and the truncation is determined
by a self-consistent way.

\subsubsection{Atomic lattices in free space}

Dyadic Green's function describing the resonant dipole-dipole interaction
in free space at resonant frequency $\omega$ has the following form:
\begin{eqnarray}
\overline{\overline{\mathbf{G}}}_{0}(\mathbf{r}) & = & \frac{e^{iqr}}{4\pi r}\left[\left(1+\frac{iqr-1}{q^{2}r^{2}}\right)\overline{\overline{\mathbf{I}}}\right.\nonumber \\
 &  & +\left.\left(-1+\frac{3-3iqr}{q^{2}r^{2}}\right)\frac{\mathbf{r}\otimes\mathbf{r}}{r^{2}}\right],\label{eq:S4}
\end{eqnarray}
where $q=\omega/c$ and $r=\left|\mathbf{r}\right|$ is the distance
between two dipole emitters, and this can be obtained either from
classical electromagnetism or quantum optical treatment~\cite{Lehmberg1970S}.
Because we are interested in the closed form of dyadic Green's function
in reciprocal space, we use Weyl decomposition and integrate out the
momentum perpendicular to 2D atomic lattice placed in the $xy$-plane
to rewrite Eq.~(\ref{eq:S4}) as
\begin{eqnarray}
\overline{\overline{\mathbf{G}}}_{0}(\mathbf{r}_{\parallel}) & = & \ensuremath{{\displaystyle \int}}\frac{d^{2}k_{\parallel}}{\left(2\pi\right)^{3}}\left[\:\overline{\overline{\mathbf{I}}}-\left(1-\frac{\mathbf{k}_{\parallel}^{2}}{q^{2}}\right)\mathbf{e}_{z}\otimes\mathbf{e}_{z}-\frac{1}{q^{2}}\mathbf{k}_{\parallel}\otimes\mathbf{k}_{\parallel}\right]\nonumber \\
 &  & \times\left[\frac{i\pi\Theta\left(q^{2}-\mathbf{k}_{\parallel}^{2}\right)}{\sqrt{q^{2}-\mathbf{k}_{\parallel}^{2}}}+\frac{\pi\Theta\left(\mathbf{k}_{\parallel}^{2}-q^{2}\right)}{\sqrt{\mathbf{k}_{\parallel}^{2}-q^{2}}}\right]e^{i\mathbf{k}_{\parallel}\cdot\mathbf{r}_{\parallel}}\nonumber \\
 & = & \ensuremath{{\displaystyle \int}}\frac{d^{2}k_{\parallel}}{\left(2\pi\right)^{2}}e^{i\mathbf{k}_{\parallel}\cdot\mathbf{r}_{\parallel}}\overline{\overline{\mathbf{g}}}_{0}(\mathbf{k}_{\parallel}),\label{eq:S5}
\end{eqnarray}
where the integrand $\overline{\overline{\mathbf{g}}}_{0}(\mathbf{k}_{\parallel})$
is the dyadic Green's function in reciprocal space and $\Theta$ denotes
the Heaviside step function. By inserting Eq.~(\ref{eq:S5}) into
Eq.~(\ref{eq:S3}) and using Poisson's identity, the effective Hamiltonian
kernel for infinite 2D Bravais lattice~$\left\{ \mathbf{R}\right\} $
with two lattice vectors $\mathbf{a}_{1}$ and $\mathbf{a}_{2}$ is
given by
\begin{eqnarray}
{\cal H}_{\text{eff}}^{\text{3D}}(\mathbf{k}_{B}) & = & \hbar\left(\omega_{0}-i\frac{\Gamma_{0}}{2}\right)\overline{\overline{\mathbf{I}}}\nonumber \\
 & - & \frac{3\pi\hbar\Gamma_{0}c}{\omega_{0}}\left[\overline{\overline{\mathbf{S}}}_{0}(\mathbf{k}_{B})-\overline{\overline{\mathbf{G}}}_{0}(\mathbf{0})\right],\label{eq:S6}
\end{eqnarray}
\begin{equation}
\overline{\overline{\mathbf{S}}}_{0}(\mathbf{k}_{B})=\sum_{\mathbf{G}}\frac{\overline{\overline{\mathbf{g}}}_{0}(\mathbf{G}+\mathbf{k}_{B})}{\left|\mathbf{a}_{1}\times\mathbf{a}_{2}\right|},\label{eq:S7}
\end{equation}
\begin{align}
\sum_{\mathbf{R}\neq\mathbf{0}} & e^{-i\mathbf{k}_{B}\cdot\mathbf{R}}\overline{\overline{\mathbf{G}}}_{0}(\mathbf{R})=\overline{\overline{\mathbf{S}}}_{0}(\mathbf{k}_{B})-\overline{\overline{\mathbf{G}}}_{0}(\mathbf{0})\nonumber \\
 & =\Biggl[\sum_{\mathbf{G}}-\ensuremath{{\displaystyle \int}}\frac{\left|\mathbf{a}_{1}\times\mathbf{a}_{2}\right|}{\left(2\pi\right)^{2}}d^{2}G\Biggr]\frac{\overline{\overline{\mathbf{g}}}_{0}(\mathbf{G}+\mathbf{k}_{B})}{\left|\mathbf{a}_{1}\times\mathbf{a}_{2}\right|},\label{eq:S8}
\end{align}
where the summation in Eq.~(\ref{eq:S6}) runs over all reciprocal
lattice vectors $\mathbf{G}$. As we can see from Eq.~(\ref{eq:S5}),
the infinite periodic sum $\overline{\overline{\mathbf{S}}}_{0}(\mathbf{k}_{B})$
in Eq.~(\ref{eq:S7}) describing how atoms in Bravais lattice~$\left\{ \mathbf{R}\right\} $
are affected by all atoms~(including the self-interaction) contains
no oscillatory terms and diverges. While $\overline{\overline{\mathbf{G}}}_{0}(\mathbf{0})$
as the self-interaction is also divergent according to Eq.~(\ref{eq:S4}),
the difference between the interactions of all atoms $\overline{\overline{\mathbf{S}}}_{0}(\mathbf{k}_{B})$
and $\overline{\overline{\mathbf{G}}}_{0}(\mathbf{0})$ in Eq.~(\ref{eq:S8}),
which describes how atoms in Bravais lattice~$\left\{ \mathbf{R}\right\} $
are affected by the other atoms~(excluding the self-interaction),
is physically meaningful and convergent, and it can be calculated
via the 2D generalization of Euler-Maclaurin formula since it is in
the form of the difference between the summation and the integration
of the same function. One of the essences of Euler-Macularin formula
is to approximate this difference over a period by the difference
of higher-order derivatives at the boundaries, such that the difference
within any duplication of the periodic region can be viewed as the
difference of higher-order derivatives at larger boundaries. Once
we push one of the boundaries to infinity in reciprocal space, the
presence of ultraviolet frequency cutoff ensures that high frequency
terms would not contribute to the low energy physics, and we can simply
approximate Eq.~(\ref{eq:S8}) out of certain boundary $\partial R$
by higher-order derivatives of dyadic Green's function in reciprocal
space $\overline{\overline{\mathbf{g}}}_{0}(\mathbf{G}+\mathbf{k}_{B})$
at this boundary.

For a non-Hermitian effective Hamiltonian given by Eq.~(\ref{eq:S6}),
the first term describes the single atom with a decay rate $\Gamma_{0}$
and the second term involves the collective RDDI without the self-interaction
such that the real and imaginary parts of dyadic Green's function
correspond to the collective Lamb shift and collective decay rate.
According to Eq.~(\ref{eq:S5}), only $\overline{\overline{\mathbf{g}}}_{0}(\mathbf{G}+\mathbf{k}_{B})$
is responsible for the collective decay when $\mathbf{G}+\mathbf{k}_{B}$
lies within the light cone defined by~$\left|\mathbf{G}+\mathbf{k}_{B}\right|\leq q$,
since the imaginary part of self-interaction $\overline{\overline{\mathbf{G}}}_{0}(\mathbf{0})$
cancels the single atom decay rate $\Gamma_{0}$. That is, this system
becomes Hermitian again if $\mathbf{G}+\mathbf{k}_{B}$ lies out of
the light cone for all reciprocal lattice vectors $\mathbf{G}$, and
the corresponding Bloch state is dissipationless and lives, in principle,
forever in the atomic lattice.

The existence of light cone can be realized from the elimination of
out-of-plane momentum in Eq.~(\ref{eq:S5}):
\begin{equation}
\overline{\overline{\mathbf{G}}}_{0}(\mathbf{r})=\ensuremath{{\displaystyle \int}}\frac{d^{2}k_{\parallel}}{\left(2\pi\right)^{3}}e^{i\mathbf{k}_{\parallel}\cdot\mathbf{r}_{\parallel}}\ensuremath{{\displaystyle \int}}dk_{z}e^{ik_{z}z}\frac{\overline{\overline{\mathbf{I}}}-\frac{1}{q^{2}}\mathbf{k}\otimes\mathbf{k}}{\mathbf{k}_{\parallel}^{2}+k_{z}^{2}-q^{2}-i\epsilon},\label{eq:S9}
\end{equation}
where $\epsilon$ is an infinitesimal positive value to ensure the
causality. The poles on complex $k_{z}$ plane are given by $\pm[\sqrt{q^{2}-\mathbf{k}_{\parallel}^{2}}+i\epsilon]\Theta(q^{2}-\mathbf{k}_{\parallel}^{2})$
and $\pm[i\sqrt{\mathbf{k}_{\parallel}^{2}-q^{2}}+\epsilon]\Theta(\mathbf{k}_{\parallel}^{2}-q^{2})$.
These respective on-shell momenta, as the real photon exchange process
in RDDI, relate to the outgoing field propagating out of atomic lattice
with a real out-of-plane momentum and evanescent wave localized in
atomic lattice due to the exponentially decay in out-of-plane direction.
It's clear to see that only those modes lie within the light cone
contribute to the collective decay and the presence of these radiative
eigenstates may lead to the radiative edge modes in finite system~\cite{Perczel2017PRLS}.

\subsubsection{Bravais and non-Bravais lattices}

Although we only discuss the Bravais lattice in the main text, here
we briefly review the formalism of effective Hamiltonian kernel for
a non-Bravais lattice in order to demonstrate that our 2D generalization
of Euler-Maclaurin formula can be used in non-Bravais lattices. Here
we take the Kagome lattice for example. Kagome lattice, as a non-Bravais
lattice composed of three Bravais triangular sublattices, has three
inequivalent sites denoted by 1, 2, and 3, then the Bloch state becomes
a $9\times1$ column vector defined in the basis of the tensor product
of sublattices and polarizations~$\bigl|\psi_{\mathbf{k}_{B}}\bigr\rangle=(\begin{array}{ccc}
\bigl|\psi_{\mathbf{k}_{B}}^{(1)}\bigr\rangle, & \bigl|\psi_{\mathbf{k}_{B}}^{(2)}\bigr\rangle, & \bigl|\psi_{\mathbf{k}_{B}}^{(3)}\bigr\rangle\end{array})^{T}$, where~$\bigl|\psi_{\mathbf{k}_{B}}^{(i)}\bigr\rangle=(\begin{array}{ccc}
c_{x}^{(i)}\hat{\sigma}_{\mathbf{k}_{B}x}^{(i)\dagger}, & c_{y}^{(i)}\hat{\sigma}_{\mathbf{k}_{B}y}^{(i)\dagger}, & c_{z}^{(i)}\hat{\sigma}_{\mathbf{k}_{B}z}^{(i)\dagger}\end{array})^{T}\left|g\right\rangle ^{\otimes N\rightarrow\infty}$. In this case, the $9\times9$ effective Hamiltonian kernel for a
2D Kagome lattice reads\pagebreak
\begin{widetext}

\begin{eqnarray}
{\cal H} & _{\text{eff}} & (\mathbf{k}_{B})=\hbar\left(\omega_{0}-i\frac{\Gamma_{0}}{2}\right)\begin{pmatrix}\overline{\overline{\mathbf{I}}} & 0 & 0\\
0 & \overline{\overline{\mathbf{I}}} & 0\\
0 & 0 & \overline{\overline{\mathbf{I}}}
\end{pmatrix}\label{eq:S10}\\
 & - & \frac{3\pi\hbar\Gamma_{0}c}{\omega_{0}}\begin{pmatrix}{\displaystyle \sum_{\mathbf{R}\neq0}e^{-i\mathbf{k}_{B}\cdot\mathbf{R}}\overline{\overline{\mathbf{G}}}_{0}(\mathbf{R})} & {\displaystyle \sum_{\mathbf{R}}e^{-i\mathbf{k}_{B}\cdot(\mathbf{R}+\mathbf{b}_{12})}\overline{\overline{\mathbf{G}}}_{0}(\mathbf{R}+\mathbf{b}_{12})} & {\displaystyle \sum_{\mathbf{R}}e^{-i\mathbf{k}_{B}\cdot(\mathbf{R}+\mathbf{b}_{13})}\overline{\overline{\mathbf{G}}}_{0}(\mathbf{R}+\mathbf{b}_{13})}\\
{\displaystyle \sum_{\mathbf{R}}e^{-i\mathbf{k}_{B}\cdot(\mathbf{R}+\mathbf{b}_{21})}\overline{\overline{\mathbf{G}}}_{0}(\mathbf{R}+\mathbf{b}_{21})} & {\displaystyle \sum_{\mathbf{R}\neq0}e^{-i\mathbf{k}_{B}\cdot\mathbf{R}}\overline{\overline{\mathbf{G}}}_{0}(\mathbf{R})} & {\displaystyle \sum_{\mathbf{R}}e^{-i\mathbf{k}_{B}\cdot(\mathbf{R}+\mathbf{b}_{23})}\overline{\overline{\mathbf{G}}}_{0}(\mathbf{R}+\mathbf{b}_{23})}\\
{\displaystyle \sum_{\mathbf{R}}e^{-i\mathbf{k}_{B}\cdot(\mathbf{R}+\mathbf{b}_{31})}\overline{\overline{\mathbf{G}}}_{0}(\mathbf{R}+\mathbf{b}_{31})} & {\displaystyle \sum_{\mathbf{R}}e^{-i\mathbf{k}_{B}\cdot(\mathbf{R}+\mathbf{b}_{32})}\overline{\overline{\mathbf{G}}}_{0}(\mathbf{R}+\mathbf{b}_{32})} & {\displaystyle \sum_{\mathbf{R}\neq0}e^{-i\mathbf{k}_{B}\cdot\mathbf{R}}\overline{\overline{\mathbf{G}}}_{0}(\mathbf{R})}
\end{pmatrix}.\nonumber 
\end{eqnarray}

The block diagonal part of Eq.~(\ref{eq:S10}) shows the effective
Hamiltonian of each triangular sublattice and its RDDI excludes the
self-interaction, while the block off-diagonal term describing the
RDDI between two sublattices keeps the whole contribution from all
triangular lattice vectors $\mathbf{R}$. For simplicity, we focus
on in-plane polarizations since it is decoupled from out-of-plane
polarization for 2D geometry according to Eq.~(\ref{eq:S5}), from
which we obtain their corresponding photonic band structures in the
main text.

\section{Euler-Maclaurin formula}

\subsubsection{2D generalization of Euler-Maclaurin formula}

We use Bernoulli polynomials $B_{n}(x)$ and Bernoulli numbers $B_{n}$~\cite{Abramowitz1972S}
to derive the 2D generalization of Euler-Maclaurin formula using a
similar procedure in the one-dimensional~(1D) case. Starting from
the summation of a function $f\left(x,y\right)$ over 2D rectangular
lattice points $\mathbf{R}=mb_{1}\mathbf{e}_{x}+nb_{2}\mathbf{e}_{y}$,
which is in the form of the linear combination of two independent
lattice vectors $\mathbf{b}_{1}=b_{1}\mathbf{e}_{x}$ and $\mathbf{b}_{2}=b_{2}\mathbf{e}_{y}$,
and the integral of $f(x,y)$ over a minimal rectangular period $S_{\square}=\left\{ (x,y)|\,k\leq x\leq k+b_{1},\,l\leq y\leq l+b_{2}\right\} $
can be written as~(assume $f\left(x,y\right)$ is a smooth function)

\begin{align}
 & \ensuremath{{\displaystyle \int_{l}^{l+b_{2}}}}dy\ensuremath{{\displaystyle \int_{k}^{k+b_{1}}}}dxf\left(x,y\right)=\ensuremath{{\displaystyle \int_{l}^{l+b_{2}}}}dy\biggl\{ b_{1}B_{1}(\tfrac{x-k}{b_{1}})f(x,y)-\sum_{i=1}^{\left\lfloor M_{x}/2\right\rfloor }\frac{b_{1}^{2i}}{(2i)!}B_{2i}(\tfrac{x-k}{b_{1}})\partial_{x}^{2i-1}f(x,y)\Bigr]\Biggr|_{k}^{k+b_{1}}+R_{M_{x}}\biggr\}\nonumber \\
= & \biggl\{ b_{1}B_{1}(\tfrac{x-k}{b_{1}})\Bigl[b_{2}B_{1}(\tfrac{y-l}{b_{2}})f(x,y)-\sum_{j=1}^{\left\lfloor M_{y}/2\right\rfloor }\frac{b_{2}^{2j}}{(2j)!}B_{2j}(\tfrac{y-l}{b_{2}})\partial_{y}^{2j-1}f(x,y)\Bigr]\biggr\}\Biggr|_{k}^{k+b_{1}}\Biggr|_{l}^{l+b_{2}}\nonumber \\
 & -\biggl\{\sum_{i=1}^{\left\lfloor M_{x}/2\right\rfloor }\frac{b_{1}^{2i}}{(2i)!}B_{2i}(\tfrac{x-k}{b_{1}})\Bigl[b_{2}B_{1}(\tfrac{y-l}{b_{2}})\partial_{x}^{2i-1}f(x,y)-\sum_{j=1}^{\left\lfloor M_{y}/2\right\rfloor }\frac{b_{2}^{2j}}{(2j)!}B_{2j}(\tfrac{y-l}{b_{2}})\partial_{x}^{2i-1}\partial_{y}^{2j-1}f(x,y)\Bigr]\biggr\}\Biggr|_{k}^{k+b_{2}}\Biggr|_{l}^{l+b_{1}}+R_{M_{x},M_{y}}\nonumber \\
= & \frac{1}{4}b_{1}b_{2}\left[f_{k+b_{1},l+b_{2}}+f_{k+b_{1},l}+f_{k,l+b_{2}}+f_{k,l}\right]\nonumber \\
 & -\frac{b_{2}}{2}\sum_{i=1}^{\left\lfloor M_{x}/2\right\rfloor }\frac{b_{1}^{2i}}{(2i)!}B_{2i}\left[\partial_{x}^{2i-1}f_{k+b_{1},l+b_{2}}+\partial_{x}^{2i-1}f_{k+b_{1},l}-\partial_{x}^{2i-1}f_{k,l+b_{2}}-\partial_{x}^{2i-1}f_{k,l}\right]\nonumber \\
 & -\frac{b_{1}}{2}\sum_{j=1}^{\left\lfloor M_{y}/2\right\rfloor }\frac{b_{2}^{2j}}{(2j)!}B_{2j}\left[\partial_{y}^{2j-1}f_{k+b_{1},l+b_{2}}+\partial_{y}^{2j-1}f_{k,l+b_{2}}-\partial_{y}^{2j-1}f_{k+b_{1},l}-\partial_{y}^{2j-1}f_{k,l}\right]\nonumber \\
 & +\sum_{i=1}^{\left\lfloor M_{x}/2\right\rfloor }\sum_{j=1}^{\left\lfloor M_{y}/2\right\rfloor }\frac{b_{1}^{2i}}{(2i)!}\frac{b_{2}^{2j}}{(2j)!}B_{2i}B_{2j}\left[\partial_{x}^{2i-1}\partial_{y}^{2j-1}f_{k+b_{1},l+b_{2}}-\partial_{x}^{2i-1}\partial_{y}^{2j-1}f_{k+b_{1},l}-\partial_{x}^{2i-1}\partial_{y}^{2j-1}f_{k,l+b_{2}}+\partial_{x}^{2i-1}\partial_{y}^{2j-1}f_{k,l}\right]\nonumber \\
 & +R_{M_{x},M_{y}}.\label{eq:S11}
\end{align}

The first equal sign in the above is nothing but 1D Euler-Maclaurin
formula expanded to the $M_{x}$th order, where $R_{M_{x}}$ is the
remainder term. Since $x$ and $y$ coordinates are linearly independent,
we directly obtain the second equal sign by treating it as the 1D
case. The last equal sign is our 2D generalization of Euler-Maclaurin
formula expanded to the $M_{x}$th order in $x$ and the $M_{y}$th
order in $y$ for 2D rectangular lattice points. Note that this formula
respects the symmetry in the interchange of~$(x,k,b_{1})$ and~$(y,l,b_{2})$,
which is important for the periodic summation.

For 2D triangular lattice points $\mathbf{R}=m\mathbf{b}_{1}+n\mathbf{b}_{2}=\frac{1}{2}b(m+n)\mathbf{e}_{x}+\frac{\sqrt{3}}{2}b(-m+n)\mathbf{e}_{y}$,
two lattice vectors are not orthogonal such that we can not directly
follow the procedure in Eq.~(\ref{eq:S11}). We try to separate the
integral over a minimal triangular period $S_{\triangle}=\bigl\{(x,y)|\,y\geq l,\,\sqrt{3}(x-k)-(y-l)\geq0,\,\sqrt{3}(x-k)+(y-l)\leq\sqrt{3}b\bigr\}$,
i.e., an equilateral triangle with vertices~$(k,l)$, $(k+b,l)$
and~$(k+\frac{1}{2}b,l+\frac{\sqrt{3}}{2}b)$, into the following
two parts

\begin{eqnarray}
\ensuremath{{\displaystyle \int_{k}^{k+\frac{b}{2}}}} & dx & \ensuremath{{\displaystyle \int_{l}^{l+\sqrt{3}(x-k)}}}dyf\left(x,y\right)+\ensuremath{{\displaystyle \int_{k+\frac{b}{2}}^{k+b}}}dx\ensuremath{{\displaystyle \int_{l}^{l+\sqrt{3}(k+b-x)}}}dyf(x,y)\nonumber \\
 & = & \ensuremath{{\displaystyle \int_{k}^{k+\frac{b}{2}}}}dx\Biggl\{\Biggl[{\textstyle \frac{\sqrt{3}}{2}}bB_{1}\Bigl(\tfrac{y-l}{\frac{\sqrt{3}}{2}b}\Bigr)f(x,y)-\sum_{i=1}^{\left\lfloor M_{y}/2\right\rfloor }\frac{\bigl(\frac{\sqrt{3}}{2}b\bigr)^{2i}}{(2i)!}B_{2i}\Bigl(\tfrac{y-l}{\frac{\sqrt{3}}{2}b}\Bigr)\partial_{y}^{2i-1}f(x,y)\Biggr]\Biggr|_{l}^{l+\sqrt{3}(x-k)}+R_{N_{y}}\Biggr\}\nonumber \\
 &  & +\ensuremath{{\displaystyle \int_{k+\frac{b}{2}}^{k+b}}}dx\Biggl\{\Biggl[{\textstyle \frac{\sqrt{3}}{2}}bB_{1}\Bigl(\tfrac{y-l}{\frac{\sqrt{3}}{2}b}\Bigr)f(x,y)-\sum_{i=1}^{\left\lfloor M_{y}/2\right\rfloor }\frac{\bigl(\frac{\sqrt{3}}{2}b\bigr)^{2i}}{(2i)!}B_{2i}\Bigl(\tfrac{y-l}{\frac{\sqrt{3}}{2}b}\Bigr)\partial_{y}^{2i-1}f(x,y)\Biggr]\Biggr|_{l}^{l+\sqrt{3}(k+b-x)}+R_{N_{y}}\Biggr\}\nonumber \\
 & = & \ensuremath{{\displaystyle \int_{k}^{k+\frac{b}{2}}}}dx\Biggl[{\textstyle \frac{\sqrt{3}}{2}}bB_{1}\Bigl(\tfrac{x-k}{b/2}\Bigr)f\left(x,l+\sqrt{3}(x-k)\right)-\sum_{i=1}^{\left\lfloor M_{y}/2\right\rfloor }\frac{\bigl(\frac{\sqrt{3}}{2}b\bigr)^{2i}}{(2i)!}B_{2i}\Bigl(\tfrac{x-k}{b/2}\Bigr)\partial_{y}^{2i-1}f\left(x,l+\sqrt{3}(x-k)\right)\Biggr]\nonumber \\
 &  & \begin{aligned}+\ensuremath{{\displaystyle \int_{k+\frac{b}{2}}^{k+b}}}dx\Biggl[ & {\textstyle \frac{\sqrt{3}}{2}}b\underbrace{B_{1}\Bigl(\tfrac{k+b-x}{b/2}\Bigr)}_{-B_{1}(\tfrac{x-k-b/2}{b/2})}f\left(x,l+\sqrt{3}(k+b-x)\right)\\
 & -\sum_{i=1}^{\left\lfloor M_{y}/2\right\rfloor }\frac{\bigl(\frac{\sqrt{3}}{2}b\bigr)^{2i}}{(2i)!}\underbrace{B_{2i}\Bigl(\tfrac{k+b-x}{b/2}\Bigr)}_{B_{2i}(\tfrac{x-k-b/2}{b/2})}\partial_{y}^{2i-1}f\left(x,l+\sqrt{3}(k+b-x)\right)\Biggr]
\end{aligned}
\nonumber \\
 &  & -\ensuremath{{\displaystyle \int_{k}^{k+b}}}dx\Biggl[{\textstyle \frac{\sqrt{3}}{2}}bB_{1}(0)f(x,l)-\sum_{i=1}^{\left\lfloor M_{y}/2\right\rfloor }\frac{\bigl(\frac{\sqrt{3}}{2}b\bigr)^{2i}}{(2i)!}B_{2i}(0)\partial_{y}^{2i-1}f(x,l)\Biggr]+R_{N_{y}}\nonumber \\
 & = & \frac{\sqrt{3}}{2}b\sum_{j=1}^{\left\lfloor M_{x}/2\right\rfloor }\frac{\bigl(\frac{b}{2}\bigr)^{2j-1}}{(2j)!}B_{2j}(\tfrac{x-k}{b/2})\left(\frac{d}{dx}\right)^{2j-2}f\left(x,l+\sqrt{3}(x-k)\right)\Biggr|_{k}^{k+\frac{b}{2}}\nonumber \\
 &  & -\sum_{i=1}^{\left\lfloor M_{y}/2\right\rfloor }\sum_{j=1}^{\left\lfloor M_{x}/2\right\rfloor }\frac{\bigl(\frac{\sqrt{3}}{2}b\bigr)^{2i}\bigl(\frac{b}{2}\bigr)^{2j}}{(2i+2j)!}B_{2i+2j}(\tfrac{x-k}{b/2})\left(\frac{d}{dx}\right)^{2j-1}\partial_{y}^{2i-1}f\left(x,l+\sqrt{3}(x-k)\right)\Biggr|_{k}^{k+\frac{b}{2}}\nonumber \\
 &  & -\frac{\sqrt{3}}{2}b\sum_{j=1}^{\left\lfloor M_{x}/2\right\rfloor }\frac{\bigl(\frac{b}{2}\bigr)^{2j-1}}{(2j)!}B_{2j}(\tfrac{x-k-b/2}{b/2})\left(\frac{d}{dx}\right)^{2j-2}f\left(x,l+\sqrt{3}(k+b-x)\right)\Biggr|_{k+\frac{b}{2}}^{k+b}\nonumber \\
 &  & -\sum_{i=1}^{\left\lfloor M_{y}/2\right\rfloor }\sum_{j=1}^{\left\lfloor M_{x}/2\right\rfloor }\frac{\bigl(\frac{\sqrt{3}}{2}b\bigr)^{2i}\bigl(\frac{b}{2}\bigr)^{2j}}{(2i+2j)!}B_{2i+2j}(\tfrac{x-k-b/2}{b/2})\left(\frac{d}{dx}\right)^{2j-1}\partial_{y}^{2i-1}f\left(x,l+\sqrt{3}(k+b-x)\right)\Biggr|_{k+\frac{b}{2}}^{k+b}\nonumber \\
 &  & +\frac{\sqrt{3}}{2}\frac{b}{2}\Biggl[bB_{1}(\tfrac{x-k}{b})f(x,l)-\sum_{j=1}^{\left\lfloor M_{x}/2\right\rfloor }\frac{b^{2j}}{(2j)!}B_{2j}(\tfrac{x-k}{b})\partial_{x}^{2j-1}f(x,l)\Biggr]\Biggr|_{k}^{k+b}\nonumber \\
 &  & +\sum_{i=1}^{\left\lfloor M_{y}/2\right\rfloor }\frac{\bigl(\frac{\sqrt{3}}{2}b\bigr)^{2i}}{(2i)!}B_{2i}\Biggl[bB_{1}(\tfrac{x-k}{b})\partial_{y}^{2i-1}f(x,l)-\sum_{j=1}^{\left\lfloor M_{x}/2\right\rfloor }\frac{b^{2j}}{(2j)!}B_{2j}(\tfrac{x-k}{b})\partial_{y}^{2i-1}\partial_{x}^{2j-1}f(x,l)\Biggr]\Biggr|_{k}^{k+b}\nonumber \\
 &  & +R_{M_{x},M_{y}}.\label{eq:S12}
\end{eqnarray}

Unlike Eq.~(\ref{eq:S11}), the integrand in the second equal sign
of Eq.~(\ref{eq:S12}) involves not only the arbitrary function $f(x,y)$
but also the Bernoulli polynomials, which leads to a more complicated
result that does not respect the mirror symmetry and the rotational
symmetry of equilateral triangle since we choose a specific orientation
and ordering to expand the integral. To ensure the mirror symmetry,
we consider the average of Eq.~(\ref{eq:S12}) and the result of
its reflection about the angle bisector of the bottom left vertex
of $S_{\triangle}$. To give a rotational symmetry-respecting formula,
we consider the average of three integrals over the same minimal triangular
period $S_{\triangle}$ written in three infinitesimal area elements,
$dxdy$, $d(-\frac{1}{2}x-\frac{\sqrt{3}}{2}y)d(\frac{1}{2}x-\frac{\sqrt{3}}{2}y)$,
and $d(-\frac{1}{2}x+\frac{\sqrt{3}}{2}y)d(-x)$. Therefore, we focus
on one vertex due to the symmetry consideration and use directional
derivatives to give a coordinate-independent form of Eq.~(\ref{eq:S12}).
Here we consider the bottom right vertex~$(x,y)=(k+b,l)$ of $S_{\triangle}$
and define $\partial_{\rightarrow}$, $\partial_{\uparrow}$, $\partial_{\rotatebox[origin=c]{-60}{\ensuremath{\rightarrow}}}$,
$\partial_{\rotatebox[origin=c]{-150}{\ensuremath{\rightarrow}}}$
and $\partial_{\rotatebox[origin=c]{-30}{\ensuremath{\rightarrow}}}$
as the directional derivatives along directions whose angular deflections
about the positive $x$ axis are $0\lyxmathsym{\textdegree}$, $90\lyxmathsym{\textdegree}$,
$-60\lyxmathsym{\textdegree}$, $-150\lyxmathsym{\textdegree}$ and
$-30\lyxmathsym{\textdegree}$, respectively, then we obtain

\begin{eqnarray}
\ensuremath{{\displaystyle \int_{S_{\triangle}}}}dydxf\left(x,y\right) & = & {\displaystyle \frac{1}{3}}{\displaystyle \frac{\sqrt{3}}{2}}{\displaystyle \frac{b^{2}}{2}}\Biggl\{ f+\sum_{i=1}^{\left\lfloor M/2\right\rfloor }\frac{b^{2i-1}}{\left(2i\right)!}B_{2i}\Biggl[-\partial_{\rightarrow}^{2i-1}f+\bigl(\frac{\sqrt{3}}{2}\partial_{\uparrow}\bigr)^{2i-1}f-\partial_{\rotatebox[origin=c]{-60}{\ensuremath{\rightarrow}}}^{2i-1}f+\bigl(\frac{\sqrt{3}}{2}\partial_{\rotatebox[origin=c]{-150}{\ensuremath{\rightarrow}}}\bigr)^{2i-1}f\Biggr]\nonumber \\
 &  & -2\sum_{i=1}^{\left\lfloor M/2\right\rfloor }\sum_{j=1}^{\left\lfloor M/2\right\rfloor }\frac{b^{2i+2j-2}}{\left(2i\right)!\left(2j\right)!}B_{2i}B_{2j}\left[\partial_{\rightarrow}^{2j-1}\bigl(\frac{\sqrt{3}}{2}\partial_{\uparrow}\bigr)^{2i-1}f+\partial_{\rotatebox[origin=c]{-60}{\ensuremath{\rightarrow}}}^{2j-1}\bigl(\frac{\sqrt{3}}{2}\partial_{\rotatebox[origin=c]{-150}{\ensuremath{\rightarrow}}}\bigr)^{2i-1}f\right]\nonumber \\
 &  & \begin{aligned}-\sum_{i=1}^{\left\lfloor M/2\right\rfloor }\sum_{j=1}^{\left\lfloor M/2\right\rfloor }\frac{b^{2i+2j-2}}{\left(2i+2j\right)!}B_{2i+2j} & \Biggl[\partial_{\rotatebox[origin=c]{-60}{\ensuremath{\rightarrow}}}^{2j-1}\bigl(\frac{\sqrt{3}}{2}\partial_{\rotatebox[origin=c]{-30}{\ensuremath{\rightarrow}}}\bigr)^{2i-1}f+\partial_{\rightarrow}^{2j-1}\bigl(\frac{\sqrt{3}}{2}\partial_{\rotatebox[origin=c]{-30}{\ensuremath{\rightarrow}}}\bigr)^{2i-1}f\\
 & +\partial_{\rotatebox[origin=c]{-60}{\ensuremath{\rightarrow}}}^{2j-1}\bigl(\frac{\sqrt{3}}{2}\partial_{\uparrow}\bigr)^{2i-1}f+\partial_{\rightarrow}^{2j-1}\bigl(\frac{\sqrt{3}}{2}\partial_{\rotatebox[origin=c]{-150}{\ensuremath{\rightarrow}}}\bigr)^{2i-1}f\Biggr]\Biggr\}
\end{aligned}
\nonumber \\
 &  & +R_{z}(\frac{2}{3}\pi)+R_{z}(\frac{4}{3}\pi)+R_{M}.\label{eq:S13}
\end{eqnarray}

The related terms at the other two vertices of $S_{\triangle}$ are
given by either $C_{3}$ rotations with respect to out-of-plane direction
or mirror reflections with respect to bisectors of the other two vertex
angles, and we choose the same expansion order $M$ in $x$ and $y$
to keep the rotational and mirror symmetries.

\end{widetext}
\pagebreak

\begin{figure*}[!tp]
\centering{}\includegraphics{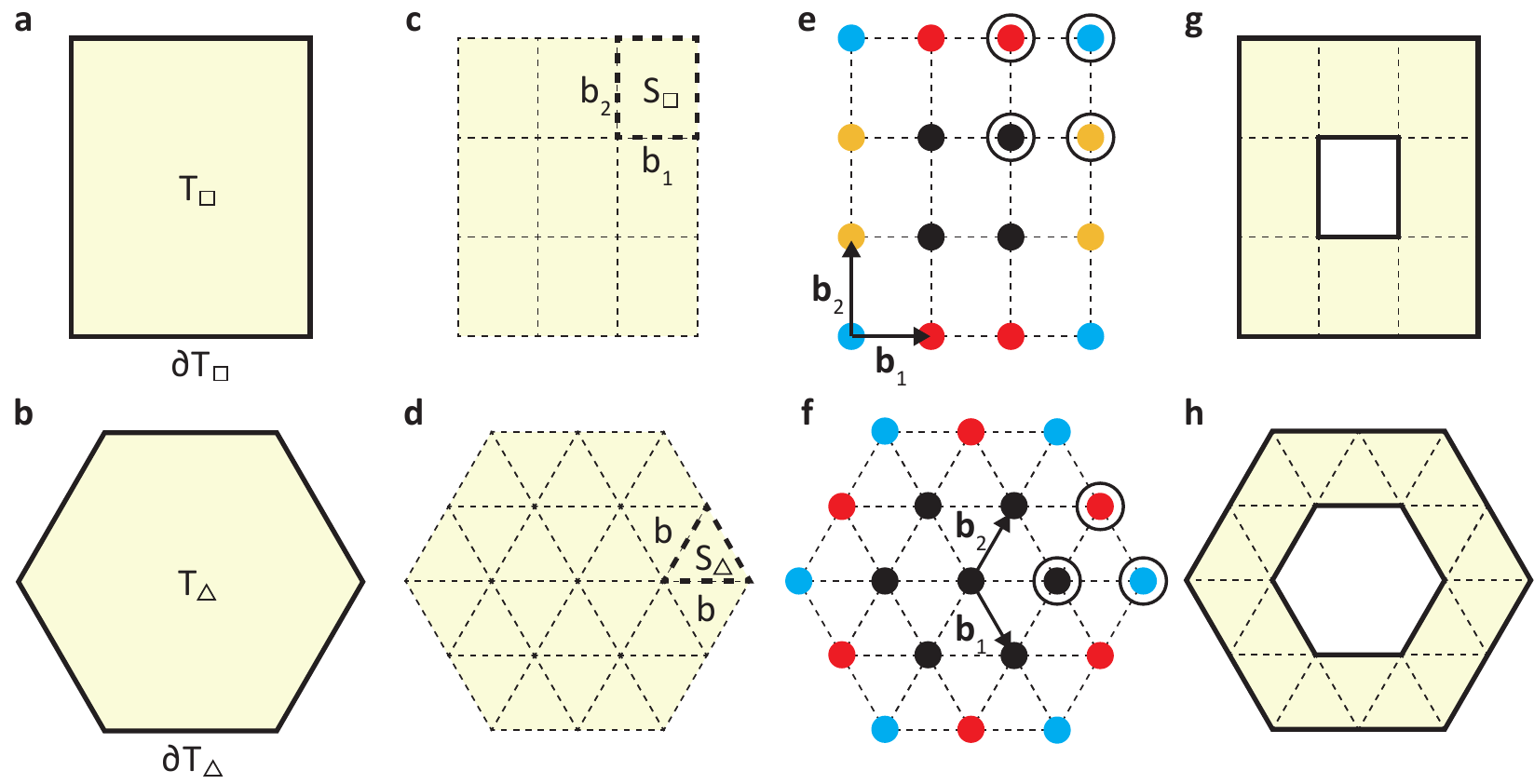}\caption{\label{fig:S1}\textbf{2D generalization of Euler-Maclaurin formula.
a},\textbf{b},~Light yellow shaded regions $T_{\square,\triangle}$
whose boundary is denoted by $\partial T_{\square,\triangle}$. \textbf{c},\textbf{d},~Tile
$T_{\square,\triangle}$ by minimal dashed rectangles $S_{\square}$
with length of sides $b_{1}$ and $b_{2}$ and minimal dashed equilateral
triangles $S_{\triangle}$ with length of side $b$, respectively.
\textbf{e},\textbf{f},~Black dots are the lattice points spanned
by two lattice vectors $\mathbf{b}_{1}$ and $\mathbf{b}_{2}$ that
are not on $\partial T_{\square,\triangle}$, and the other dots are
those lattice points on $\partial T_{\square,\triangle}$. Points
with different color means they have different contributions to the
expansion of integral of arbitrary function $f(x,y)$ via Eqs.~(\ref{eq:S11})
and~(\ref{eq:S13}). \textbf{g},\textbf{h},~Hollow regions $T_{\square,\triangle}$
as tilings of $S_{\square,\triangle}$ have outer and inner boundaries.}
\end{figure*}

\subsubsection{Correction terms}

The 2D generalization of Euler-Maclaurin formula is similar to the
1D case, but the configurations of the boundaries of 2D regions differ
from the 1D case. Take simple finite regions $T_{\square,\triangle}$
tiled by $S_{\square,\triangle}$ in FIG.~\ref{fig:S1}a-d for example.
We treat the integral of $f(x,y)$ over $T_{\square,\triangle}$ as
the summation of that over all $S_{\square,\triangle}$, and the expansion
of integral of $f(x,y)$ over $S_{\square,\triangle}$ can be viewed
as $f(x,y)$ and its higher-order derivatives on the vertices of $S_{\square,\triangle}$
according to Eqs.~(\ref{eq:S11}) and~(\ref{eq:S13}). By collecting
all the contributions from the vertices of all $S_{\square,\triangle}$,
the integral of $f(x,y)$ over $T_{\square,\triangle}$ becomes $f(x,y)$
and its higher-order derivatives on the lattice points in FIG.~\ref{fig:S1}e,
f. Since there are several $S_{\square,\triangle}$ containing the
same lattice point, the contribution from each lattice point can be
categorized according to the configuration of its surroundings. For
the rectangular tilings, we neglect the remainder term in Eq.~(\ref{eq:S11})
such that the contributions from those circled dots in FIG.~\ref{fig:S1}e
colored in black, yellow, red, and blue are given by

\begin{equation}
b_{1}b_{2}f,\label{eq:S14}
\end{equation}

\begin{equation}
\frac{1}{2}b_{1}b_{2}f-b_{2}\sum_{i=1}^{\left\lfloor M_{x}/2\right\rfloor }\frac{b_{1}^{2i}}{\left(2i\right)!}B_{2i}\partial_{x}^{2i-1}f,\label{eq:S15}
\end{equation}

\begin{equation}
\frac{1}{2}b_{1}b_{2}f-b_{1}\sum_{i=1}^{\left\lfloor M_{y}/2\right\rfloor }\frac{b_{2}^{2i}}{\left(2i\right)!}B_{2i}\partial_{y}^{2i-1}f,\label{eq:S16}
\end{equation}

\begin{eqnarray}
\frac{1}{4} &  & b_{1}b_{2}f+\sum_{i=1}^{\left\lfloor M_{x}/2\right\rfloor }\sum_{j=1}^{\left\lfloor M_{y}/2\right\rfloor }\frac{b_{1}^{2i}b_{2}^{2j}}{\left(2i\right)!\left(2j\right)!}B_{2i}B_{2j}\partial_{x}^{2i-1}\partial_{y}^{2j-1}f\nonumber \\
- &  & \frac{b_{2}}{2}\sum_{i=1}^{\left\lfloor M_{x}/2\right\rfloor }\frac{b_{1}^{2i}}{\left(2i\right)!}B_{2i}\partial_{x}^{2i-1}f-\frac{b_{1}}{2}\sum_{i=1}^{\left\lfloor M_{y}/2\right\rfloor }\frac{b_{2}^{2i}}{\left(2i\right)!}B_{2i}\partial_{y}^{2i-1}f,\nonumber \\
\label{eq:S17}
\end{eqnarray}
respectively. Note that all the other black dots in FIG.~\ref{fig:S1}e
have the same result as Eq.~(\ref{eq:S14}), while the contributions
from other yellow, red and blue dots on the boundary $\partial T_{\square,\triangle}$
are given by the related mirror reflections of Eqs.~(\ref{eq:S15}),~(\ref{eq:S16})
and~(\ref{eq:S17}), respectively. Following the same procedure as
the rectangular tiling, then the contributions from those circled
dots in FIG.~\ref{fig:S1}f colored in black, red and blue are given
by
\begin{equation}
\frac{\sqrt{3}}{2}b^{2}f,\label{eq:S18}
\end{equation}

\begin{eqnarray}
\frac{1}{2} &  & \frac{\sqrt{3}}{2}b^{2}f+{\displaystyle \frac{1}{3}}{\displaystyle \frac{\sqrt{3}}{2}}{\displaystyle \frac{b^{2}}{2}}\sum_{i=1}^{\left\lfloor M/2\right\rfloor }\frac{b^{2i-1}}{\left(2i\right)!}B_{2i}\nonumber \\
\times &  & \Biggl[-2\partial_{\rightarrow}^{2i-1}f-2\partial_{\rotatebox[origin=c]{60}{\ensuremath{\rightarrow}}}^{2i-1}f+2\bigl(\frac{\sqrt{3}}{2}\partial_{\rotatebox[origin=c]{-150}{\ensuremath{\rightarrow}}}\bigr)^{2i-1}f\Biggr],\label{eq:S19}
\end{eqnarray}

\begin{align}
{\displaystyle \frac{1}{3}} & \frac{\sqrt{3}}{2}b^{2}\Biggl\{ f+{\displaystyle \frac{1}{2}}\sum_{i=1}^{\left\lfloor M/2\right\rfloor }\frac{b^{2i-1}}{\left(2i\right)!}B_{2i}\Biggl[-2\partial_{\rightarrow}^{2i-1}f-\partial_{\rotatebox[origin=c]{-60}{\ensuremath{\rightarrow}}}^{2i-1}f\nonumber \\
- & \partial_{\rotatebox[origin=c]{60}{\ensuremath{\rightarrow}}}^{2i-1}f+\bigl(\frac{\sqrt{3}}{2}\partial_{\rotatebox[origin=c]{-150}{\ensuremath{\rightarrow}}}\bigr)^{2i-1}f+\bigl(\frac{\sqrt{3}}{2}\partial_{\rotatebox[origin=c]{150}{\ensuremath{\rightarrow}}}\bigr)^{2i-1}f\Biggr]\nonumber \\
- & \sum_{i=1}^{\left\lfloor M/2\right\rfloor }\sum_{j=1}^{\left\lfloor M/2\right\rfloor }\frac{b^{2i+2j-2}}{\left(2i\right)!\left(2j\right)!}B_{2i}B_{2j}\nonumber \\
 & \times\left[\partial_{\rotatebox[origin=c]{-60}{\ensuremath{\rightarrow}}}^{2j-1}\bigl(\frac{\sqrt{3}}{2}\partial_{\rotatebox[origin=c]{-150}{\ensuremath{\rightarrow}}}\bigr)^{2i-1}f+\partial_{\rotatebox[origin=c]{60}{\ensuremath{\rightarrow}}}^{2j-1}\bigl(\frac{\sqrt{3}}{2}\partial_{\rotatebox[origin=c]{150}{\ensuremath{\rightarrow}}}\bigr)^{2i-1}f\right]\nonumber \\
- & {\displaystyle \frac{1}{2}}\sum_{i=1}^{\left\lfloor M/2\right\rfloor }\sum_{j=1}^{\left\lfloor M/2\right\rfloor }\frac{b^{2i+2j-2}}{\left(2i+2j\right)!}B_{2i+2j}\nonumber \\
 & \times\Biggl[\partial_{\rotatebox[origin=c]{-60}{\ensuremath{\rightarrow}}}^{2j-1}\bigl(\frac{\sqrt{3}}{2}\partial_{\rotatebox[origin=c]{-30}{\ensuremath{\rightarrow}}}\bigr)^{2i-1}f+\partial_{\rotatebox[origin=c]{-60}{\ensuremath{\rightarrow}}}^{2j-1}\bigl(\frac{\sqrt{3}}{2}\partial_{\uparrow}\bigr)^{2i-1}f\nonumber \\
 & +\partial_{\rotatebox[origin=c]{60}{\ensuremath{\rightarrow}}}^{2j-1}\bigl(\frac{\sqrt{3}}{2}\partial_{\rotatebox[origin=c]{30}{\ensuremath{\rightarrow}}}\bigr)^{2i-1}f+\partial_{\rotatebox[origin=c]{60}{\ensuremath{\rightarrow}}}^{2j-1}\bigl(\frac{\sqrt{3}}{2}\partial_{\downarrow}\bigr)^{2i-1}f\Biggr]\Biggr\},\nonumber \\
\label{eq:S20}
\end{align}
respectively. We use the directional derivatives again for simplicity,
and $\partial_{\rotatebox[origin=c]{60}{\ensuremath{\rightarrow}}}$
and $\partial_{\rotatebox[origin=c]{30}{\ensuremath{\rightarrow}}}$
in Eqs.~(\ref{eq:S19}) and (\ref{eq:S20}) are the directional derivatives
along directions whose angular deflections about the positive $x$
axis are $60\lyxmathsym{\textdegree}$ and $30\lyxmathsym{\textdegree}$.

\subsubsection{Applications on infinite summations}

In order to simplify the infinite summation and integral over the
whole 2D reciprocal space in Eq.~(\ref{eq:S8}), we now try to transform
Eq.~(\ref{eq:S8}) as the boundary terms in the same manner as 1D
Euler-Maclaurin formula. Notice that the expansion of integral of
$f(x,y)$ over $T_{\square,\triangle}$ is linear in $f(x,y)$ in
Eqs.~(\ref{eq:S14}) and~(\ref{eq:S18})~(except for the lattice
points on the boundary of $T_{\square,\triangle}$), and its coefficient
is determined by the area of unit cell~($b_{1}b_{2}$ and $\frac{\sqrt{3}}{2}b^{2}$),
i.e.,~$\left|\mathbf{b}_{1}\times\mathbf{b}_{2}\right|$, which means
the difference between the integral of $f(x,y)$ over $T_{\square,\triangle}$
divided by the area of unit cell~$\left|\mathbf{b}_{1}\times\mathbf{b}_{2}\right|$
and the summation of $f(x,y)$ over all lattice points in $T_{\square,\triangle}$
only depends on $f(x,y)$ and its higher-order derivatives on the
boundary $\partial T_{\square,\triangle}$. Therefore, we formally
write our 2D generalization of Euler-Maclaurin formula for rectangular
and triangular lattices in arbitrary $\mathbb{R}^{2}$ space as
\begin{align}
 & \sum_{\substack{\mathbf{r}=m_{1}\mathbf{b}_{1}+m_{2}\mathbf{b}_{2}\\
\mathbf{r}\,\in\,T_{\square,\triangle}
}
}f(\mathbf{r})-\ensuremath{{\displaystyle \int_{T_{\square,\triangle}}}}\frac{d^{2}r}{\left|\mathbf{b}_{1}\times\mathbf{b}_{2}\right|}f(\mathbf{r})\nonumber \\
= & \text{correction terms on }\partial T_{\square,\triangle}\nonumber \\
 & -\text{remainder terms}.\label{eq:S21}
\end{align}
Notice that the finite region $T_{\square,\triangle}$ tiled by $S_{\square,\triangle}$
is not restricted to those shapes in FIG.~\ref{fig:S1}a,b. In fact,
Eq.~(\ref{eq:S21}) can be applied to arbitrary $T_{\square,\triangle}$
such as the hollow regions in FIG.~\ref{fig:S1}g, h since Eqs.~(\ref{eq:S14})
and~(\ref{eq:S18}) do not depend on the boundary, while different
$T_{\square,\triangle}$ would result in the correction terms on $\partial T_{\square,\triangle}$
that differs from the above results. In general, we evaluate the correction
terms on $\partial T_{\square,\triangle}$ by Eqs.~(\ref{eq:S11})
and~(\ref{eq:S13}), but we usually perform our 2D generalization
of Euler-Maclaurin formula on the simple geometry like FIG.~\ref{fig:S1}a,
b, g, and h for the symmetry consideration.

Therefore, we calculate Eq.~(\ref{eq:S8}) in reciprocal space by
connecting its right hand side to Eq.~(\ref{eq:S21}). We choose
the arbitrary function $f(\mathbf{r})$ as the dyadic Green's function
in reciprocal space $\overline{\overline{\mathbf{g}}}_{0}(\mathbf{G}+\mathbf{k}_{B})$
divided by the area of unit cell in real space~$\left|\mathbf{a}_{1}\times\mathbf{a}_{2}\right|$.
Since the reciprocal lattice points are spanned by two reciprocal
lattice vectors $\mathbf{b}_{1}=2\pi\frac{\mathbf{a}_{2}\times\mathbf{e}_{z}}{\mathbf{e}_{z}\cdot(\mathbf{a}_{1}\times\mathbf{a}_{2})}$
and $\mathbf{b}_{2}=2\pi\frac{\mathbf{e}_{z}\times\mathbf{a}_{1}}{\mathbf{e}_{z}\cdot(\mathbf{a}_{1}\times\mathbf{a}_{2})}$,
the self-interaction reads
\begin{eqnarray}
\overline{\overline{\mathbf{G}}}_{0}(\mathbf{0}) & = & \ensuremath{{\displaystyle \int}}\frac{d^{2}G}{\left(2\pi\right)^{2}}\overline{\overline{\mathbf{g}}}_{0}(\mathbf{G})=\ensuremath{{\displaystyle \int}}\frac{\left|\mathbf{a}_{1}\times\mathbf{a}_{2}\right|}{\left(2\pi\right)^{2}}d^{2}G\frac{\overline{\overline{\mathbf{g}}}_{0}(\mathbf{G})}{\left|\mathbf{a}_{1}\times\mathbf{a}_{2}\right|}\nonumber \\
 & = & \ensuremath{{\displaystyle \int}}\frac{d^{2}G}{\left|\mathbf{b}_{1}\times\mathbf{b}_{2}\right|}\underbrace{\frac{\overline{\overline{\mathbf{g}}}_{0}(\mathbf{G})}{\left|\mathbf{a}_{1}\times\mathbf{a}_{2}\right|}}_{f(\mathbf{G})},\label{eq:S22}
\end{eqnarray}
which means Eq.~(\ref{eq:S8}) is in the form of Eq.~(\ref{eq:S21})
such that we can utilize Eqs.~(\ref{eq:S11}) and~(\ref{eq:S13})
to evaluate Eq.~(\ref{eq:S8}) for square and triangular lattices
by choosing an appropriate $T_{\square,\triangle}$. The absence of
oscillatory term in $\overline{\overline{\mathbf{g}}}_{0}(\mathbf{G})$
results in the good convergence of Euler-Maclaurin formula even if
we consider a small expansion order, and this is the reason we try
to calculate the RDDI in the reciprocal space.

In fact, equation~(\ref{eq:S21}) can be applied in real space by
adjusting the relationship among the lattice summation, integral and
the correction terms in Eq.~(\ref{eq:S21}), but the existence of
the oscillatory terms requires larger expansion order for the convergence.
Taking the RDDI without the self-interaction such as Eq.~(\ref{eq:S8})
for example, the left-hand side of Eq.~(\ref{eq:S8}) is the summation
on the whole lattice~(without the origin) in real space, and it can
be written in the following form by applying Eq.~(\ref{eq:S21})
on a 2D real lattice spanned by two direct lattice vectors $\mathbf{a}_{1}$
and $\mathbf{a}_{2}$

\begin{eqnarray}
\sum_{\mathbf{R}\neq0} &  & e^{-i\mathbf{k}_{B}\cdot\mathbf{R}}\overline{\overline{\mathbf{G}}}_{0}(\mathbf{R})\nonumber \\
= &  & \Biggl[\sum_{\substack{\mathbf{R}=m_{1}\mathbf{a}_{1}+m_{2}\mathbf{a}_{2}\\
\mathbf{R}\,\in\,T_{\square,\triangle}
}
}+\sum_{\substack{\mathbf{R}\,\notin\,T_{\square,\triangle}\\
\mathbf{R}\neq0
}
}\Biggr]e^{-i\mathbf{k}_{B}\cdot\mathbf{R}}\overline{\overline{\mathbf{G}}}_{0}(\mathbf{R})\nonumber \\
= &  & \Biggl[\ensuremath{{\displaystyle \int_{T_{\square,\triangle}}}}\frac{d^{2}R}{\left|\mathbf{a}_{1}\times\mathbf{a}_{2}\right|}+\sum_{\substack{\mathbf{R}\,\notin\,T_{\square,\triangle}\\
\mathbf{R}\neq0
}
}\Biggr]e^{-i\mathbf{k}_{B}\cdot\mathbf{R}}\overline{\overline{\mathbf{G}}}_{0}(\mathbf{R})\nonumber \\
 &  & +\text{ correction terms on }\partial T_{\square,\triangle}\nonumber \\
 &  & -\text{ remainder terms},\label{eq:S23}
\end{eqnarray}
where we need to choose a region $T_{\square,\triangle}$ that is
periodically-tiled without the origin of real space~(similar to FIG.~\ref{fig:S1}g
but its outer boundary extends infinitely) in avoid of the ill-defined
unrenormalized self-interaction $\overline{\overline{\mathbf{G}}}_{0}(\mathbf{0})$.
The last second line in Eq.~(\ref{eq:S23}) includes the integral
of RDDI over $T_{\square,\triangle}$ and the lattice summation~(without
the origin) over a finite region, i.e., the complement of $T_{\square}$,
and the former one requires model-dependent way to simplify the numerical
computation. For instance, we separate the integral of RDDI over $T_{\square,\triangle}$
into the following two parts
\begin{eqnarray}
\ensuremath{{\displaystyle \int_{T_{\square,\triangle}}}} &  & \frac{d^{2}R}{\left|\mathbf{a}_{1}\times\mathbf{a}_{2}\right|}e^{-i\mathbf{k}_{B}\cdot\mathbf{R}}\overline{\overline{\mathbf{G}}}_{0}(\mathbf{R})\nonumber \\
= &  & \underbrace{\ensuremath{{\displaystyle \int}}\frac{d^{2}R}{\left|\mathbf{a}_{1}\times\mathbf{a}_{2}\right|}e^{-i\mathbf{k}_{B}\cdot\mathbf{R}}\overline{\overline{\mathbf{G}}}_{0}(\mathbf{R})}_{\frac{1}{\left|\mathbf{a}_{1}\times\mathbf{a}_{2}\right|}\overline{\overline{\mathbf{g}}}_{0}(\mathbf{k}_{B})}\nonumber \\
 &  & -\ensuremath{{\displaystyle \int_{\mathbf{R}\,\notin\,T_{\square,\triangle}}}}\frac{d^{2}R}{\left|\mathbf{a}_{1}\times\mathbf{a}_{2}\right|}e^{-i\mathbf{k}_{B}\cdot\mathbf{R}}\overline{\overline{\mathbf{G}}}_{0}(\mathbf{R}),\label{eq:S24}
\end{eqnarray}
where the last term has a singularity at the origin~(self-interaction
$\overline{\overline{\mathbf{G}}}_{0}(\mathbf{0})$) that needs to
be treated individually for different electromagnetic environments.
Also, the convergence of Euler-Maclaurin formula is determined by
whether we can neglect the remainder terms in Eqs.~(\ref{eq:S11})
and~(\ref{eq:S13}) or not, and the existence of the oscillatory
terms~($e^{-i\mathbf{k}_{B}\cdot\mathbf{R}}\overline{\overline{\mathbf{G}}}_{0}(\mathbf{R})$)
usually requires a larger expansion order $M$ to achieve the desired
precision. These issues imply that it is not that efficient to deal
with Eq.~(\ref{eq:S8}) in real space. 

\begin{figure*}[!tp]
\centering{}\includegraphics{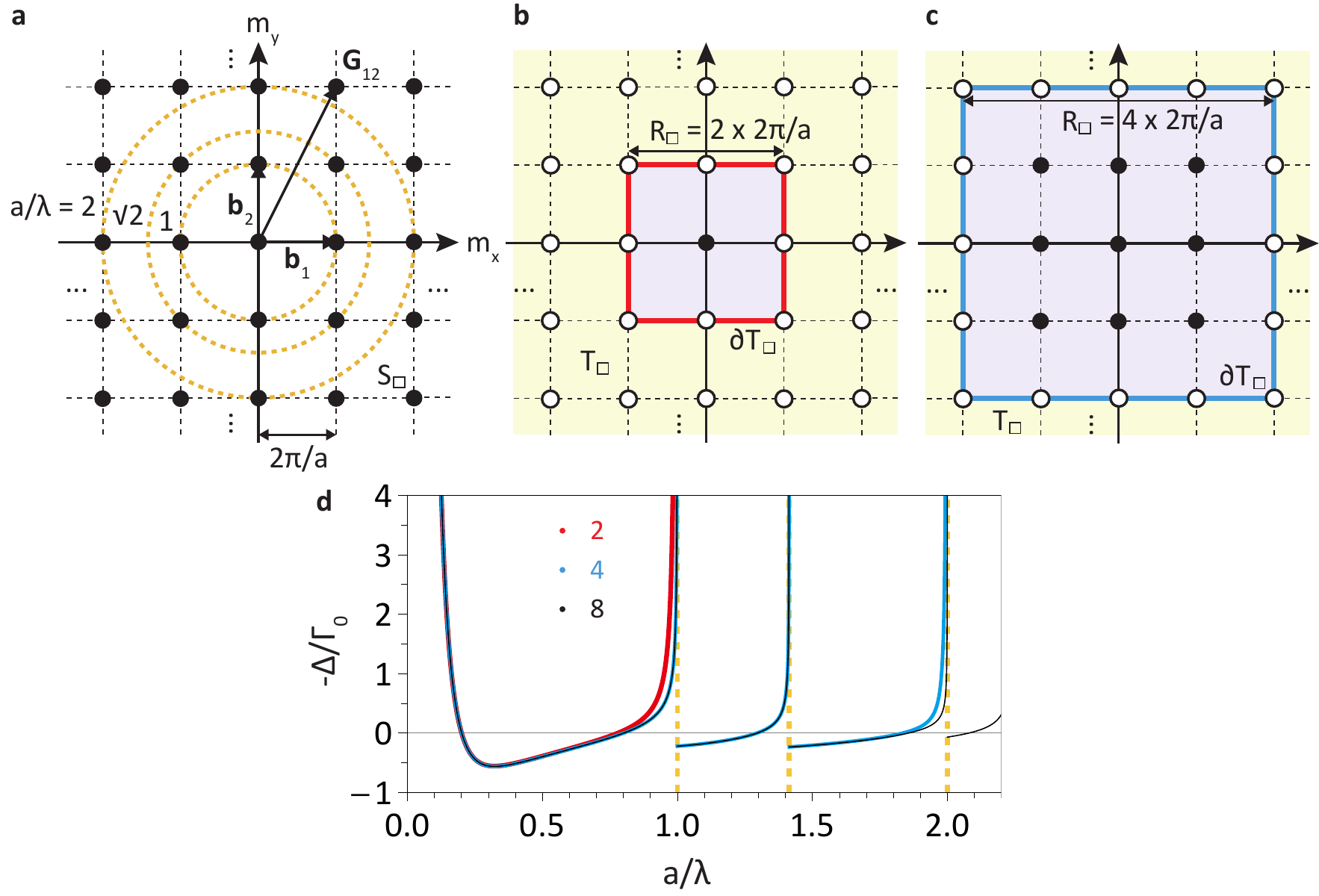}\caption{\label{fig:S2}\textbf{Demonstration of 2D generalization of Euler-Maclaurin
formula for 2D free-space square lattice at normal incidence.} \textbf{a},~2D
reciprocal space periodically tiled by the minimal period $S_{\square}$~(dashed
square). Reciprocal lattice points $\mathbf{G}_{m_{x}m_{y}}=m_{x}\mathbf{b}_{1}+m_{y}\mathbf{b}_{2}$
are spanned by two reciprocal lattice vectors $\mathbf{b}_{1}$ and
$\mathbf{b}_{2}$ and marked as black dots. Yellow dashed circles
indicate the smallest three light cones, defined by the ratio of atomic
constant $a$ to the resonant wavelength $\lambda$, intersecting
the reciprocal lattice points. \textbf{b},\textbf{c},~Light yellow
shaded region is the tilings $T_{\square}$ whose outer boundary extends
infinitely and inner boundary $\partial T_{\square}$ is shown as
a red~(blue) square centered at the origin with the length of sides
$R_{\square}$ 2~(4) in unit of $2\pi/a$. Lattice points and the
region of integration belong to $T_{\square}$ in the second line
of Eq.~(\ref{eq:S26}) are marked as black circles and light yellow
shaded region, and the rest are marked as black dots and light violet
shaded region. \textbf{d},~Collective Lamb shift calculated within
three different approximation boundaries $\partial T_{\square}$ whose
length of sides $R_{\square}$ are 2~(red), 4~(blue) and 8~(black)
in unit of $2\pi/a$, respectively. Red and blue curves associate
with different choices of $T_{\square}$ in \textbf{b},\textbf{c},
and three veritcal yellow dashed lines correspond to the smallest
three light cones in \textbf{a}.}
\end{figure*}

\subsubsection{Collective Lamb shift at normal incidence}

As a demonstration of our method, we consider the collective Lamb
shift of 2D square lattice in free space at normal incidence~\cite{Shahmoon2017S,Javanainen2019S}.
The collective Lamb shift $\Delta_{\mathbf{k}_{B}}$ and collective
decay rate $\Gamma_{\mathbf{k}_{B}}$ for a 2D square lattice are
given by
\begin{eqnarray}
\Delta & _{\mathbf{k}_{B}} & -\frac{i}{2}\Gamma_{\mathbf{k}_{B}}\Biggr|_{\mathbf{k}_{B}=0}\nonumber \\
 & = & -\frac{3\pi\Gamma_{0}c}{\omega_{0}a^{2}}\mathbf{e}_{d}\cdot\Biggl[\sum_{\mathbf{G}}\overline{\overline{\mathbf{g}}}_{0}(\mathbf{G})-\frac{a^{2}}{\left(2\pi\right)^{2}}\ensuremath{{\displaystyle \int}}d^{2}G\overline{\overline{\mathbf{g}}}_{0}(\mathbf{G})\Biggr]\cdot\mathbf{e}_{d}\nonumber \\
 & = & -\frac{3\lambda^{2}\Gamma_{0}}{8\pi a^{2}}\Biggl[\sum_{\mathbf{G}}-\frac{a^{2}}{\left(2\pi\right)^{2}}\ensuremath{{\displaystyle \int}}d^{2}G\Biggr]\nonumber \\
 & \times & \Biggl(1-\frac{\mathbf{G}^{2}}{2q^{2}}\Biggr)\Biggl[\frac{iq\Theta(q^{2}-\mathbf{G}^{2})}{\sqrt{q^{2}-\mathbf{G}^{2}}}+\frac{q\Theta(\mathbf{G}^{2}-q^{2})}{\sqrt{\mathbf{G}^{2}-q^{2}}}\Biggr],\label{eq:S25}
\end{eqnarray}
where $\mathbf{e}_{d}$ is the unit vector of arbitrary in-plane polarization.
Since the validity of 2D Euler-Maclaurin formula requires the smooth
function and free-space dyadic Green's function in reciprocal space
is ill-defined on the light cone, we choose a hollow region $T_{\square}$
that is periodically-tiled out of the light cone in which $\overline{\overline{\mathbf{g}}}_{0}(\mathbf{G}+\mathbf{k}_{B})$
is smooth and centered at the origin of reciprocal space~(see FIG.~\ref{fig:S2}b,c)
to ensure the $C_{4}$ rotational symmetry of 2D square lattice to
perform Eqs.~(\ref{eq:S11}) and~(\ref{eq:S21}) on the calculation
of Eq.~(\ref{eq:S25}), then Eq.~(\ref{eq:S25}) can be divided
into two parts by considering
\begin{eqnarray}
\sum_{\mathbf{G}} & - & \frac{a^{2}}{\left(2\pi\right)^{2}}\ensuremath{{\displaystyle \int}}d^{2}G\nonumber \\
 & = & \Biggl[\sum_{\mathbf{G}\notin T_{\square}}-\ensuremath{{\displaystyle \int_{\mathbf{G}\notin T_{\square}}}}\frac{a^{2}d^{2}G}{\left(2\pi\right)^{2}}\Biggr]+\underbrace{\Biggl[\sum_{\mathbf{G}\in T_{\square}}-\ensuremath{{\displaystyle \int_{T_{\square}}}}\frac{a^{2}d^{2}G}{\left(2\pi\right)^{2}}\Biggr]}_{=\text{ correction terms on }\partial T_{\square}}.\nonumber \\
\label{eq:S26}
\end{eqnarray}
Similar to Eq.~(\ref{eq:S23}), the first term in Eq.~(\ref{eq:S26})
is the finite lattice summation and the integral over a finite region
enclosed by the inner boundary of $T_{\square}$. Because the outer
boundary of $T_{\square}$ extends infinitely in which the correction
terms given by Eq.~(\ref{eq:S21}) vanish thanks to the ultraviolet
frequency cutoff, we approximate the second term in Eq.~(\ref{eq:S26})
by the correction terms on the inner boundary $\partial T_{\square}$
and discard the negligible remainder term in Eq.~(\ref{eq:S11}),
then we choose $\partial T_{\square}$ centered at the origin of reciprocal
space~(see FIG.~\ref{fig:S2}b,c) with the same length of sides
of $\partial T_{\square}$ in $m_{x}$ and $m_{y}$ directions, denoted
by $R_{\square}$, to ensure the $C_{4}$ rotational symmetry of 2D
square lattice.

Now we discuss how to decide a proper $\partial T_{\square}$ by comparing
the prediction given by different lengths of sides $R_{\square}$,
where we fix the expansion order $M$ at 2 in both $m_{x}$ and $m_{y}$
to demonstrate the efficiency of our method. Note that the validity
of the approximation given by Eq.~(\ref{eq:S11}) requires the small
enough remainder term. Once the different $R_{\square}$ give the
similar predictions within the desired error tolerance, we claim that
our approximation is valid since the difference between Eqs.~(\ref{eq:S26})
given by two different $R_{\square}$ and $R'_{\square}$~($R_{\square}<R'_{\square}$)
is
\begin{eqnarray}
 & - & \Biggl[\sum_{\mathbf{G}\in T_{\square}/T'_{\square}}-\ensuremath{{\displaystyle \int_{T_{\square}/T'_{\square}}}}\frac{a^{2}d^{2}G}{\left(2\pi\right)^{2}}\Biggr]\nonumber \\
 & + & \Biggl[\begin{alignedat}{1} & \text{correction terms}\\
 & \text{on }\partial T_{\square}
\end{alignedat}
\Biggr]-\Biggl[\begin{alignedat}{1} & \text{correction terms}\\
 & \text{on }\partial T'_{\square}
\end{alignedat}
\Biggr],\label{eq:S27}
\end{eqnarray}
which is nothing but the difference between the exact expression and
the related approximation, i.e., the remainder term, within the region
$T_{\square}/T'_{\square}$ enclosed by the two approximation boundaries
$\partial T_{\square}$ and $\partial T'_{\square}$.

In FIG.~\ref{fig:S2}d, we find that different approximation regions
$T_{\square}$ result in the same prediction of Eq.~(\ref{eq:S25})
except for the divergent behaviors. The reason why $R_{\square}=2\times\frac{2\pi}{a}$
deviates from the correct divergent behavior at $\frac{a}{\lambda}=1$
is that the divergence is well described by the dyadic Green's function
at reciprocal lattice points while the approximation given by Eq.~(\ref{eq:S11})
with $R_{\square}=2\times\frac{2\pi}{a}$ leaves only part of $\overline{\overline{\mathbf{g}}}_{0}(\mathbf{G})$
at reciprocal lattice points on $\partial T_{\square}$ in FIG.~\ref{fig:S2}b.
In addition, $R_{\square}=2\times\frac{2\pi}{a}$ can only be used
in subwavelength regime $\frac{a}{\lambda}<1$ since the inner boundary
of $T_{\square}$ needs to enclose the whole light cone whose diameter
is $2\times\frac{2\pi}{\lambda}$ in order to make the remainder term
negligible. Similarly, the calculation using $R_{\square}=4\times\frac{2\pi}{a}$
coincides with the case of $R_{\square}=8\times\frac{2\pi}{a}$ and
starts to deviate from the correct result when $\frac{a}{\lambda}>2$.
In conclusion, we need to choose an appropriate $\partial T_{\square}$
whose interior contains the light cone and all potentially divergent
terms determined by specific lattice constant and lattice geometry,
and then we use larger $R_{\square}$ to check the convergence. In
our experience, the calculation in the subwavelength regime using
$R_{\square}=4\times\frac{2\pi}{a}$ to $8\times\frac{2\pi}{a}$ is
good enough~(see FIG.~\ref{fig:S2}d).\\

\subsubsection{Off-diagonal terms}

Although the RDDI between the same sublattice is present in the form
of Eq.~(\ref{eq:S26}) in reciprocal space, it's not trivial to see
whether we can still approximate the RDDI between different sublattices
such as the off-diagonal terms in Eq.~(\ref{eq:S10}). We take honeycomb
lattice with nearest separation $a$ for example. The RDDI between
two sublattices has the following form\\
\begin{figure}[!t]
\centering{}\includegraphics[width=8.5cm]{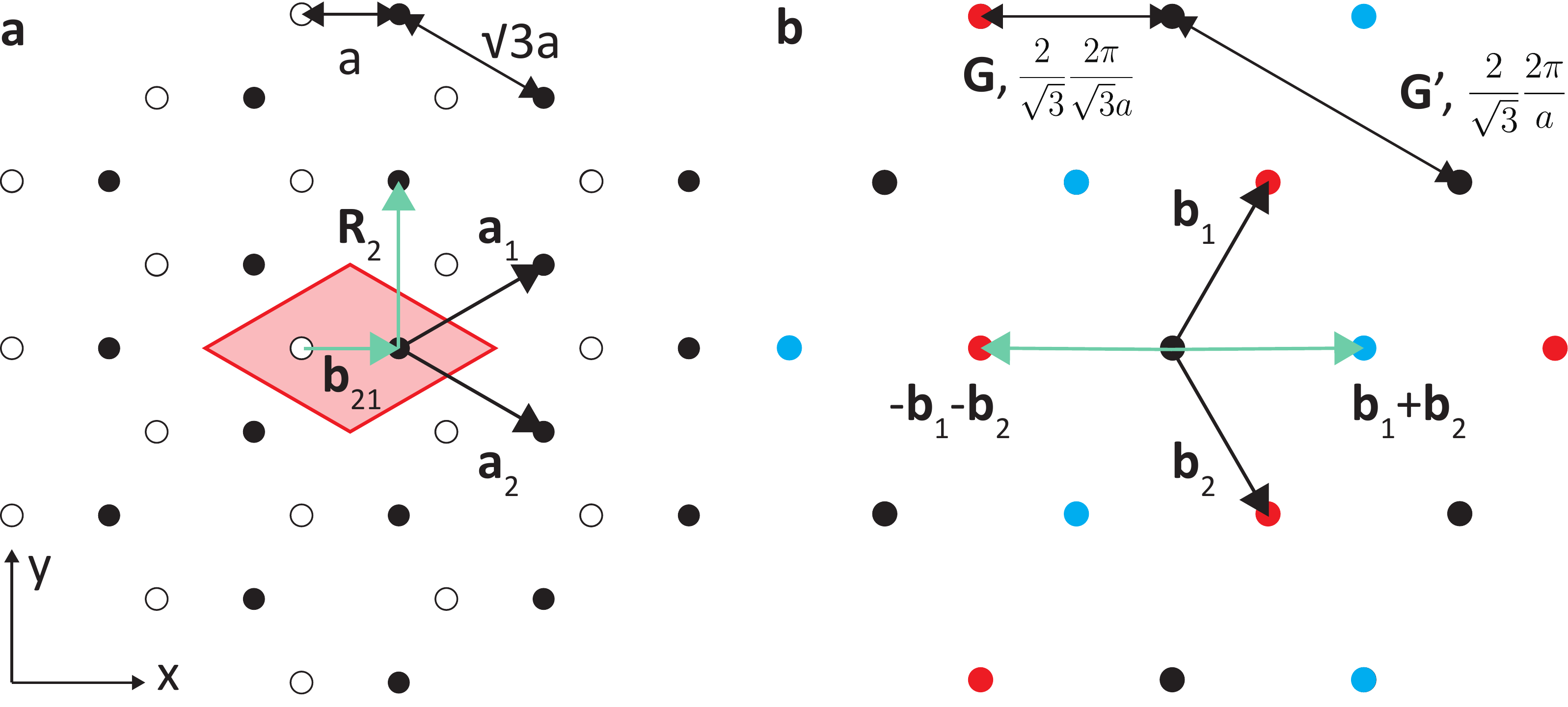}\caption{\label{fig:S3}\textbf{Honeycomb lattice.} \textbf{a}, Honeycomb lattice
with a nearest separation $a$ in real space consists of two triangular
sublattices 1~(white) and 2~(black) with the same lattice constant
$\sqrt{3}a$. $\mathbf{R}_{2}$ means the direct lattice vectors of
sublattice 2 and $\mathbf{b}_{21}$ is the nearest neighbor vector
within the periodic unit cell~(red shaded region). \textbf{b},~Reciprocal
lattice vectors $\mathbf{G}$ of honeycomb lattices are spanned by
$\mathbf{b}_{1}$ and $\mathbf{b}_{2}$ with the same length $\frac{2}{\sqrt{3}}\frac{2\pi}{\sqrt{3}a}$.
Black, blue and red points form three triangular sublattices categorized
by the global phase factors $1$, $e^{i\frac{2\pi}{3}}$ and $e^{i\frac{4\pi}{3}}$
in Eq.~(\ref{eq:S29}), respectively.}
\end{figure}
\begin{eqnarray}
 &  & \sum_{\mathbf{R}_{2},\sqrt{3}a}e^{-i\mathbf{k}_{B}\cdot(\mathbf{R}_{2}+\mathbf{b}_{21})}\overline{\overline{\mathbf{G}}}(\mathbf{R}_{2}+\mathbf{b}_{21})\nonumber \\
= &  & \frac{2}{\sqrt{3}}\frac{1}{(\sqrt{3}a)^{2}}\sum_{\mathbf{G},\frac{2}{\sqrt{3}}\frac{2\pi}{\sqrt{3}a}}e^{i\mathbf{G}\cdot\mathbf{b}_{21}}\overline{\overline{\mathbf{g}}}(\mathbf{G}+\mathbf{k}_{B}),\label{eq:S28}
\end{eqnarray}
where the subscripts of summation indicate the lengths of direct and
reciprocal lattice vectors~(see FIG.~\ref{fig:S3}). Since the reciprocal
lattice vectors $\mathbf{G}=m\mathbf{b}_{1}+n\mathbf{b}_{2}$ result
in three possible phase factors $e^{i\mathbf{G}\cdot\mathbf{b}_{21}}=e^{i\frac{2\pi}{3}(m+n)}$,
Eq.~(\ref{eq:S28}) can be decomposed into three parts
\begin{eqnarray}
 & {\displaystyle \sum_{\mathbf{G},\frac{2}{\sqrt{3}}\frac{2\pi}{\sqrt{3}a}}} & e^{i\mathbf{G}\cdot\mathbf{b}_{21}}\overline{\overline{\mathbf{g}}}(\mathbf{G}+\mathbf{k}_{B})\nonumber \\
= & {\displaystyle \sum_{\mathbf{G}',\frac{2}{\sqrt{3}}\frac{2\pi}{a}}} & \left[\overline{\overline{\mathbf{g}}}(\mathbf{G}'+\mathbf{k}_{B})+e^{i\frac{2\pi}{3}}\overline{\overline{\mathbf{g}}}(\mathbf{G}'-\mathbf{b}_{1}-\mathbf{b}_{2}+\mathbf{k}_{B})\right.\nonumber \\
 &  & +\left.e^{i\frac{4\pi}{3}}\overline{\overline{\mathbf{g}}}(\mathbf{G}'+\mathbf{b}_{1}+\mathbf{b}_{2}+\mathbf{k}_{B})\right].\label{eq:S29}
\end{eqnarray}
Note that $1+e^{i\frac{2\pi}{3}}+e^{i\frac{4\pi}{3}}=0$, which means
we can assign the self-interactions for each dyadic Green's function
in Eq.~(\ref{eq:S29}). That is, all infinite sums of RDDI in Eq.~(\ref{eq:S10})
can be approximated by our 2D generalization of Euler-Maclaurin formula,
and the efficiency helps us explore the photonic band structures over
the whole Brillouin zone.

\section{Non-Hermitian band structures}

Recently, bulk Fermi arcs and paired exceptional points~(EPs) have
been theoretically proposed and experimentally realized in minimal
two-band photonic crystals~\cite{Zhou2018S} by adding the radiation
loss to generate EPs split from a Dirac point in the Hermitian regime,
which arises from a $C_{4v}$ symmetry-breaking perturbation, such
as the change in lattice constant in one direction. However, we can
not obtain EPs in the same way due to the inherent non-Hermiticity
of RDDI in atomic arrays. Here we start from a non-Hermitian viewpoint
to discuss the criteria for the emergence of EPs.

We consider a rectangular atomic lattice in free space. Each atom
has a V-type energy level composed of one ground state $\left|g\right\rangle $
and two circularly-polarized excited states $\left|\pm\right\rangle =\mp(\left|x\right\rangle \pm i\left|y\right\rangle )/\sqrt{2}$
such that the system supports two in-plane polarizations. In the circularly-polarized
basis, the 2 $\times$ 2 effective Hamiltonian kernel is
\begin{widetext}
\begin{eqnarray}
{\cal H}{}_{\text{eff}}(\mathbf{k}_{B}) & = & \hbar\left(\omega_{0}-i\frac{\Gamma_{0}}{2}\right)\begin{pmatrix}1 & 0\\
0 & 1
\end{pmatrix}-\frac{3\pi\hbar\Gamma_{0}c}{2\omega_{0}}\sum_{\mathbf{R}\neq0}e^{-i\mathbf{k}_{B}\cdot\mathbf{R}}\nonumber \\
 &  & \times\begin{pmatrix}G_{0,xx}(\mathbf{R})+G_{0,yy}(\mathbf{R})+i\left[G_{0,xy}(\mathbf{R})-G_{0,yx}(\mathbf{R})\right] & G_{0,yy}(\mathbf{R})-G_{0,xx}(\mathbf{R})+i\left[G_{0,xy}(\mathbf{R})+G_{0,yx}(\mathbf{R})\right]\\
G_{0,yy}(\mathbf{R})-G_{0,xx}(\mathbf{R})-i\left[G_{0,xy}(\mathbf{R})+G_{0,yx}(\mathbf{R})\right] & G_{0,xx}(\mathbf{R})+G_{0,yy}(\mathbf{R})-i\left[G_{0,xy}(\mathbf{R})-G_{0,yx}(\mathbf{R})\right]
\end{pmatrix}\nonumber \\
 & = & \hbar(\omega_{0}+\Omega_{\mathbf{k}_{B}})\begin{pmatrix}1 & 0\\
0 & 1
\end{pmatrix}+\hbar\Gamma_{0}\begin{pmatrix}0 & \kappa_{+-}(\mathbf{k}_{B})\\
\kappa_{-+}(\mathbf{k}_{B}) & 0
\end{pmatrix}.\label{eq:S30}
\end{eqnarray}
\end{widetext}

Due to the dyadic form of Eq.~(\ref{eq:S4}), $G_{0,xy}(\mathbf{R})=G_{0,yx}(\mathbf{R})$
leads to the identical diagonal terms for $\left|\pm\right\rangle $.
In the absence of out-of-plane magnetic field $B\mathbf{e}_{z}$ that
lifts the degeneracy in $\left|\pm\right\rangle $ by a Zeeman shift
$\pm\hbar\mu B$, where $\mu$ is the magnetic moment, a real-valued
$\hbar\mu B\sigma_{z}$ chiral term would not be induced in the effective
Hamiltonian, and the emergence of degeneracy point only relies on
the off-diagonal term in Eq.~(\ref{eq:S30}), as discussed in the
main text.

To determine the criteria for the coalescence of two eigenstates of
Eq.~(\ref{eq:S30}), we first use Eqs.~(\ref{eq:S5}) and~(\ref{eq:S8})
to express the non-Hermitian couplings $\kappa_{+-(-+)}(\mathbf{k}_{B})$
in reciprocal space as~(assume the Bloch quasi-momentum $\mathbf{k}_{B}$
lies within the light cone)
\begin{align}
\kappa & _{+-(-+)}(\mathbf{k}_{B})=-\frac{3}{16\pi}\frac{\lambda^{2}}{\left|\mathbf{a}_{1}\times\mathbf{a}_{2}\right|}\Biggl[\sum_{\mathbf{G}}-\ensuremath{{\displaystyle \int}}\frac{\left|\mathbf{a}_{1}\times\mathbf{a}_{2}\right|}{\left(2\pi\right)^{2}}d^{2}G\Biggr]\nonumber \\
 & \times\left[\frac{i\Theta\left(q^{2}-(\mathbf{G}+\mathbf{k}_{B})^{2}\right)}{\sqrt{1-\frac{1}{q^{2}}(\mathbf{G}+\mathbf{k}_{B})^{2}}}+\frac{\Theta\left((\mathbf{G}+\mathbf{k}_{B})^{2}-q^{2}\right)}{\sqrt{\frac{1}{q^{2}}(\mathbf{G}+\mathbf{k}_{B})^{2}-1}}\right]\nonumber \\
 & \times\frac{1}{q^{2}}\left[(\mathbf{G}+\mathbf{k}_{B})_{x}\pm i(\mathbf{G}+\mathbf{k}_{B})_{y}\right]^{2},\label{eq:S31}
\end{align}
which can be calculated by our 2D generalization of Euler-Maclaurin
formula. The second line of Eq.~(\ref{eq:S31}) distinguishes between
the contributions within and out of the light cone, which results
in the anti-Hermitian and Hermitian couplings in Eq.~(\ref{eq:S30})
since only those modes lie within the light cone couple to outgoing
photons in free space. For a square lattice, non-Hermitian couplings
$\kappa_{+-(-+)}(\mathbf{k}_{B})$ vanish when $\mathbf{k}_{B}$ locates
at high symmetry points $\Gamma$~($\mathbf{k}_{B}=0$) in the first
Brillouin zone due to the $C_{4}$ rotational symmetry, which means
square lattices with in-plane polarizations have a non-Hermitian NDP
at $\Gamma$ and the corresponding eigenstates are distinct. This
NDP is unstable since its existence relies on more conditions than
that of EP in 2D system~\cite{Yang2021S}. In our case, this NDP
can be deformed into one or several EPs by breaking the $C_{4}$ rotational
symmetry, which can be realized by deforming a square lattice into
a rectangular or rhombic lattice.

According to the above discussion, the existence of NDP requires two
things: one is the crystalline $C_{4}$ rotational symmetry, and the
other is the absence of chiral term in the effective Hamiltonian.
In fact, the latter is protected by the following symmetry
\begin{eqnarray}
\sigma_{x}[{\cal H}{}_{\text{eff}}(\mathbf{k}_{B})]^{T}\sigma_{x} & = & {\cal H}{}_{\text{eff}}(\mathbf{k}_{B}),\label{eq:32}
\end{eqnarray}
and a 2 $\times$ 2 matrix $A$ satisfying Eq.~(\ref{eq:32}) is
generally in the form of $A=a_{0}I+a_{x}\sigma_{x}+a_{y}\sigma_{y}$,
where $a_{0,x,y}$ can be complex. Combining Eq.~(\ref{eq:32}) with
the in-plane parity~(inversion) symmetry $\mathcal{P}{\cal H}{}_{\text{eff}}(\mathbf{k}_{B})\mathcal{P}^{-1}=(-I){\cal H}{}_{\text{eff}}(-\mathbf{k}_{B})(-I)={\cal H}{}_{\text{eff}}(\mathbf{k}_{B})$
in two dimensions, which reverses the direction of polarization vectors
and in-plane momentum, we obtain
\begin{equation}
\sigma_{x}[{\cal H}{}_{\text{eff}}(-\mathbf{k}_{B})]^{T}\sigma_{x}={\cal H}{}_{\text{eff}}(\mathbf{k}_{B}),\label{eq:33}
\end{equation}
whose left-hand side is the Hermitian-conjugated time-reversal operation
$\mathcal{T}_{+}{\cal H}{}_{\text{eff}}(\mathbf{k}_{B})\mathcal{T}_{+}^{-1}$
with $\mathcal{T}_{+}^{\text{ }}\mathcal{T}_{+}^{*}=\sigma_{x}^{\text{ }}\sigma_{x}^{*}=1$,
and this equality means that ${\cal H}{}_{\text{eff}}(\mathbf{k}_{B})$
respects this Hermitian-conjugated time-reversal symmetry~\cite{Kawabata2019S}.
That is, both square and rectangular lattices respect this symmetry
since it requires no higher crystalline symmetry, such that our system
belongs to class $\text{AI}^{\dagger}$~\cite{Kawabata2019S}. As
a result, NDP in a square lattice is protected by the $C_{4}$ rotational
symmetry and the Hermitian-conjugated time-reversal symmetry, and
NDP would be annihilated if one of these two symmetries is broken.
Therefore, a Hermitian-conjugated time-reversal symmetry-breaking
perturbation induced by an out-of-plane magnetic field can break the
degeneracy at NDP, which is similar to the way an magnetic field opens
a gap at the time-reversal symmetry-protected Dirac point in the Hermitian
system.

\section{Atomic lattice in a ribbon geometry}

To calculate the OBC spectra in a ribbon geometry, we need to rewrite
the RDDI in 1D reciprocal space~(along the direction parallel to
the open boundary). We integrate out the other two directions in Eq.~(\ref{eq:S9})
to obtain
\begin{widetext}
\begin{eqnarray}
\overline{\overline{\mathbf{G}}}_{0}(\mathbf{r}_{\parallel},\mathbf{r}_{\perp}) & = & \ensuremath{{\displaystyle \int}}\frac{dk_{\parallel}}{\left(2\pi\right)^{2}}e^{i\mathbf{k}_{\parallel}\cdot\mathbf{r}_{\parallel}}\Biggl\{\Bigl(\overline{\overline{\mathbf{I}}}-\frac{\mathbf{k}_{\parallel}^{2}}{q^{2}}\mathbf{e}_{\parallel}\otimes\mathbf{e}_{\parallel}\Bigr)\left[\Theta(q^{2}-\mathbf{k}_{\parallel}^{2})\frac{i\pi}{2}H_{0}^{(1)}(\sqrt{q^{2}-\mathbf{k}_{\parallel}^{2}}|\mathbf{r}_{\perp}|)+\Theta(\mathbf{k}_{\parallel}^{2}-q^{2})K_{0}(\sqrt{\mathbf{k}_{\parallel}^{2}-q^{2}}|\mathbf{r}_{\perp}|)\right]\nonumber \\
 &  & -\Bigl(\mathbf{k}_{\parallel}\otimes\mathbf{r}_{\perp}+\mathbf{r}_{\perp}\otimes\mathbf{k}_{\parallel}\Bigr)\frac{1}{|\mathbf{r}_{\perp}|}\frac{1}{q^{2}}\left[\begin{aligned}- & \Theta(q^{2}-\mathbf{k}_{\parallel}^{2})\frac{\pi}{2}\sqrt{q^{2}-\mathbf{k}_{\parallel}^{2}}H_{1}^{(1)}(\sqrt{q^{2}-\mathbf{k}_{\parallel}^{2}}|\mathbf{r}_{\perp}|)\\
+ & \Theta(\mathbf{k}_{\parallel}^{2}-q^{2})i\sqrt{\mathbf{k}_{\parallel}^{2}-q^{2}}K_{1}(\sqrt{\mathbf{k}_{\parallel}^{2}-q^{2}}|\mathbf{r}_{\perp}|)
\end{aligned}
\right]\nonumber \\
 &  & -\mathbf{e}_{\perp}\otimes\mathbf{e}_{\perp}\left[\begin{aligned} & \Theta(q^{2}-\mathbf{k}_{\parallel}^{2})\frac{i\pi}{4}(1-\frac{\mathbf{k}_{\parallel}^{2}}{q^{2}})\biggl(H_{0}^{(1)}(\sqrt{q^{2}-\mathbf{k}_{\parallel}^{2}}|\mathbf{r}_{\perp}|)-H_{2}^{(1)}(\sqrt{q^{2}-\mathbf{k}_{\parallel}^{2}}|\mathbf{r}_{\perp}|)\biggr)\\
+ & \Theta(\mathbf{k}_{\parallel}^{2}-q^{2})\frac{1}{2}(1-\frac{\mathbf{k}_{\parallel}^{2}}{q^{2}})\biggl(K_{0}(\sqrt{\mathbf{k}_{\parallel}^{2}-q^{2}}|\mathbf{r}_{\perp}|)+K_{2}(\sqrt{\mathbf{k}_{\parallel}^{2}-q^{2}}|\mathbf{r}_{\perp}|)\biggr)
\end{aligned}
\right]\Biggr\}\nonumber \\
 & = & \ensuremath{{\displaystyle \int}}\frac{dk_{\parallel}}{2\pi}e^{i\mathbf{k}_{\parallel}\cdot\mathbf{r}_{\parallel}}\overline{\overline{\mathbf{g}}}_{0}(\mathbf{k}_{\parallel},\mathbf{r}_{\perp}),\label{eq:S32}
\end{eqnarray}
\end{widetext}

where $H_{n}^{(1)}$ is the Hankel function of the first kind, and
$K_{n}$ is the modified Bessel function of the second kind. Note
that Eq.~(\ref{eq:S32}) is valid for the nonzero atomic separation
$\mathbf{r}_{\perp}$ normal to the open boundary and used for calculating
the off-diagonal blocks of effective Hamiltonian~(see Eq.~(\ref{eq:S34})).
Owing to the exponentially decaying asymptotic behavior of the modified
Bessel functions, the coupling between two 1D infinite atomic chains
separated by $\mathbf{r}_{\perp}$ in a ribbon geometry~(denoting
their separation within the periodic unit cell, with a period $a_{\parallel}$,
along the parallel direction as $\mathbf{b}_{\parallel}$),
\begin{align}
\overline{\overline{\mathbf{t}}} & (\mathbf{k}_{\parallel},\mathbf{b}_{\parallel},\mathbf{r}_{\perp})\nonumber \\
 & =-\frac{3\pi\hbar\Gamma_{0}c}{\omega_{0}}\sum_{\mathbf{R}_{\parallel}}e^{-i\mathbf{k}_{\parallel}\cdot(\mathbf{R}_{\parallel}+\mathbf{b}_{\parallel})}\overline{\overline{\mathbf{G}}}_{0}(\mathbf{R}_{\parallel}+\mathbf{b}_{\parallel},\mathbf{r}_{\perp})\nonumber \\
 & =-\frac{3\pi\hbar\Gamma_{0}c}{\omega_{0}}\sum_{\mathbf{G}_{\parallel}}\frac{1}{a_{\parallel}}e^{i\mathbf{G}_{\parallel}\cdot\mathbf{b}_{\parallel}}\overline{\overline{\mathbf{g}}}_{0}(\mathbf{G}_{\parallel}+\mathbf{k}_{\parallel},\mathbf{r}_{\perp}),\label{eq:S33}
\end{align}
can be easily addressed by summing over several reciprocal lattice
vectors. For the self-interaction term $\overline{\overline{\mathbf{t}}}(\mathbf{k}_{\parallel},\mathbf{b}_{\parallel}=0,\mathbf{r}_{\perp}=0)$,
we should exclude $\mathbf{R}_{\parallel}=0$ term in the second line
of Eq.~(\ref{eq:S33}) besides setting $\mathbf{r}_{\perp},\mathbf{b}_{\parallel}=0$.
We can determine whether the atomic lattice in a ribbon geometry has
nonreciprocal coupling or not by Eq.~(\ref{eq:S33}). Since the reciprocity
of RDDI results in $\overline{\overline{\mathbf{g}}}_{0}(\mathbf{k}_{\parallel},\mathbf{r}_{\perp})=\overline{\overline{\mathbf{g}}}_{0}{}^{T}(-\mathbf{k}_{\parallel},-\mathbf{r}_{\perp})$,
this reciprocity leads to the reciprocal couplings in a ribbon geometry,
$\overline{\overline{\mathbf{t}}}(\mathbf{k}_{\parallel},\mathbf{b}_{\parallel},\mathbf{r}_{\perp})=\overline{\overline{\mathbf{t}}}{}^{T}(\mathbf{k}_{\parallel},-\mathbf{b}_{\parallel},-\mathbf{r}_{\perp})$,
only when $\mathbf{k}_{\parallel}=0$. Other reciprocal couplings
comes from spatial symmetries in the real lattice space, such as the
mirror symmetry.

\subsubsection{Energy spectra topology}

Following the same procedure as the 2D infinite atomic lattice, the
effective non-Hermitian Hamiltonian of $L$ 1D infinite atomic chains
that supports a collective excitation with a fixed parallel momentum
$\mathbf{k}_{\parallel}$ in a ribbon geometry is given by 1D Fourier
transformation of Eq.~(\ref{eq:S1})
\begin{figure}[H]
\centering{}\includegraphics{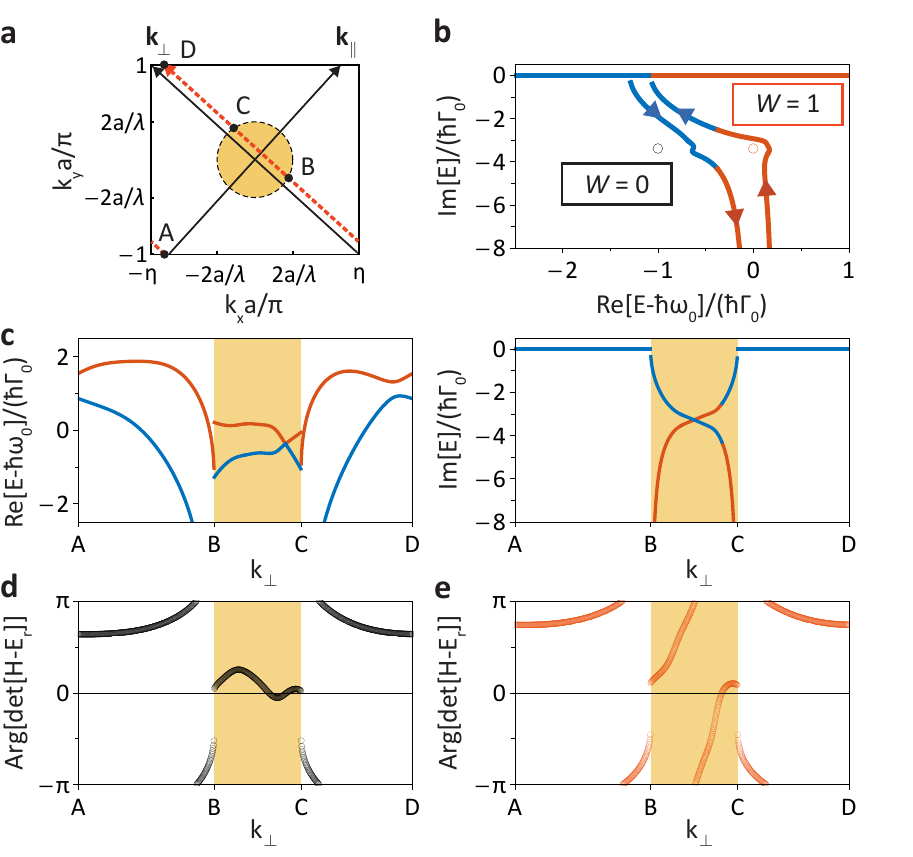}\caption{\label{fig:S4}\textbf{Bulk spectrum, band structure and winding numbers
in a ribbon geometry.} \textbf{a},\textbf{b},~Momentum slice~(orange
dashed line, which forms a closed loop $C$ on the $T^{2}$, 2D torus,
Brillouin zone) and bulk spectrum of a 2D rectangular~($\eta=1.1$)
lattice at fixed $\mathbf{k}_{\parallel}=0.1\pi/a$~(same as the
top panel of FIG.2b in the main text). Black and orange circles indicate
two topologically-distinct choices of the reference energy points
$E_{r}$, and the corresponding spectral winding numbers are shown
in FIG.~\ref{fig:S4}d,e, respectively. \textbf{c},~Bulk eigenenergy
as a function of $\mathbf{k}_{\perp}$, where the asymmetric behaviors
at the opposite $\mathbf{k}_{\perp}$ reflect the nonreciprocity.
Colors of two bands follow the same rule as those in \textbf{b}. \textbf{d},\textbf{e},~Except
for the discontinuities at the intersections of the light cone and
the closed loop $C$, i.e., $\mathbf{k}_{\perp,\text{B}}$ and $\mathbf{k}_{\perp,\text{C}}$,
the accumulated change in the $\text{arg}[\det[\mathcal{H}(\mathbf{k}_{\parallel},\mathbf{k}_{\perp})-E_{r}]]$
along the closed loop $C$ represents the corresponding topological
invariant $2\pi W(\mathbf{k}_{\parallel},E_{r})$.}
\end{figure}
\begin{eqnarray}
\hat{{\cal H}}_{L}(\mathbf{k}_{\parallel},\mathbf{r}^{\perp}) & = & \sum_{m=1}^{L}\sum_{\nu=x,y}\hbar\left(\omega_{0}-i\frac{\Gamma_{0}}{2}\right)\hat{\sigma}_{m\mathbf{k}_{\parallel}\nu}^{\dagger}\hat{\sigma}_{m\mathbf{k}_{\parallel}\nu}\nonumber \\
 & + & \sum_{m,n}\sum_{\mu,\nu=x,y}t_{\mu\nu}(\mathbf{k}_{\parallel},\mathbf{b}_{mn}^{\parallel},\mathbf{r}_{mn}^{\perp})\hat{\sigma}_{m\mathbf{k}_{\parallel}\mu}^{\dagger}\hat{\sigma}_{n\mathbf{k}_{\parallel}\nu},\nonumber \\
\label{eq:S34}
\end{eqnarray}
whose kernel is a $2L\times2L$ matrix. Numerical diagonalization
of the Hamiltonian kernel of Eq.~(\ref{eq:S34}) $\mathcal{H}_{L}(\mathbf{k}_{\parallel},\mathbf{r}_{\perp})$
gives the OBC spectra $\sigma[\mathcal{H}_{L}(\mathbf{k}_{\parallel},\mathbf{r}_{\perp})]$
and OBC eigenstates~$(\psi_{x}(\mathbf{r}_{\perp}),\psi_{y}(\mathbf{r}_{\perp}))^{T}$
of a single excited 2D atomic lattice in a ribbon geometry at a fixed
$\mathbf{k}_{\parallel}$ in the main text.

The bulk spectrum of a 2D atomic lattice in a ribbon geometry at a
fixed $\mathbf{k}_{\parallel}$ is obtained by considering its open
boundary extends infinitely such that it becomes a 2D infinite lattice
at a fixed $\mathbf{k}_{\parallel}$. Since the perpendicular direction
also becomes infinite, its Hamiltonian can be characterized by a perpendicular
momentum $\mathbf{k}_{\perp}$, and the bulk spectrum of this ribbon
geometry is nothing but the 2D photonic band structures $\sigma[\mathcal{H}_{\text{eff}}(\mathbf{k}_{\parallel},\mathbf{k}_{\perp})]$
at a fixed $\mathbf{k}_{\parallel}$, e.g., a momentum slice denoted
as the orange dashed line in FIG.~\ref{fig:S4}a.

In FIG.~\ref{fig:S4}, we discuss the details in the calculation
of bulk band properties of a ribbon geometry, where we use the same
parameter as FIG.~2b in the main text. Figure~\ref{fig:S4}b presents
the bulk energy spectrum parameterized by the perpendicular momentum
$\mathbf{k}_{\perp}$, and FIG.~\ref{fig:S4}c shows the $\mathbf{k}_{\perp}$
dependence of real and imaginary part of eigenenergy~(band structures),
where one of two bands diverges~(see the denominators in Eqs.~(\ref{eq:S5})
and~(\ref{eq:S34})) at the intersections of the light cone and the
momentum slice, i.e., $\mathbf{k}_{\perp,\text{B}}$ and $\mathbf{k}_{\perp,\text{C}}$.
These singularities are artifacts that arise from the infinite system,
while it does not matter if we are interested in the near-resonant
regime. Here we focus on the effect of RDDI and still keep these artifacts
for the consistency of the theoretical model.

As shown in FIG.~\ref{fig:S4}d,e, the discontinuities in the argument
of $\det[\mathcal{H}_{\text{eff}}(\mathbf{k}_{\parallel},\mathbf{k}_{\perp})-E_{r}\mathds{1}]$
at $\mathbf{k}_{\perp,\text{B}}$ and $\mathbf{k}_{\perp,\text{C}}$
make the determination of spectral winding number for a reference
energy $E_{r}$ ambiguous. Here we prove that these discontinuities
at $\mathbf{k}_{\perp,\text{B}}$ and $\mathbf{k}_{\perp,\text{C}}$
cancel each other out. First notice that these discontinuities lead
to a minus infinity to finite value jump in $\text{Re}[E_{2}(\mathbf{k}_{\parallel},\mathbf{k}_{\perp,\text{B}})]$,
a finite value to minus infinity jump in $\text{Re}[E_{2}(\mathbf{k}_{\parallel},\mathbf{k}_{\perp,\text{C}})]$,
a finite value to minus infinity jump in $\text{Im}[E_{1}(\mathbf{k}_{\parallel},\mathbf{k}_{\perp,\text{B}})]$,
and a minus infinity to finite value jump in $\text{Im}[E_{1}(\mathbf{k}_{\parallel},\mathbf{k}_{\perp,\text{C}})]$
when $\mathbf{k}_{\perp}$ is swept from $\text{A}$ to $\text{D}$
in FIG.~\ref{fig:S4}c. By rewriting $\det[\mathcal{H}_{\text{eff}}(\mathbf{k}_{\parallel},\mathbf{k}_{\perp})-E_{r}\mathds{1}]$
as~$(E_{1}(\mathbf{k}_{\parallel},\mathbf{k}_{\perp})-E_{r})(E_{2}(\mathbf{k}_{\parallel},\mathbf{k}_{\perp})-E_{r})$,
the discontinuity in $\text{arg}\det[\mathcal{H}_{\text{eff}}(\mathbf{k}_{\parallel},\mathbf{k}_{\perp,\text{B}})-E_{r}\mathds{1}]$
is given by
\begin{widetext}
\begin{eqnarray}
 &  & \underset{\mathbf{k}_{\perp}\rightarrow\mathbf{k}_{\perp,\text{B}}^{+}}{\text{lim}}\text{arg}\det[\mathcal{H}_{\text{eff}}(\mathbf{k}_{\parallel},\mathbf{k}_{\perp})-E_{r}\mathds{1}]-\underset{\mathbf{k}_{\perp}\rightarrow\mathbf{k}_{\perp,\text{B}}^{-}}{\text{lim}}\text{arg}\det[\mathcal{H}_{\text{eff}}(\mathbf{k}_{\parallel},\mathbf{k}_{\perp})-E_{r}\mathds{1}]\nonumber \\
 & = & \text{arg}\underset{\mathbf{k}_{\perp}\rightarrow\mathbf{k}_{\perp,\text{B}}^{+}}{\text{lim}}[E_{1}(\mathbf{k}_{\parallel},\mathbf{k}_{\perp})-E_{r}]-\text{arg}\underset{\mathbf{k}_{\perp}\rightarrow\mathbf{k}_{\perp,\text{B}}^{-}}{\text{lim}}[E_{2}(\mathbf{k}_{\parallel},\mathbf{k}_{\perp})-E_{r}]\nonumber \\
 &  & +\bigl\{\text{arg}\underset{\mathbf{k}_{\perp}\rightarrow\mathbf{k}_{\perp,\text{B}}^{+}}{\text{lim}}[E_{2}(\mathbf{k}_{\parallel},\mathbf{k}_{\perp})-E_{r}]-\text{arg}\underset{\mathbf{k}_{\perp}\rightarrow\mathbf{k}_{\perp,\text{B}}^{-}}{\text{lim}}[E_{1}(\mathbf{k}_{\parallel},\mathbf{k}_{\perp})-E_{r}]\bigr\}\nonumber \\
 & = & \text{arg}\underset{\mathbf{k}_{\perp}\rightarrow\mathbf{k}_{\perp,\text{B}}^{+}}{\text{lim}}i\text{Im}[E_{1}(\mathbf{k}_{\parallel},\mathbf{k}_{\perp})]-\text{arg}\underset{\mathbf{k}_{\perp}\rightarrow\mathbf{k}_{\perp,\text{B}}^{-}}{\text{lim}}\text{Re}[E_{2}(\mathbf{k}_{\parallel},\mathbf{k}_{\perp})]=\pi/2.\label{eq:S35}
\end{eqnarray}
\end{widetext}

Since there is one continuous band, the first equality in Eq.~(\ref{eq:S35})
reduces to the discontinuity in the other band, where the divergent
parts outside and within the light cone are a minus infinity in $\text{Re}[E_{2}(\mathbf{k}_{\parallel},\mathbf{k}_{\perp,\text{B}})]$
and a minus infinity in $\text{Im}[E_{1}(\mathbf{k}_{\parallel},\mathbf{k}_{\perp,\text{B}})]$,
respectively. Similarly, the discontinuity in $\text{arg}\det[\mathcal{H}_{\text{eff}}(\mathbf{k}_{\parallel},\mathbf{k}_{\perp})-E_{r}\mathds{1}]$
is $-\pi/2$. These two discontinuities in $\text{arg}\det[\mathcal{H}_{\text{eff}}(\mathbf{k}_{\parallel},\mathbf{k}_{\perp,\text{C}})-E_{r}\mathds{1}]$
are consistent with the jumps in FIG.~\ref{fig:S4}d,e at $\mathbf{k}_{\perp,\text{B}}$
and $\mathbf{k}_{\perp,\text{C}}$ such that these discontinuities
cancel each other out exactly, which makes the spectral winding number
integer-valued. Therefore, we can easily determine the respective
spectral winding numbers in FIG.~\ref{fig:S4}d,e as $W(\mathbf{k}_{\parallel},E_{r})=0$
and $W(\mathbf{k}_{\parallel},E_{r})=1$ by the overall change in
the $\text{arg}[\det[\mathcal{H}_{\text{eff}}(\mathbf{k}_{\parallel},\mathbf{k}_{\perp})-E_{r}]]$
along the closed loop $C$. We note that these discontinuities at
the light cone cancel each other out no matter what lattice configurations
we considered.

\subsubsection{Ribbon geometry with a finite width}

Besides the numerical diagonalization, OBC eigenstates can be solved
by the following recursive sequence~\cite{Yao2018S,Yokomizo2019S}~(here
we review the previous method in terms of our system)
\begin{eqnarray}
E_{\text{OBC}}\psi_{m\mu} & = & \hbar\left(\omega_{0}-i\frac{\Gamma_{0}}{2}\right)\psi_{m\mu}+\sum_{\nu=x,y}t_{\mu\nu}(\mathbf{k}_{\parallel},0,0)\psi_{m,\nu}\nonumber \\
 & + & \sum_{j=1}^{L-1}\sum_{\nu=x,y}t_{\mu\nu}(\mathbf{k}_{\parallel},\mathbf{b}_{m,m+j}^{\parallel},-j\boldsymbol{a}_{\perp})\psi_{m+j,\nu}\nonumber \\
 & + & \sum_{j=1}^{L-1}\sum_{\nu=x,y}t_{\mu\nu}(\mathbf{k}_{\parallel},\mathbf{b}_{m,m-j}^{\parallel},j\boldsymbol{a}_{\perp})\psi_{m-j,\nu},\nonumber \\
\label{eq:S36}
\end{eqnarray}
where $\psi_{m\mu}=\psi_{\mu}(m\boldsymbol{a}_{\perp})$ for $1\leq m\leq L,\ \mu=x,y$
and $\boldsymbol{a}_{\perp}$ is the lattice vector normal to the
open boundary. The ansatz to Eq.~(\ref{eq:S36}) is
\begin{eqnarray}
\psi_{m\mu} & = & \beta^{m}A_{\mu}(\beta),\ (\mu=x,y)\label{eq:S37}
\end{eqnarray}
By inserting Eq.~(\ref{eq:S37}) into Eq.~(\ref{eq:S36}), nontrivial
solutions of the coefficients $A_{\mu}(\beta)$ correspond to the
following characteristic equation
\begin{eqnarray}
\mathcal{H}_{\mu\nu}(\beta) & = & \hbar\left(\omega_{0}-i\frac{\Gamma_{0}}{2}\right)\delta_{\mu\nu}+\sum_{\nu=x,y}t_{\mu\nu}(\mathbf{k}_{\parallel},0,0)\nonumber \\
 & + & \sum_{j=1}^{L-1}t_{\mu\nu}(\mathbf{k}_{\parallel},\mathbf{b}_{0j}^{\parallel},-j\boldsymbol{a}_{\perp})\beta^{j}\nonumber \\
 & + & \sum_{j=1}^{L-1}t_{\mu\nu}(\mathbf{k}_{\parallel},\mathbf{b}_{j0}^{\parallel},j\boldsymbol{a}_{\perp})\beta^{-j},\label{eq:S39}
\end{eqnarray}
\begin{equation}
\det[\mathcal{H}(\beta)-E_{\text{OBC}}]=0,\label{eq:S38}
\end{equation}
\begin{equation}
\sum_{\nu=x,y}\mathcal{H}_{\mu\nu}(\beta)A_{\nu}(\beta)=E_{\text{OBC}}A_{\mu}(\beta),\label{eq:S40}
\end{equation}
where $\mathcal{H}(\beta)$ is a $2\times2$ matrix~(two in-plane
polarizations). Since there are $2\times[(L-1)+(L-1)]=4(L-1)$~(two
in-plane polarizations $\times$~(leftward interaction range $+$
rightward interaction range)) solutions of $\beta$ to Eq.~(\ref{eq:S38}),
OBC eigenstate is composed of the linear combination of these $4(L-1)$
modes
\begin{eqnarray}
\psi_{m\mu}^{\text{OBC}} & = & \sum_{i=1}^{4(L-1)}(\beta_{i})^{m}A_{i\mu},\label{eq:S41}
\end{eqnarray}
where $\beta_{i}$ are sorted by their moduli in the ascending order
and $A_{i\mu}=A_{\mu}(\beta_{i})$, and the OBCs are $\psi_{m\mu}^{\text{OBC}}=0$
for $-L+2\leq m\leq0$, $L+1\leq m\leq2L-1$, and $\mu=x,y$. According
to these OBCs, the existence of nontrivial solutions of $A_{i\mu}$
requires
\begin{widetext}
\begin{equation}
\left|\begin{array}{ccc}
(\beta_{1})^{-L+2} & \cdots & (\beta_{4(L-1)})^{-L+2}\\
(\beta_{1})^{-L+2}\frac{E_{\text{OBC}}-\mathcal{H}_{xx}(\beta_{1})}{\mathcal{H}_{xy}(\beta_{1})} & \cdots & (\beta_{4(L-1)})^{-L+2}\frac{E_{\text{OBC}}-\mathcal{H}_{xx}(\beta_{4(L-1)})}{\mathcal{H}_{xy}(\beta_{4(L-1)})}\\
\vdots & \vdots & \vdots\\
(\beta_{1})^{0} & \cdots & (\beta_{4(L-1)})^{0}\\
(\beta_{1})^{0}\frac{E_{\text{OBC}}-\mathcal{H}_{xx}(\beta_{1})}{\mathcal{H}_{xy}(\beta_{1})} & \cdots & (\beta_{4(L-1)})^{0}\frac{E_{\text{OBC}}-\mathcal{H}_{xx}(\beta_{4(L-1)})}{\mathcal{H}_{xy}(\beta_{4(L-1)})}\\
(\beta_{1})^{L+1} & \cdots & (\beta_{4(L-1)})^{L+1}\\
(\beta_{1})^{L+1}\frac{E_{\text{OBC}}-\mathcal{H}_{xx}(\beta_{1})}{\mathcal{H}_{xy}(\beta_{1})} & \cdots & (\beta_{4(L-1)})^{L+1}\frac{E_{\text{OBC}}-\mathcal{H}_{xx}(\beta_{4(L-1)})}{\mathcal{H}_{xy}(\beta_{4(L-1)})}\\
\vdots & \vdots & \vdots\\
(\beta_{1})^{2L-1} & \cdots & (\beta_{4(L-1)})^{2L-1}\\
(\beta_{1})^{2L-1}\frac{E_{\text{OBC}}-\mathcal{H}_{xx}(\beta_{1})}{\mathcal{H}_{xy}(\beta_{1})} & \cdots & (\beta_{4(L-1)})^{2L-1}\frac{E_{\text{OBC}}-\mathcal{H}_{xx}(\beta_{4(L-1)})}{\mathcal{H}_{xy}(\beta_{4(L-1)})}
\end{array}\right|=0.\label{eq:S42}
\end{equation}
\end{widetext}

Note that $\beta_{i}$, $E_{\text{OBC}}$ and $A_{i\mu}$ are self-consistently
determined by Eqs.~(\ref{eq:S38}) to~(\ref{eq:S42}). Rewriting
$\beta$ as $e^{i(\mathbf{k}_{\perp}+i\boldsymbol{\kappa}_{\perp})\cdot\boldsymbol{a}_{\perp}}$,
the modes with $|\beta|=1$~($|\beta|\neq1$) in the thermodynamic
limit are extended Bloch modes~(non-Bloch modes), and $\mathcal{H}(\beta)$
becomes the non-Bloch real-space Hamiltonian.

To see how the long-range interactions modify the non-Bloch band theory,
here we briefly review the celebrated generalized Brillouin zone~\cite{Yao2018S,Yokomizo2019S,Kawabata2020S}
in the systems with one degree of freedom in the unit cell and finite-range~($\mu$)
couplings. The leading term in Eq.~(\ref{eq:S42}) now becomes $\beta_{1}^{-\mu+1}\beta_{2}^{-\mu+2}\cdots\beta_{2\mu-1}^{L+\mu-1}\beta_{2\mu}^{L+\mu}$.
If each power function of $\beta_{i}$ in $\beta_{1}^{-\mu+1}\beta_{2}^{-\mu+2}\cdots\beta_{2\mu-1}^{L+\mu-1}\beta_{2\mu}^{L+\mu}$
is exponentially growing in $L$, i.e., $\beta_{i}$ has a fixed length
scale, the term closest to $\beta_{1}^{-\mu+1}\beta_{2}^{-\mu+2}\cdots\beta_{2\mu-1}^{L+\mu-1}\beta_{2\mu}^{L+\mu}$
would be its interchange of two adjacent $\beta_{\mu}$ and $\beta_{\mu+1}$
whose magnitudes are closest to unity. To make Eq.~(\ref{eq:S42})
vanish in the thermodynamic limit, these two leading terms should
cancel each other out, which results in the condition~$|\beta_{\mu}|=|\beta_{\mu+1}|$.
The trajectories of $\beta_{\mu}$ and $\beta_{\mu+1}$, parameterized
by their phases and known as generalized Brillouin zone, form the
continuum bands $\sigma[\mathcal{H}(\beta)]$ in the thermodynamic
limit. The conditions for the continuum bands depend on the symmetry
class of the non-Hermitian matrices. For instance, the non-Bloch modes
$\beta$ and $\beta^{-1}$ appear in pair in the presence of reciprocal
couplings, where the non-Hermitian Hamiltonians belong to class $\text{AI}^{\dagger}$
or $\text{AII}^{\dagger}$~\cite{Kawabata2019S}, and the conditions
for the continuum bands would be modified~\cite{Kawabata2020S}.
\begin{figure*}[!t]
\centering{}\includegraphics{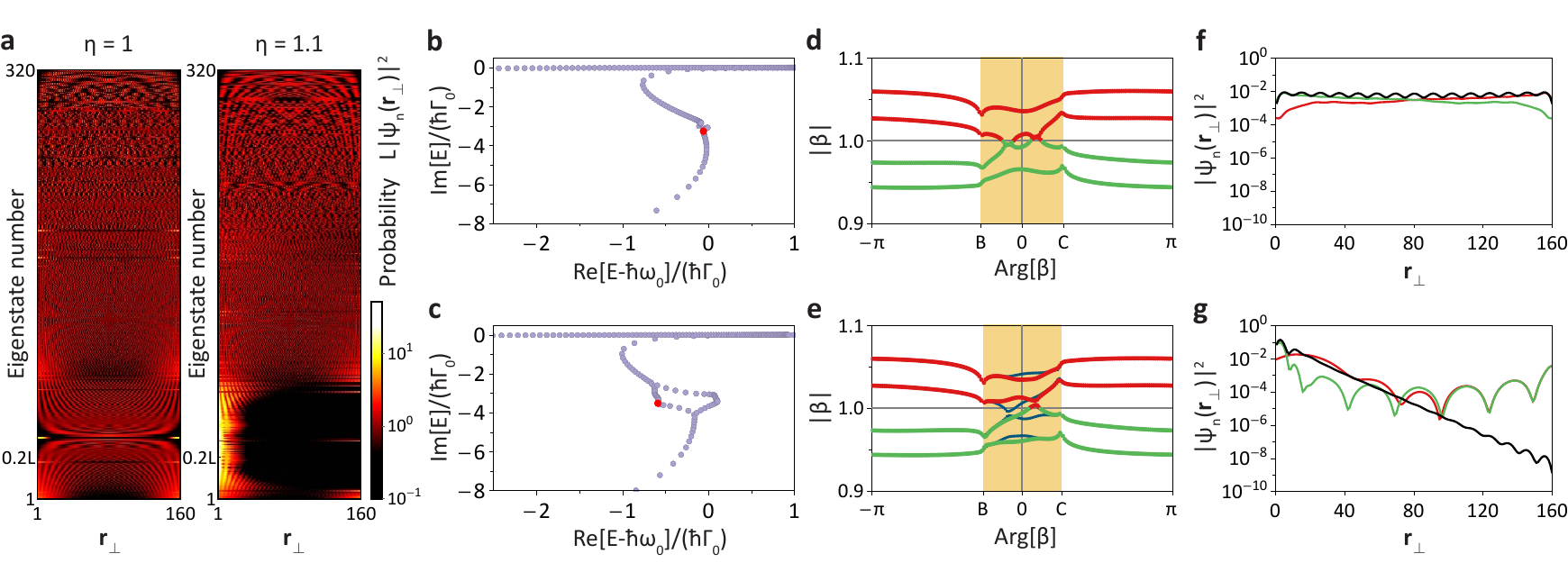}\caption{\label{fig:S5}\textbf{Open boundary eigenstates and $\boldsymbol{\beta}$.
}Open boundary eigenstates~(\textbf{a}) and spectra~(\textbf{b},\textbf{c})
of 2D square and rectangular atomic lattices in ribbon geometries
at $160$ uint cells. The plots are obtained with the same parameters
as FIG.~3 in the main text. In the left~(right) panel of \textbf{a},
the modes with complex eigenenergies are delocalized~(localized)
in 2D square~(rectangular) atomic lattices, where the eigenenergy
of the~$(0.2L)$th eigenstate is colored in red in \textbf{b}~(\textbf{c}).
\textbf{d}-\textbf{g},~The $4(L-1)$ solutions of $\beta$~(\textbf{d},\textbf{e})
and OBC eigenstate profiles~(\textbf{f},\textbf{g}) of the~$(0.2L)$th
eigenstates of 2D square~(\textbf{d},\textbf{f}) and rectangular~(\textbf{e},\textbf{g})
atomic lattices in ribbon geometries. The modes with $|\beta|>1$~($|\beta|<1$)
that are localized at the boundary in increasing~(decreasing) $\mathbf{r}_{\perp}$
direction are colored in red~(green). Owing to the mirror symmetry,
those $\beta$ and $\beta^{-1}$~(blue) in \textbf{d} are paired,
while $\beta$ in \textbf{e} are not paired in the presence of non-reciprocal
coupling. In contrast to the tight-binding model, OBC eigenstates~(black)
in \textbf{f},\textbf{g} are composed of several $\beta$ in the presence
of long-range interaction, where the contributions from the modes
with $|\beta|>1$ and $|\beta|<1$ are colored in red and green, respectively.}
\end{figure*}

However, the size independence of the decay length of $\beta$ breaks
down in the presence of long-range interactions. The reason that this
size independence of $\beta$ is reasonable in the standard non-Bloch
band theory is the size independence of both the characteristic equation~(Eq.~(\ref{eq:S38}))
and the numerically calculated OBC spectra. Once we have the size-independent
characteristic equation and the size-dependent OBC spectra, Eq.~(\ref{eq:S38})
predicts the presence of $\beta$ with size-dependent decay length,
where the competition between the $\beta$ with constant and size-dependent
decay lengths results in the critical non-Hermitian skin effect~\cite{Li2020S,Yokomizo2021S}.
In our case, both the characteristic equation and OBC spectra are
size-dependent, which leads to $\beta$ with no fixed scales, and
the number of $\beta$ is also size-dependent. As a result, each OBC
eigenstate is not dominated by two but several $\beta$, which means
the decaying behaviors of $\beta$ do not directly reflect the decaying
behaviors of OBC eigenstates due to the interference among several
$\beta$. We note that the size dependence of OBC spectra and $\beta$
should be self-consistently determined by Eqs.~(\ref{eq:S38}) to~(\ref{eq:S42}),
while we can first numerically diagonalize the Hamiltonian kernel
of Eq.~(\ref{eq:S34}) $\mathcal{H}_{L}(\mathbf{k}_{\parallel},\mathbf{r}_{\perp})$
to see the size dependence of OBC spectra such that we can qualitatively
determined the size dependence of $\beta$.

Here we choose the~$(0.2L)$th eigenstate of 2D square and rectangular
lattices in a ribbon geometry open on the~$(\bar{1}1)$ plane~(see
FIG.~\ref{fig:S5}a-c) to numerically demonstrate the above discussions.
In FIG.~\ref{fig:S5}d-g, we numerically calculate the $4(L-1)$
solutions of $\beta$ for square and rectangular lattices in ribbon
geometries at $L=160$ unit cells. Due to the mirror symmetry over
the open boundary of a square lattice in a ribbon geometry, the effective
coupling is reciprocal and the Hamiltonian kernel of Eq.~(\ref{eq:S34})
$\mathcal{H}_{L}(\mathbf{k}_{\parallel},\mathbf{r}_{\perp})$ respects
the Hermitian-conjugated time-reversal symmetry~\cite{Kawabata2019S}
\begin{eqnarray}
\mathcal{T}_{+}^{\text{ }}[\mathcal{H}_{L}(\mathbf{k}_{\parallel},\mathbf{r}_{\perp})]^{T}\mathcal{T}_{+}^{-1} & = & \mathcal{H}_{L}(\mathbf{k}_{\parallel},\mathbf{r}_{\perp}),\label{eq:S43}\\
\mathcal{T}_{+}^{\text{ }}\mathcal{T}_{+}^{*} & = & +1,\label{eq:S44}
\end{eqnarray}
where a unitary matrix $\mathcal{T}_{+}^{\text{ }}=\mathds{1}\otimes\sigma_{z}$
when $\mathcal{H}_{L}(\mathbf{k}_{\parallel},\mathbf{r}_{\perp})$
is defined in the linearly-polarized basis along the parallel and
perpendicular directions. The left-hand side of Eq.~(\ref{eq:S43})
is the Hamiltonian kernel after the mirror reflection with respect
to the open boundary in a ribbon geometry, which reverses the lattice
site $\mathbf{r}_{\perp}$, leaves the parallel polarization unchanged,
and flips the perpendicular polarization. That is, a non-Hermitian
Hamiltonian kernel $\mathcal{H}_{L}(\mathbf{k}_{\parallel},\mathbf{r}_{\perp})$
of a square lattice in a ribbon geometry that respects the mirror
symmetry over the open boundary belongs to class $\text{AI}^{\dagger}$~\cite{Kawabata2019S}.
Since each $\beta$ becomes $\beta^{-1}$ when the lattice sites $\mathbf{r}_{\perp}$
are reversed, the corresponding Hamiltonian $\mathcal{H}(\beta)$
respects
\begin{eqnarray}
\sigma_{z}[\mathcal{H}(\beta)]^{T}\sigma_{z} & = & \mathcal{H}(\beta^{-1}).\label{eq:S45}
\end{eqnarray}
Accordingly, when $\beta$ is a solution to the characteristic equation
Eq.~(\ref{eq:S38}), we have
\begin{eqnarray}
\det[\mathcal{H}(\beta^{-1})-E_{\text{OBC}}] & = & \det[\sigma_{z}[\mathcal{H}(\beta)]^{T}\sigma_{z}-E_{\text{OBC}}]\nonumber \\
 & = & \det[\mathcal{H}(\beta)-E_{\text{OBC}}]\nonumber \\
 & = & 0,\label{eq:S46}
\end{eqnarray}
which means $\beta^{-1}$ is also a solution to Eq.~(\ref{eq:S38}).
Therefore, the degenerate $\beta$ and $\beta^{-1}$ appear in pair
in the presence of reciprocal couplings~\cite{Kawabata2020S}, which
is also shown in FIG.~\ref{fig:S5}d, where the numerically calculated
$4(L-1)$ solutions of $\beta$ coincide with their reciprocal $\beta^{-1}$
colored in blue~(right behind those red and green dots). By contrast,
$4(L-1)$ soultions of $\beta$ are not paired in FIG.~\ref{fig:S5}e
since the effective coupling in a rectangular lattice in a ribbon
geometry is nonreciprocal.

We can also reconstruct each OBC eigenstate by these $\beta$. In
FIG.~\ref{fig:S5}h, it is clear to see that the decaying behavior
of OBC eigenstate~(skin mode) does not arise from those decaying
modes with $|\beta|<1$~(green) but from the interference of the
growing and decaying $\beta$. Also, delocalized OBC eigenstate in
FIG.~\ref{fig:S5}g is not composed of two degenerate $\beta$ but
several $\beta$. Therefore, we conclude that the OBC eigenstates
of a non-Hermitian Hamiltonian in the presence of long-range interactions
can not be described by two dominant $\beta$, which implies that
the generalized Brillouin zone is inapplicable when the coupling range
is not fixed.

\section{Simulations on optical responses of atomic lattices}

In this section, we start from the review of light scattering off
a collection of point-like dipoles~\cite{Shahmoon2017S}, which shows
the collective optical responses and the single diffraction order
from the normal process~(no change in the in-plane momentum) in the
2D Brillouin zone. By eliminating the only detuning-dependent quantity,
i.e., the Green's function or the collective polarizability, we are
able to extract the two-band effective bulk Hamiltonian from the scattering
matrices at only two detunings, and its eigenenergy spectrum reflects
the emergence of EP by tuning the lattice constant ratio. This scattering
method requires no detuning-resolved measurement followed by the fitting
with Lorentzian functions~\cite{Su2021S} or the resonance enhancement~\cite{Zhou2018S},
which is quite different from the existing techniques to the best
of our knowledge.

\subsubsection{Scattering problem}

In the weak driven limit, each atom labeled by lattice site $n$ in
a 2D finite atomic array composed of $N$ identical atoms driven at
detuning $|\delta|\ll\omega_{0}=cq$~(near resonance) can be modeled
by a classical point-like dipole emitter $\mathbf{p}(\mathbf{r}'_{m})$
located at $\mathbf{r}'_{n}$ on the 2D atomic plane~($z'_{n}=0$).
The scattering problem with a given incident field $\mathbf{E}_{\text{inc}}(\mathbf{r})$
is now described by the solution to the Maxwell's curl equation $\nabla\times\nabla\times\mathbf{E}(\mathbf{r})-q^{2}\mathbf{E}(\mathbf{r})=q^{2}\epsilon_{0}^{-1}\sum_{m=1}^{N}\mathbf{p}(\mathbf{r}'_{m})$
\begin{eqnarray}
\mathbf{E}_{\text{tot}}(\mathbf{r}) & = & \mathbf{E}_{\text{inc}}(\mathbf{r})+\frac{q^{2}}{\epsilon_{0}}\sum_{m=1}^{N}\overline{\overline{\mathbf{G}}}_{0}(\mathbf{r}-\mathbf{r}'_{m})\cdot\mathbf{p}(\mathbf{r}'_{m})\nonumber \\
 & = & \mathbf{E}_{\text{inc}}(\mathbf{r})+\mathbf{E}_{\text{sc}}(\mathbf{r}),\label{eq:S47}
\end{eqnarray}
\begin{equation}
\mathbf{p}(\mathbf{r}'_{m})=\alpha(\delta)\mathbf{E}_{\text{tot}}(\mathbf{r}'_{m}),\label{eq:S48}
\end{equation}
where $\alpha(\delta)$ is the bare polarizability of an atom near
resonance~(ignore the non-radiative decay rate)
\begin{equation}
\alpha(\delta)=-\frac{3}{4\pi^{2}}\epsilon_{0}\lambda^{3}\frac{\Gamma_{0}/2}{\delta+i\Gamma_{0}/2}.\label{eq:S49}
\end{equation}
The spatial distribution of induced dipoles $\mathbf{p}(\mathbf{r}'_{m})$
can be obtained from Eq.~(\ref{eq:S47}) by setting the field position
$\mathbf{r}$ on the 2D atomic plane and excluding the self-interaction
$\overline{\overline{\mathbf{G}}}_{0}(\mathbf{0})$. Inserting Eq.~(\ref{eq:S48})
into Eq.~(\ref{eq:S47}), we have

\begin{eqnarray}
\mathbf{p} & = & \Bigl[\alpha^{-1}(\delta)\mathds{1}-\frac{q^{2}}{\epsilon_{0}}\overline{\overline{\mathds{G}}}_{0}\Bigr]^{-1}\cdot\mathbf{E}_{\text{inc}}\nonumber \\
 & = & -\frac{2}{3\Gamma_{0}}\frac{4\pi^{2}}{\epsilon_{0}\lambda^{3}}\frac{1}{(\delta+\omega_{0})\mathds{1}-{\cal H}{}_{\text{eff}}^{\text{3D}}/\hbar}\cdot\mathbf{E}_{\text{inc}}\nonumber \\
 & = & \overline{\overline{\mathds{G}}}(\delta)\cdot\mathbf{E}_{\text{inc}},\label{eq:S50}
\end{eqnarray}
where the induced dipoles $\mathbf{p}=[p_{\mu}(\mathbf{r}'_{m})]_{3N\times1}$,
the dyadic Green's function $\overline{\overline{\mathds{G}}}_{0}=[G_{0,\mu\nu}(\mathbf{r}'_{mn})]_{3N\times3N}$,
the incident field $\mathbf{E}_{\text{inc}}=[E_{\text{inc},\mu}(\mathbf{r}'_{m})]_{3N\times1}$,
the effective Hamiltonian kernel ${\cal H}{}_{\text{eff}}^{\text{3D}}$
in Eq.~(\ref{eq:S3}), and the collective polarizability $\overline{\overline{\mathds{G}}}(\delta)=[G_{\mu\nu}(\delta,\mathbf{r}'_{mn})]_{3N\times3N}$
on the atomic lattice plane are defined in the tensor product space
between the atomic site and three independent polarizations. Collective
polarizability $\overline{\overline{\mathds{G}}}(\delta)$ can be
also regarded as the Green's function in the linear response theory.
The scattered field at $\mathbf{r}$ generated by these induced dipoles
at $\mathbf{r}'$ is given by the dyadic Green's function according
to Eq.~(\ref{eq:S47}).
\begin{eqnarray}
\mathbf{E}_{\text{sc}}(\mathbf{r}) & = & \frac{q^{2}}{\epsilon_{0}}\sum_{m=1}^{N}\overline{\overline{\mathbf{G}}}_{0}(\mathbf{r}-\mathbf{r}'_{m})\cdot\mathbf{p}(\mathbf{r}'_{m})\nonumber \\
 & = & \sum_{m=1}^{N}\overline{\overline{\mathbf{g}}}_{\text{sc}}(\mathbf{r}-\mathbf{r}'_{m})\cdot\sum_{n\neq m}\overline{\overline{\mathbf{G}}}(\delta,\mathbf{r}'_{mn})\cdot\mathbf{E}_{\text{inc}}(\mathbf{r}'_{n})\nonumber \\
 & = & \sum_{n=1}^{N}\overline{\overline{\mathbf{S}}}(\delta,\mathbf{r}-\mathbf{r}'_{n})\cdot\mathbf{E}_{\text{inc}}(\mathbf{r}'_{n}),\label{eq:S51}
\end{eqnarray}
where $\overline{\overline{\mathbf{S}}}(\delta,\mathbf{r}-\mathbf{r}'_{n})$,
scattering matrix associated with the phase shift from the propagation,
is a $3\times3$ matrix.

For an infinite 2D atomic array shined by an incident plane wave $\mathbf{E}_{\text{inc}}(\mathbf{r})=\mathbf{E}_{0,\mathbf{k}}e^{i\mathbf{k}\cdot\mathbf{r}}$,
the scattered field from the atomic array is now given by Eq.~(\ref{eq:S51})
\begin{eqnarray}
\mathbf{E}_{\text{sc}}(\mathbf{r}) & = & \sum_{m}\overline{\overline{\mathbf{g}}}_{\text{sc}}(\mathbf{r}-\mathbf{r}'_{m})\cdot\sum_{n\neq m}\overline{\overline{\mathbf{G}}}(\delta,\mathbf{r}'_{mn})\cdot\mathbf{E}_{0,\mathbf{k}}e^{i\mathbf{k}_{\parallel}\cdot\mathbf{r}'_{n}}\nonumber \\
 & = & \overline{\overline{\mathbf{g}'}}_{\text{sc}}(\mathbf{k},\mathbf{r})\cdot\overline{\overline{\mathbf{G}'}}(\delta,\mathbf{k}_{\parallel})\cdot\mathbf{E}_{0,\mathbf{k}}\nonumber \\
 & = & \overline{\overline{\mathbf{S}'}}(\delta,\mathbf{k},\mathbf{r})\cdot\mathbf{E}_{0,\mathbf{k}},\label{eq:S52}
\end{eqnarray}
with
\begin{equation}
\overline{\overline{\mathbf{g}'}}_{\text{sc}}(\mathbf{k},\mathbf{r})=\sum_{m}\frac{q^{2}}{\epsilon_{0}}\overline{\overline{\mathbf{G}}}_{0}(\mathbf{r}-\mathbf{r}'_{m})e^{i\mathbf{k}_{\parallel}\cdot\mathbf{r}'_{m}},\label{eq:S53}
\end{equation}
and
\begin{equation}
\overline{\overline{\mathbf{G}'}}(\delta,\mathbf{k}_{\parallel})=\sum_{n\neq m}\overline{\overline{\mathbf{G}}}(\delta,\mathbf{r}'_{mn})e^{-i\mathbf{k}_{\parallel}\cdot\mathbf{r}'_{mn}},\label{eq:S54}
\end{equation}
where the Fourier component $\overline{\overline{\mathbf{g}'}}_{\text{sc}}(\mathbf{k},\mathbf{r})$
is obtained from Eq.~(\ref{eq:S9}) by integrating out its $k_{z}$
component
\begin{widetext}
\begin{eqnarray}
\overline{\overline{\mathbf{G}}}_{0}(\mathbf{r}-\mathbf{r}'_{m}) & = & \ensuremath{{\displaystyle \int}}\frac{d^{2}k_{\parallel}}{(2\pi)^{3}}\Biggl\{\biggl[\overline{\overline{\mathbf{I}}}-\Bigl(1-\frac{\mathbf{k}_{\parallel}^{2}}{q^{2}}\Bigr)\mathbf{e}_{z}\otimes\mathbf{e}_{z}-\frac{1}{q^{2}}\mathbf{k}_{\parallel}\otimes\mathbf{k}_{\parallel}\biggr]\biggl[i\pi\Theta(q^{2}-\mathbf{k}_{\parallel}^{2})\frac{e^{i\sqrt{q^{2}-\mathbf{k}_{\parallel}^{2}}|z|}}{\sqrt{q^{2}-\mathbf{k}_{\parallel}^{2}}}+\pi\Theta(\mathbf{k}_{\parallel}^{2}-q^{2})\frac{e^{-\sqrt{\mathbf{k}_{\parallel}^{2}-q^{2}}|z|}}{\sqrt{\mathbf{k}_{\parallel}^{2}-q^{2}}}\biggr]\nonumber \\
 &  & +\biggl[-\frac{1}{q^{2}}(\mathbf{k}_{\parallel}\otimes\mathbf{e}_{z}+\mathbf{e}_{z}\otimes\mathbf{k}_{\parallel})\text{sgn}(z)\biggr]\biggl[i\pi\Theta(q^{2}-\mathbf{k}_{\parallel}^{2})e^{i\sqrt{q^{2}-\mathbf{k}_{\parallel}^{2}}|z|}+\pi\Theta(\mathbf{k}_{\parallel}^{2}-q^{2})e^{-\sqrt{\mathbf{k}_{\parallel}^{2}-q^{2}}|z|}\biggr]\Biggr\} e^{i\mathbf{k}_{\parallel}\cdot(\mathbf{r}-\mathbf{r}'_{m})}\nonumber \\
 & = & \ensuremath{{\displaystyle \int}}\frac{d^{2}k_{\parallel}}{(2\pi)^{2}}e^{i\mathbf{k}_{\parallel}\cdot(\mathbf{r}-\mathbf{r}'_{m})}\overline{\overline{\mathbf{g}}}_{0}(\mathbf{k}_{\parallel},z).\label{eq:S55}
\end{eqnarray}
\end{widetext}

Similar to Eq.~(\ref{eq:S8}), the Fourier component $\overline{\overline{\mathbf{g}'}}_{\text{sc}}(\mathbf{k},\mathbf{r})$
can be regarded as the infinite summation on the Fourier transformed
$\overline{\overline{\mathbf{g}}}_{0}(\mathbf{k}_{\parallel},\mathbf{r})$
\begin{equation}
\overline{\overline{\mathbf{g}'}}_{\text{sc}}(\mathbf{k},\mathbf{r})=\sum_{\mathbf{G}}\frac{q^{2}}{\epsilon_{0}}\frac{\overline{\overline{\mathbf{g}}}_{0}(\mathbf{G}+\mathbf{k}_{\parallel},z)}{\left|\mathbf{a}_{1}\times\mathbf{a}_{2}\right|}e^{i\mathbf{k}_{\parallel}\cdot\mathbf{r}}.\label{eq:S56}
\end{equation}
By identifying the $z$ dependence of Eq.~(\ref{eq:S56}), we can
categorize the collective emission into the propagating wave $e^{i\sqrt{q^{2}-\mathbf{k}_{\parallel}^{2}}|z|}$
and the evanescent wave $e^{-\sqrt{\mathbf{k}_{\parallel}^{2}-q^{2}}|z|}$,
in the same manner as what we have discussed in Eq.~(\ref{eq:S9}),
and only the propagating wave contributes to far-field radiations.
Also, the existence of $|z|$ ensures the symmetric scattered field
with respect to the atomic plane~(see the top panels of FIG.~\ref{fig:S6}a,b).
When there is only one mode~($\mathbf{G}=0$) in reciprocal space
that contributes to the decay~(single diffraction order when $a/\lambda<0.5$
for square lattice), the Fourier component $\overline{\overline{\mathbf{g}'}}_{\text{sc}}(\mathbf{k},\mathbf{r})$
in 3D reciprocal space is
\begin{align}
\overline{\overline{\mathbf{g}'}}_{\text{sc}}(\mathbf{k},\mathbf{r}) & \bigl|_{q|z|\gg1}=\frac{q^{2}}{\epsilon_{0}}\frac{ie^{i\mathbf{k}_{\parallel}\cdot\mathbf{r}}e^{i\sqrt{q^{2}-\mathbf{k}_{\parallel}^{2}}|z|}}{2\left|\mathbf{a}_{1}\times\mathbf{a}_{2}\right|\sqrt{q^{2}-\mathbf{k}_{\parallel}^{2}}}\nonumber \\
 & \times\left[\begin{aligned} & \overline{\overline{\mathbf{I}}}-\Bigl(1-\frac{\mathbf{k}_{\parallel}^{2}}{q^{2}}\Bigr)\mathbf{e}_{z}\otimes\mathbf{e}_{z}-\frac{1}{q^{2}}\mathbf{k}_{\parallel}\otimes\mathbf{k}_{\parallel}\\
- & \frac{1}{q^{2}}(\mathbf{k}_{\parallel}\otimes\mathbf{e}_{z}+\mathbf{e}_{z}\otimes\mathbf{k}_{\parallel})\text{sgn}(z)\sqrt{q^{2}-\mathbf{k}_{\parallel}^{2}}
\end{aligned}
\right]\nonumber \\
 & =g'_{\text{sc}}(\mathbf{k},\mathbf{r})\bigl|_{q|z|\gg1}(\overline{\overline{\mathbf{I}}}-\mathbf{e}_{k}\otimes\mathbf{e}_{k})\nonumber \\
 & =g'_{\text{sc}}(\mathbf{k},\mathbf{r})\bigl|_{q|z|\gg1}(\mathbf{e}_{p}\otimes\mathbf{e}_{p}+\mathbf{e}_{s}\otimes\mathbf{e}_{s}),\label{eq:S57}
\end{align}
which projects the scattered field on the subspace spanned by $p$-
and $s$-polarized basis vectors due to the dyadic operator $\overline{\overline{\mathbf{I}}}+q^{-2}\nabla\otimes\nabla$
of RDDI and $k_{z}=\text{sgn}(z)\sqrt{q^{2}-\mathbf{k}_{\parallel}^{2}}$.
Here the $p$- and $s$-polarized basis vectors $\mathbf{e}_{p}=\text{cos}\theta\text{cos}\phi\mathbf{e}_{x}+\text{cos}\theta\text{sin}\phi\mathbf{e}_{y}-\text{sin}\theta\mathbf{e}_{z}$
and $\mathbf{e}_{s}=\text{sin}\phi\mathbf{e}_{x}-\text{cos}\phi\mathbf{e}_{y}$
are defined for the scattered far-field plane wave indicated by a
wave vector $\mathbf{k}$ with magnitude $|\mathbf{k}|=q$, the polar
angle $\theta=\text{tan}^{-1}(k_{z}/q)$, and azimuthal angle $\phi=\text{tan}^{-1}(k_{y}/k_{x})$.

\subsubsection{Extracted effective Hamiltonian}

If a 2D finite atomic array exhibits the corresponding bulk properties,
we can assume $\overline{\overline{\mathbf{g}}}_{\text{sc}}(\mathbf{r}-\mathbf{r}')$
also serves as a projector such that the scattering matrix $\overline{\overline{\mathbf{S}}}(\delta,\mathbf{r}-\mathbf{r}')$
on the transverse polarization subspace at each incident angle can
be determined by calculating the scattered fields arising from $p$-
and $s$-polarized incident fields. In this case, we are able to extract
the information of the counterpart of the Green's function $\overline{\overline{\mathbf{G}}}(\delta,\mathbf{r}')$
in 2D reciprocal space~(Brillouin zone) $\overline{\overline{\mathbf{G}'}}(\delta,\mathbf{k}_{\parallel})$.
The reason we are interested in this $\overline{\overline{\mathbf{G}'}}(\delta,\mathbf{k}_{\parallel})$
is that $\overline{\overline{\mathbf{G}'}}(\delta,\mathbf{k}_{\parallel})$
involves the effective Hamiltonian kernel, which is obtained from
the Fourier transformation of Eq.~(\ref{eq:S50})

\begin{eqnarray}
\mathbf{p}_{\mathbf{k}_{\parallel}} & = & \Bigl[\alpha^{-1}(\delta)\overline{\overline{\mathbf{I}}}-\frac{q^{2}}{\epsilon_{0}}\sum_{n\neq m}\overline{\overline{\mathbf{G}_{0}}}(\mathbf{r}'_{mn})e^{-i\mathbf{k}_{\parallel}\cdot\mathbf{r}'_{mn}}\Bigr]^{-1}\cdot\mathbf{E}_{\text{inc},\mathbf{k}_{\parallel}}\nonumber \\
 & = & -\frac{3}{4\pi^{2}}\epsilon_{0}\lambda^{3}\frac{\Gamma_{0}}{2}\Bigl[(\delta+\omega_{0})\overline{\overline{\mathbf{I}}}-\frac{1}{\hbar}{\cal H}_{\text{eff}}^{\text{3D}}(\mathbf{k}_{\parallel})\Bigr]^{-1}\cdot\mathbf{E}_{\text{inc},\mathbf{k}_{\parallel}}\nonumber \\
 & = & \overline{\overline{\mathbf{G}'}}(\delta,\mathbf{k}_{\parallel})\cdot\mathbf{E}_{\text{inc},\mathbf{k}_{\parallel}},\label{eq:S58}
\end{eqnarray}
where ${\cal H}_{\text{eff}}^{\text{3D}}(\mathbf{k}_{\parallel})$
is defined in Eq.~(\ref{eq:S6}). Comparing the denominators of Eq.~(\ref{eq:S49})
and $\overline{\overline{\mathbf{G}'}}(\delta,\mathbf{k}_{\parallel})$,
we can see the real and imaginary parts of ${\cal H}_{\text{eff}}^{\text{3D}}(\mathbf{k}_{\parallel})$
actually represent the collective Lamb shift and the overall decay
rate $E_{1,2}(\mathbf{k})=\hbar(\omega_{0}+\Delta_{\mathbf{k}})-\frac{i}{2}\hbar\Gamma_{\mathbf{k}}$,
such that the information of EPs of a two-band effective Hamiltonian
kernel ${\cal H}_{\text{eff}}(\mathbf{k}_{\parallel})$ in Eq.~(\ref{eq:S30})
could be extracted from the Fourier transformed Green's function $\overline{\overline{\mathbf{G}'}}(\delta,\mathbf{k}_{\parallel})$
if we can discard the out-of-plane components.
\begin{figure*}[!t]
\centering{}\includegraphics{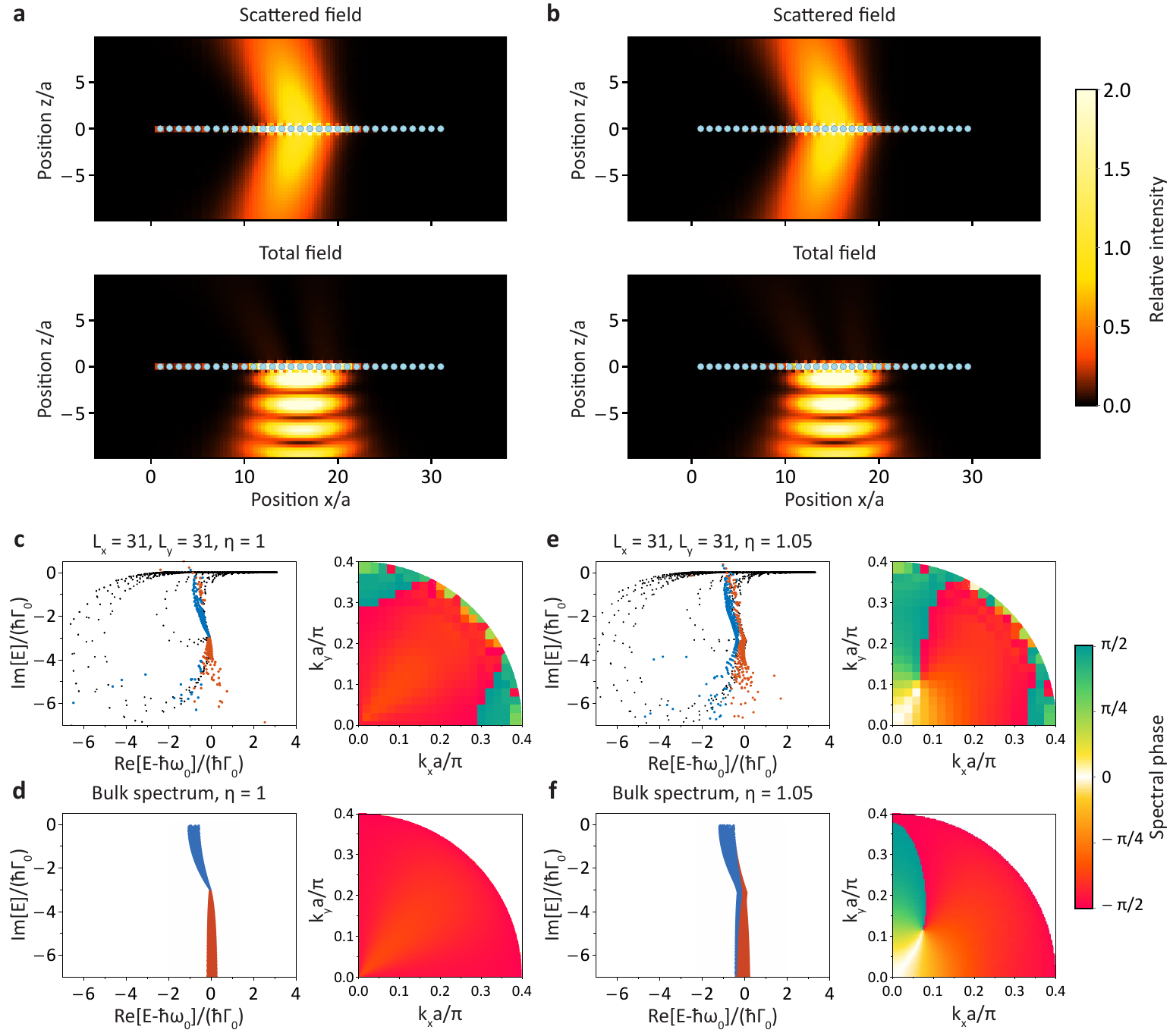}\caption{\label{fig:S6}\textbf{Bulk properties extracted from scattering problems
in finite systems. }High-reflectivity optical mirror formed by 2D
square~(\textbf{a},~$\eta=1$) and rectangular~(\textbf{b},~$\eta=1.05$)
subwavelength atomic lattices~($a/\lambda=0.2$) with rectangle-shaped
open boundaries. The probe beams in the simulations are $p$-polarized
Gaussian beams with in-plane momenta $\mathbf{k}_{\parallel}=-0.1\pi/a\mathbf{e}_{x}$~(at
oblique incident angle $\theta\approx14.5\lyxmathsym{\protect\textdegree}$),
which propagate toward positive $z$ direction and shine at the centers~$(x,y,z)=(16a/\eta,16a,0a)$
of atomic lattices with beam waists $w_{0}=4a=0.8\lambda$ and detuning
$\delta=0\Gamma_{0}$. Here we consider $y/a=16$ plane to draw the
relative intensity~$|\mathbf{E}_{\text{sc,tot}}(\mathbf{r})|^{2}/|\mathbf{E}_{\text{inc}}(\mathbf{r})|^{2}$
of scattered field $\mathbf{E}_{\text{sc}}(\mathbf{r})$~(top panels)
and total field $\mathbf{E}_{\text{tot}}(\mathbf{r})$~(bottom panels),
and the blue dots represent the configuration of atoms on $y/a=16$
plane. \textbf{c}-\textbf{f},~Eigenenergy spectra~(left panels)
and spectral windings of eigenenergies indicated by a spectral phase
$\text{Arg}[E_{1}(\mathbf{k})-E_{2}(\mathbf{k})]$~(right panels)
extracted from the simulations in finite systems~(\textbf{c},\textbf{e},
we consider $\delta=\pm0.5\Gamma_{0}$ to use Eq.~(\ref{eq:S61}))
and obtained from bulk band structures~(\textbf{d},\textbf{f}), respectively.
At small incident angles, finite system simulations in both lattice
configurations capture the bulk properties well, where the spectral
phase presents the half-integer spectral winding around EP, has a
discontinuity at the bulk Fermi arc, and becomes zero at imaginary
bulk Fermi arc in the right panel of \textbf{e}. In the numerical
simulations, we consider a far-detuned $\pi$ transition with detuning
$30\Gamma_{0}$ with respect to two in-plane polarizations $\left|\pm\right\rangle $.}
\end{figure*}

Another thing to be addressed in order to extract bulk Hamiltonians
on 2D atomic arrays is that the scattering matrix lives in the transverse
polarization subspace, defined in 3D real space $\mathbf{r}$ spanned
by $\mathbf{e}_{p,s}$, while the two-band effective Hamiltonian lives
in the in-plane polarization subspace, defined on 2D atomic array
$\mathbf{r}'$ spanned by $\mathbf{e}_{x,y}$. To transfer the transverse
polarization subspace to the in-plane polarization subspace at single
diffraction order, notice that Eq.~(\ref{eq:S58}) means that far-field
radiations always live in the transverse polarization subspace, and
${\cal P}_{p,s}\overline{\overline{\mathbf{S}'}}(\delta,\mathbf{k},\mathbf{r})\bigl|_{q|z|\gg1}=g'_{\text{sc}}(\mathbf{k},\mathbf{r})\bigl|_{q|z|\gg1}{\cal P}_{p,s}\overline{\overline{\mathbf{G}'}}(\delta,\mathbf{k}_{\parallel})$
are both projected onto the transverse polarization subspace. Since
in-plane polarizations and out-of-plane polarization in $\overline{\overline{\mathbf{G}'}}(\delta,\mathbf{k}_{\parallel})$
are decoupled due to the 2D geometry, only the $pp$ component of
${\cal P}_{p,s}\overline{\overline{\mathbf{G}'}}(\delta,\mathbf{k}_{\parallel})$
contains the $zz$ component of $\overline{\overline{\mathbf{G}'}}(\delta,\mathbf{k}_{\parallel})$,
and it becomes negligible when out-of-plane polarization is far-detuned
from in-plane polarizations or the level structure is V-type. For
instance, we can express the $pp$ component of ${\cal P}_{p,s}\overline{\overline{\mathbf{G}'}}(\delta,\mathbf{k}_{\parallel})$
in terms of ${\cal P}_{\text{IP}}\overline{\overline{\mathbf{G}'}}(\delta,\mathbf{k}_{\parallel})$
and ${\cal P}_{z}\overline{\overline{\mathbf{G}'}}(\delta,\mathbf{k}_{\parallel})$~(where
IP stands for in-plane)
\begin{eqnarray}
\bigl[{\cal P}_{p,s}\overline{\overline{\mathbf{G}'}}(\delta,\mathbf{k}_{\parallel})\bigr]_{pp} & = & ({\cal P}_{\text{IP}}\mathbf{e}_{p})^{\dagger}\cdot\bigl[{\cal P}_{\text{IP}}\overline{\overline{\mathbf{G}'}}(\delta,\mathbf{k}_{\parallel})\bigr]\cdot({\cal P}_{\text{IP}}\mathbf{e}_{p})\nonumber \\
 & + & \underbrace{({\cal P}_{z}\mathbf{e}_{p})^{\dagger}\cdot\bigl[{\cal P}_{z}\overline{\overline{\mathbf{G}'}}(\delta,\mathbf{k}_{\parallel})\bigr]\cdot({\cal P}_{z}\mathbf{e}_{p})}_{\text{negligible}},\nonumber \\
\label{eq:S59}
\end{eqnarray}
where
\begin{alignat}{1}
{\cal P}_{\text{IP},z} & \overline{\overline{\mathbf{G}'}}(\delta,\mathbf{k}_{\parallel})\label{eq:S60}\\
= & -\frac{3}{4\pi^{2}}\epsilon_{0}\lambda^{3}\frac{\Gamma_{0}}{2}\Bigl[(\delta+\omega_{0}){\cal P}_{\text{IP},z}\overline{\overline{\mathbf{I}}}-\frac{1}{\hbar}{\cal P}_{\text{IP},z}{\cal H}_{\text{eff}}^{\text{3D}}(\mathbf{k}_{\parallel})\Bigr]^{-1}.\nonumber 
\end{alignat}
In this case, ${\cal P}_{p,s}\overline{\overline{\mathbf{G}'}}(\delta,\mathbf{k}_{\parallel})$
involves only the in-plane part of $\overline{\overline{\mathbf{G}'}}(\delta,\mathbf{k}_{\parallel})$
such that we can obtain ${\cal P}_{\text{IP}}\overline{\overline{\mathbf{G}'}}(\delta,\mathbf{k}_{\parallel})$
and ${\cal H}_{\text{eff}}(\mathbf{k}_{\parallel})={\cal P}_{\text{IP}}{\cal H}{}_{\text{eff}}^{\text{3D}}(\mathbf{k}_{\parallel})$
from two scattering matrices at different detunings $\delta_{1}$
and $\delta_{2}$. Specifically, we consider scattering matrices at
two detunings $\delta_{1}$ and $\delta_{2}$ to eliminate their identical
$\overline{\overline{\mathbf{g}'}}_{\text{sc}}(\mathbf{k},\mathbf{r})\bigl|_{q|z|\gg1}$
parts, from which we obtain the following detuning-independent effective
Hamiltonian kernel composed of only two in-plane polarizations
\begin{widetext}
\begin{eqnarray}
{\cal H}_{\text{eff}}(\mathbf{k}_{\parallel}) & = & (\delta_{1}+\omega_{0}){\cal P}_{\text{IP}}\overline{\overline{\mathbf{I}}}\label{eq:S61}\\
 & - & (\delta_{1}-\delta_{2})\Bigl[{\cal P}_{\text{IP}}\overline{\overline{\mathbf{S}'}}(\delta_{1},\mathbf{k},\mathbf{r})\bigl|_{q|z|\gg1}\Bigr]^{-1}\cdot\biggl\{\Bigl[{\cal P}_{\text{IP}}\overline{\overline{\mathbf{S}'}}(\delta_{1},\mathbf{k},\mathbf{r})\bigl|_{q|z|\gg1}\Bigr]^{-1}-\Bigl[{\cal P}_{\text{IP}}\overline{\overline{\mathbf{S}'}}(\delta_{2},\mathbf{k},\mathbf{r})\bigl|_{q|z|\gg1}\Bigr]^{-1}\biggr\}^{-1}.\nonumber 
\end{eqnarray}
\end{widetext}

To demonstrate that a 2D finite atomic array can exhibit the bulk
properties, we use Eq.~(\ref{eq:S61}) to extract and estimate the
bulk Hamiltonian from the scattering problem of a 2D finite atomic
array. In FIG.~\ref{fig:S6}a,b, we consider the scattering problems
of 2D finite square and rectangular atomic lattices with rectangle-shaped
open boundaries shined by an oblique incident Gaussian beam towards
the positive $z$ direction. We calculate the ratio between the scattered
field and the incident field at $z>0$ region at the intersection
between the central line of incident field and $z=5a$ plane in the
far-field region~($q|z|=10\pi a/\lambda=5\pi\gg1$) to discard the
evanescent wave corresponding to the subradiant modes. The reason
we choose the fields at the same point is to eliminate the traveling
phase in $\overline{\overline{\mathbf{S}}}(\delta,\mathbf{r}-\mathbf{r}')$.
By treating the far-field part of $\overline{\overline{\mathbf{S}}}(\delta,\mathbf{r})$
obtained from the finite-size simulation as the far-field part of
$\overline{\overline{\mathbf{S}}}(\delta,\mathbf{k},\mathbf{r})$
in the infinite system, where $\mathbf{k}$ corresponds to the same
oblique incident angle, we are able to use Eq.~(\ref{eq:S61}) to
extract and estimate the bulk Hamiltonian ${\cal H}{}_{\text{eff}}(\mathbf{k}_{\parallel})={\cal P}_{\text{IP}}{\cal H}{}_{\text{eff}}^{\text{3D}}(\mathbf{k}_{\parallel})$
since both $\overline{\overline{\mathbf{S}}}(\delta,\mathbf{r})$
and $\overline{\overline{\mathbf{S}'}}(\delta,\mathbf{k},\mathbf{r})$
are $3\times3$ matrices and describe the relation between the scattered
and incident field. Note that the in-plane momentum here always lies
within the light cone since the in-plane momentum provided by a resonant
light is bounded by the magnitude of wave vector $q=2\pi/\lambda$.

The main reason we can use Eq.~(\ref{eq:S61}) to extract and estimate
the bulk Hamiltonian from the scattering matrices is that the dyadic
Green's function $\overline{\overline{\mathbf{g}}}_{\text{sc}}(\mathbf{k},\mathbf{r})$
is approximately detuning independent when the detuning $\delta$
is not too large compared to the resonant frequency $\omega_{0}=cq\gg|\delta|$~(near
resonance). Once $\overline{\overline{\mathbf{g}}}_{\text{sc}}(\mathbf{k},\mathbf{r})$
depends on the detuning $\delta$, we are not able to easily eliminate
the $\overline{\overline{\mathbf{g}}}_{\text{sc}}(\mathbf{k},\mathbf{r})$
part in the scattering matrix to extract the effective Hamiltonian
${\cal H}{}_{\text{eff}}(\mathbf{k}_{\parallel})={\cal P}_{\text{IP}}{\cal H}{}_{\text{eff}}^{\text{3D}}(\mathbf{k}_{\parallel})$.
However, the collective optical response in Eq.~(\ref{eq:S58}) shows
that the atomic array becomes transparent to the incoming light at
large detuning since the dispersion shifts in the left panels of FIG.~\ref{fig:S6}c-f
are small. That is, we conclude that Eq.~(\ref{eq:S61}) works well
for the 2D atomic array with photon-mediated dipole-dipole interaction.
We note that the scattering matrices in Eq.~(\ref{eq:S61}) can be
obtained in both the transmitted and reflected fields.

\subsubsection{Non-Hermitian topological invariant: vorticity}

With a two-band effective Hamiltonian, we can directly calculate the
non-Hermitian topological invariant of its non-Hermitian degeneracy
point, called the vorticity or spectral winding~\cite{Leykam2017S,Gong2018S,Shen2018S,Su2021S}
\begin{equation}
v=-\oint_{C}\frac{d\mathbf{k}}{2\pi}\cdot\nabla_{\mathbf{k}}\text{arg}[E_{1}(\mathbf{k})-E_{2}(\mathbf{k})],\label{eq:S62}
\end{equation}
where $C$ is a closed loop~(counterclockwise) in the 2D Brillouin
zone and spectral phase $\text{arg}[E_{1}(\mathbf{k})-E_{2}(\mathbf{k})]$
is well-defined except at non-Hermitian degeneracy point. In the right
panels of FIG.~\ref{fig:S6}f, there is a $-\pi$ spectral phase
acquired around the loop $C$ enclosed the EP associated with a discontinuity
at the bulk Fermi arc and a zero spectral phase at imaginary bulk
Fermi arc, which manifests the existence of EP. This nonzero spectral
phase change around the loop $C$ enclosed the EP also reflects that
EP is a branch point singularity, and such a singular point as some
point $\mathbf{k}$ means that $|E_{1}(\mathbf{k})-E_{2}(\mathbf{k})|$
is zero, i.e., a degeneracy point. We note that here the dispersive
bulk Fermi arc is not truncated by the light cone since the change
in the lattice constant ratio $\eta$ is small. Comparing the spectral
phases of atomic lattices with $\eta=1$, $1.05$, and $1.1$~(in
the main text), we can see the way EPs and dispersive Fermi arcs evolve
as $\eta$ changes.

In FIG.~\ref{fig:S6}c,d, both the eigenenergy spectrum and the spectral
phase deviate from those in an infinite system at large oblique incidence
due to the diffraction from the edge. At a small incident angle, our
simulation results show that a 2D finite atomic array exhibits its
bulk properties and behaves like a mirror at cooperative resonance
when the incident light shines in the vicinity of the center of arrays.
We also check that smaller 2D finite atomic arrays~($14\times14$)
show similar spectral phases as FIG.~\ref{fig:S6}c,e.

Spectral phases in FIG.~\ref{fig:S6}c,e also help us directly identify
the vorticity of each non-Hermitian degeneracy point in square and
rectangular lattices. According to the mirror symmetries with respect
to the $k_{x}$ and $k_{y}$ axes, we can construct the full spectral
phases over Brillouin zone within the light cone, and the vorticity
corresponding to NDP at the origin in the right panel of FIG.~\ref{fig:S6}c
is zero. In contrast, the vorticity of the EP in the first quadrant
of Brillouin zone in FIG.~\ref{fig:S6}e is $1/2$, and the vorticity
of other mirror reflected EPs are $\pm1/2$ according to the required
number of mirror reflections with respect to the EP in the first quadrant.

\begin{figure*}
\centering{}\includegraphics{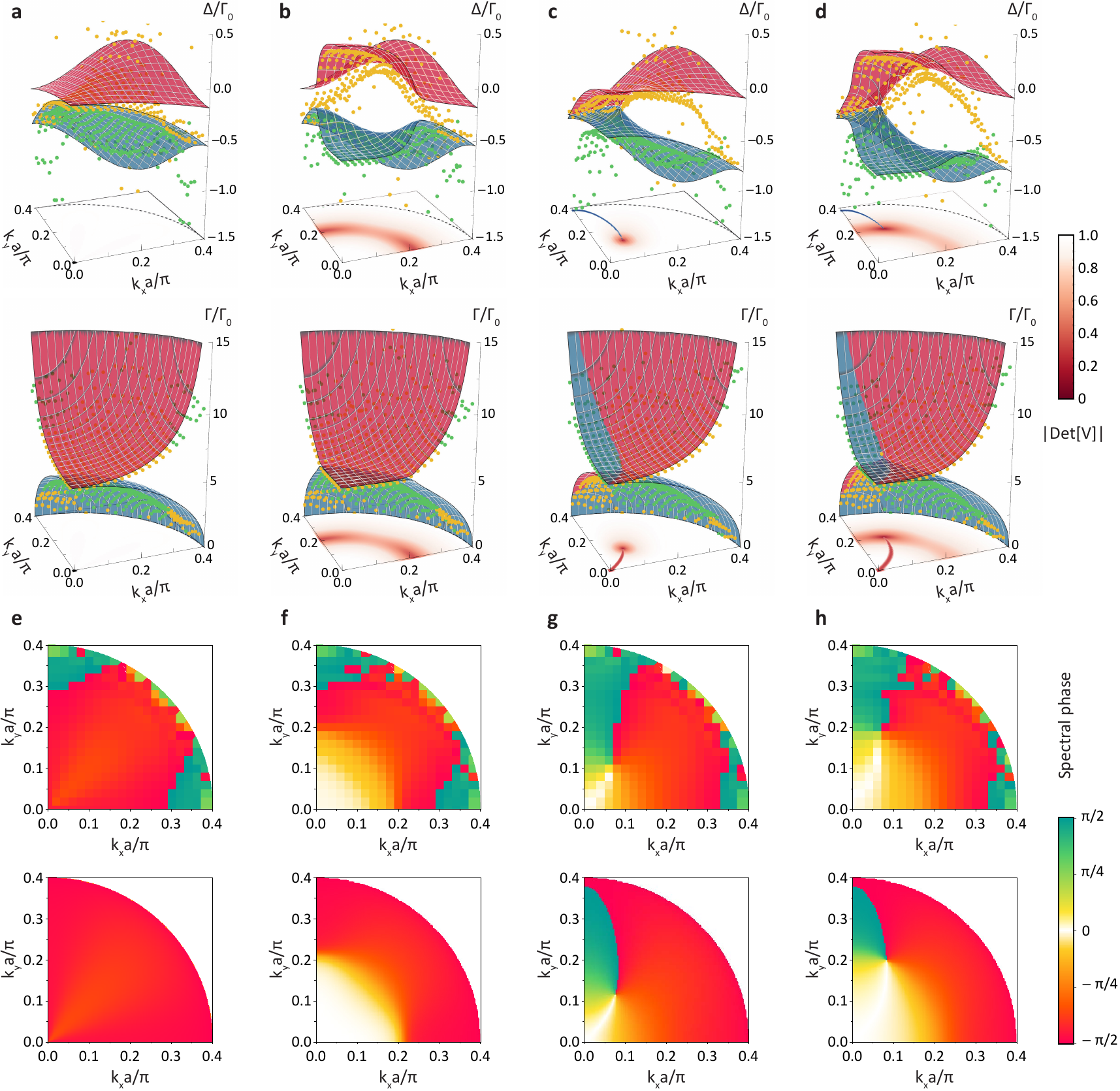}\caption{\label{fig:S7}\textbf{Topologically stable exceptional points. a}-\textbf{d},~Collective
frequency shift $\Delta_{\mathbf{k}}$~(top) and overall decay rate
$\Gamma_{\mathbf{k}}$~(bottom) of infinite square~(\textbf{a},\textbf{b},
$\eta=1$) and rectangular~(\textbf{c},\textbf{d}, $\eta=1.05$)
lattices with in-plane polarization within the light cone~(black
dashed circle). Two bulk energy bands $E_{1,2}(\mathbf{k})=\hbar(\omega_{0}+\Delta_{\mathbf{k}})-\frac{i}{2}\hbar\Gamma_{\mathbf{k}}$
obtained from ${\cal H}_{\text{eff}}(\mathbf{k}_{\parallel})$ are
colored in red and blue, and those extracted from the simulations
in finite systems are denoted by yellow and green dots. By applying
an out-of-plane magnetic field $B\mathbf{e}_{z}$ with $\mu B=0.5\Gamma_{0}$
in \textbf{b},\textbf{d}, NDP in \textbf{a} is lifted by a $2\hbar\mu|B|$
shift in its energy difference between two bands. However, EPs in
\textbf{c} still exist and are shifted within the light cone. \textbf{e}-\textbf{h},~Spectral
phases $\text{Arg}[E_{1}(\mathbf{k})-E_{2}(\mathbf{k})]$ extracted
from the simulations in finite systems~(top) and obtained from bulk
band structures~(bottom) in atomic lattices with the same lattice
configuration and external magnetic field in \textbf{a}-\textbf{d}.
All simulation parameters in \textbf{a}-\textbf{h}, except for the
magnetic fields, are the same as FIG.~\ref{fig:S6}.}
\end{figure*}

For the stability of these non-Hermitian degeneracy points, we can
consider a nonzero out-of-plane magnetic field $B\mathbf{e}_{z}$
to see that NDP would be lifted~(FIG.~\ref{fig:S7}a,b) due to the
broken Hermitian-conjugated time-reversal symmetry~(Eq.~\ref{eq:33})
while EPs still exist~(FIG.~\ref{fig:S7}c,d). In FIG.~\ref{fig:S7}b,
only the real part of bulk eigenenergies are lifted since the degeneracy
is determined by the discriminant $\hbar\sqrt{\mu^{2}B^{2}+\Gamma_{0}^{2}\kappa_{+-}(\mathbf{k})\kappa_{-+}(\mathbf{k})}$
of the effective Hamiltonian, where magnetic field results in a $2\hbar\mu|B|$
difference in two bulk eigenenergies. Since the real part of bulk
eigenenergies in the top panel of FIG.~\ref{fig:S7}b are two non-intersecting
bands, there is no other non-Hermitian degeneracy point within the
light cone. The topological stability of EP against a weak real-valued
$\hbar\mu B\sigma_{z}$ term arises from a nontrivial vorticity~(FIG.~\ref{fig:S7}g,h).
That is, unless $\hbar\mu B\sigma_{z}$ modifies the range of the
discriminant of the effective Hamiltonian significantly, the positions
of EPs within the light cone could be shifted, while EPs still persist.
